\documentclass[twoside,twocolumn,english,aps,prb
]{revtex4}
\usepackage{amsmath}
\usepackage{graphicx}
\usepackage{amssymb}

\makeatletter
\usepackage{graphicx}
\usepackage{amsmath}

\usepackage{color}
\usepackage{babel}
\makeatother

 \newcommand{\Eq}[1]{Eq.~(\ref{#1})} 
\newcommand{\Eqs}[1]{Eqs.~(\ref{#1})}

 \newcommand{\EqII}[1]{Eq.~(II.\ref{#1})} 
 \newcommand{\EqsII}[1]{Eqs.~(II.\ref{#1})}

\newcommand{\toh}{{\textstyle \frac{1}{2}}}

\newcommand{\eg}{\emph{e.g.}} 
\newcommand{\ie}{\emph{i.e.}}
\newcommand{\ve}{\varepsilon}

\newcommand{\bC}{\bar C}

\newcommand{\tCO}{{\tilde C^0}}
\newcommand{\tDO}{{\tilde D^0}}
\newcommand{\bP}{{\bar P}}
\newcommand{\tP}{{\tilde P}}
\newcommand{\cP}{{{\cal P}}}

\newcommand{\cF}{{{\cal F}}}
\newcommand{\ctF}{{\tilde {\cal F}}}
\newcommand{\qb}{{\bar{q}}}
\newcommand{\kb}{{\bar{k}}}
\newcommand{\hhbar}{{\hbar}}

\newcommand{\shbar}{{/\hbar}}
\newcommand{\oneoverhbar}{{1 \over \hbar}}
\newcommand{\smalloneoverhbar}{{\textstyle {1 \over \hbar}}}
\newcommand{\oneoverhbarsq}{{1 \over \hbar^2}}
\newcommand{\staroverhbar}[1]{{{#1} \over \hbar}}
\newcommand{\staroverhbarsq}[1]{{{#1} \over \hbar^2}}
\newcommand{\hbaroverstar}[1]{{ \hbar \over {#1} }}
\newcommand{\wb}{\bar{\omega}}
\newcommand{\wbu}{\bar{\omega}_{\rm u}}

\newcommand{\cW}{{\mathcal{W}}}
\newcommand{\tC}{{\tilde C}}
\newcommand{\LL}{\mathcal{L}}
\newcommand{\class}{\textrm{cl}}
\newcommand{\eff}{\textrm{eff}}
\newcommand{\AAK}{\textrm{AAK}}
\newcommand{\sAAK}{{\scriptscriptstyle {\rm AAK}}}
\newcommand{\sAAG}{{\scriptscriptstyle {\rm AAG}}}

\newcommand{\qqph}{\qquad \phantom{.}}
\newcommand{\qph}{\quad \phantom{.}}

\newcommand{\urw}{\textrm{urw}}
\newcommand{\sqn}{\textrm{sqn}}
\newcommand{\crw}{\textrm{crw}}
\newcommand{\rw}{\textrm{rw}}
\newcommand{\deco}{\textrm{dec}}
\newcommand{\tauphi}{{\tau_\varphi}}
\newcommand{\WL}{{\rm WL}}

\newcommand{\pp}{\textrm{pp}}
\newcommand{\app}{{\overline { \hspace{-0.1mm} \pp \hspace{-0.3mm}} 
\hspace{0.4mm} }}

\newcommand{\tildet}{\widetilde t}

\newcommand{\GZ}{\textrm{GZ}}

\newcommand{\tauH}{{\tau_H}}

\newcommand{\gammaphi}{{\gamma_\varphi}}
\newcommand{\gammaH}{{\gamma_H}}

\newcommand{\tauel}{{\tau_\textrm{el}}}

\newcommand{\kF}{k_\textrm{F}}

\newcommand{\vF}{v_\textrm{F}}

\newcommand{\Sec}[1]{Sec.~\ref{#1}}        
\newcommand{\App}[1]{App.~\ref{#1}}        
\newcommand{\Ref}[1]{Ref.~[\onlinecite{#1}]}  

\newcommand{\bmq}{{\bm{q}}}

\newcommand{\G}{{\tilde G}}

\newcommand{\bcL}{\bar {\cal L}{}}

\newcommand{\ttau}{{\widetilde \tau}}

\newcommand{\tcC}{{\tilde {\cal C}}}

\newcommand{\bcC}{{\overline {\cal C}}{}}

\newcommand{\bomega}{{\bar \omega}}

\newcommand{\bbmq}{{\bar \bmq}}

\newcommand{\bq}{{\bar q}}

\newcommand{\gammaone}{{\gamma_1}}

\newcommand{\weightingFunction}{{spectrum}}

\begin{document}

\include{labelsII}
%
%

\title{Decoherence in weak localization I: Pauli principle in 
influence functional}


\author{
Florian Marquardt,${}^1$ Jan von Delft,${}^1$ 
R. A.   Smith,${}^2$ Vinay Ambegaokar${}^3$} \affiliation{$^1$ 
  Physics Department, Arnold Sommerfeld Center for Theoretical
  Physics, and Center for NanoScience,
  Ludwig-Maximilians-Universit\"at M\"unchen, 80333 M\"unchen,
  Germany\\
  $^2$School of Physics and Astronomy, University of Birmingham,
  Edgbaston, Birmingham B15 2TT, England\\
  $^3$Laboratory of Atomic and Solid State Physics Cornell University Ithaca, New York 14850, USA}

\date{October 21, 2005}

\begin{abstract} This is the first in a series of two papers (I and
  II), in which we revisit the problem of decoherence in weak
  localization. The basic challenge addressed in our work is to
  calculate the decoherence of electrons interacting with a
  quantum-mechanical environment, while taking proper account of the
  Pauli principle.  First, we review the usual influence functional
  approach valid for decoherence of electrons due to classical noise,
  showing along the way how the quantitative accuracy can be improved
  by properly averaging over closed (rather than unrestricted) random
  walks. We then use a heuristic approach to show how the Pauli
  principle may be incorporated into a path-integral description of
  decoherence in weak localization. This is accomplished by
  introducing an effective modification of the quantum noise spectrum,
  after which the calculation proceeds in analogy to the case of
  classical noise. Using this simple but efficient method, which is
  consistent with much more laborious diagrammatic calculations, we
  demonstrate how the Pauli principle serves to suppress the
  decohering effects of quantum fluctuations of the environment, and
  essentially confirm the classic result of Altshuler, Aronov and
  Khmelnitskii for the energy-averaged decoherence rate, which
  vanishes at zero temperature.  Going beyond that, we employ our
  method to calculate explicitly the leading quantum corrections to
  the classical decoherence rates, and to provide a detailed analysis
  of the energy-dependence of the decoherence rate.  The basic idea of
  our approach is general enough to be applicable to decoherence of
  degenerate Fermi systems in contexts other than weak localization as
  well. --- Paper II will provide a more rigorous diagrammatic basis
  for our results, by rederiving them from a Bethe-Salpeter equation
  for the Cooperon.

\end{abstract}
\maketitle

\section{Introduction}

The weak localization of electrons by coherent backscattering in a
disordered conductor, which manifests itself via a characteristic
contribution to the magnetoconductivity, is a unique, particulary
robust interference effect\cite{Abrahams79,AltshulerLee80,AltshulerLarkin82,Bergmann84,Lee85,CS,Kramer93}.  It is not suppressed by thermal
averaging, and the temperature dependence of the effect arises only
due to the destruction of quantum coherence by inelastic scattering
events, whose likelihood increases with rising temperature. The study
of decoherence, and in particular of the temperature dependence of the
decoherence rate $\gammaphi (T)$ governing the magnetoconductivity,
therefore plays a central role in this subject.

There are two features which make this problem nontrivial, related to
the influence of low- and high-frequency environmental fluctuations on
the propagating electron, respectively: On the one hand, the
environmental fluctuations at the \emph{lowest} frequencies do not contribute
to decoherence, since they are so slow that they resemble an elastic
impurity potential: for trajectories of duration $t$, environmental frequencies
$\bomega \lesssim 1/t$ do not contribute to decoherence.
 This fact is most easily accounted for in an
\emph{influence functional} or path-integral description in the
\emph{time domain}, which works well as long as the fluctuations are
classical.  On the other
hand, environmental modes with frequencies much \emph{higher} than the
temperature do not contribute either, since an (electron- or hole-like)
quasi-particle propagating with energy $\varepsilon \simeq T$ relative
to the Fermi surface does not have enough energy to excite them: Pauli
blocking forbids the quasi-particle to loose an energy $\bomega$
larger than $\simeq T$ to the environment.
This fact is obvious in the frequency domain, where Pauli factors such
as $f(\varepsilon)[1- f(\varepsilon - \bomega)]$ become explicit ($f$
being the Fermi function); hence, a proper treatment of high
frequencies is most easily achieved in a \emph{perturbative many-body
  calculation} in the \emph{frequency domain}, which allows for a
fully quantum-mechanical treatment of the environment.

Although the essential physics of both the low- and high-frequency
environmental modes is well understood, 
it is rather difficult to explicitly and accurately treat both regimes
on an equal footing within a single, unified framework.  On the one
hand, standard influence functional approaches usually do not
incorporate the Pauli principle explicitly (a notable exception being
the work of Golubev and Zaikin,\cite{GZ1,GZ2,GZ3,GZ4,GZ5,GZ6,GZ7}
which is, however,
controversial\cite{AAG,EH97,AV,KirkpatrickBelitz11,AAV,Imry02,FMSR,AleinerVavilov02,vonDelft03,vonDelft04} and whose results for
$\gammaphi (T)$ we disagree with). On the other hand, diagrammatic
approaches in the present context have difficulties in accurately
dealing with infrared divergencies, which are often simply cut off by
hand, with the cutoff being determined self-consistently (or else the
presence of an external cutoff is assumed\cite{AAG}, as provided by an
applied magnetic field).  In the present series of two papers (I and
II\cite{paperII}), we fill in the respective ``gaps'' in both the influence
functional approach (paper I) and the diagrammatic approach (paper
II), by showing how each can be extended to achieve an accurate,
explicit treatment of both low- and high-frequency modes.  The
resulting two approaches, though thoroughly different in style and
detail, yield the same result for the Cooperon decoherence rate
$\gammaphi (T)$, for which we evaluate both the leading and
next-to-leading terms (in an expansion in which the dimensionless
conductance is the small parameter). The leading terms coincide with
that found by Altshuler, Aronov and Khmelnitskii \cite{AAK} (AAK) for
decoherence due to the thermal part of Nyquist noise (which we shall
call classical white Nyquist noise below).  The next-to-leading terms
are checked against and found to be consistent with results for the
magnetoconductivity in large magnetic fields of Aleiner, Altshuler and
Gerzhenson \cite{AAG} (AAG).

Paper I is intended for a wide audience and will hopefully be
accessible to nonexperts. It presents a path integral analysis of
decoherence by \emph{quantum} Nyquist noise, achieving not only a
natural infrared cutoff, but also incorporating the Pauli principle in
a physically transparent way, by suitably modifying the interaction
propagator.  In particular, we offer an elementary but quantitatively
accurate explanation for why and how the Fermi function enters the
decoherence rate. As a byproduct of our analysis we show (i) how the
accuracy of the path-integral approach can be improved by performing
trajectory averages over \emph{closed} (as opposed to unrestricted)
random walks\cite{Montambaux04}, (ii) calculate the leading quantum
corrections to the classical results for the decoherence rate, and
(iii) are also able to analyse explicitly the energy dependence of the
decoherence rate.

We reach our goal by a series of steps, whose main arguments and
results are summarized concisely in
Section~\ref{sec:overviewofresults} in a type of overview for the
benefit of readers not interested in the details of the derivations.

The price paid for our avoidance of a large formal apparatus in favor
of simple, transparent arguments is that paper I does not entirely
stand on its own feet: its discussion of Pauli blocking relies in part
on heuristic arguments and/or influence functional results derived
elsewhere\cite{GZ2,vonDelft04}.  In paper II\cite{paperII}, 
addressed to experts,
we aim to put the heuristic arguments of paper I on a solid
footing, by rederiving the main results for the Cooperon propagation
in a completely different manner, using purely diagrammatic means. To
this end, we use Keldysh perturbation theory to set up a
Bethe-Salpeter equation for the Cooperon, which includes both
self-energy and vertex contributions to the Cooperon self-energy and
whose leading terms are free from both ultraviolet and infrared
divergencies.  This equation is then converted to the time domain and
solved approximately with an exponential Ansatz, which, remarkably,
turns out to reproduce the results of the paper I.

Our work is built on a foundation laid over many years by many
different authors.  The influence of classical (purely thermal)
white Nyquist 
noise was first studied in the seminal work of AAK \cite{AAK},
 where they derived a
path-integral description and were able to solve exactly the
quasi 1-dimensional case. Chakravarty and Schmid elaborated on this
approach in their review\cite{CS}, which also includes a detailed
discussion of electron-phonon scattering.  More recently, Voelker and
Kopietz\cite{VoelkerKopietz00} provided an alternative to
path-integration, an {}``Eikonal'' ansatz for the time-evolution of
the Cooperon, which also includes the correct infrared behaviour. 

All of these works, explicitly or implicitly, deal with the Pauli
principle by using a classical noise spectrum that is derived from the
physical quantum-mechanical spectrum by eliminating the possibility of
spontaneous emission into the bath (see our discussion in Section
\ref{sub:Previous-approaches}).  This prescription was consistent with
perturbative diagrammatic calculations (such as the calculation of the
inelastic electron scattering rate in Ref.~\onlinecite{Abrahams}), and
it was recently reconfirmed by AAG \cite{AAG} via a detailed
diagrammatic calculation of the short-time behaviour of the Cooperon,
to leading order in the interaction. An expansion in the quantum
corrections to the picture of decoherence by purely classical noise,
performed by Vavilov and Ambegaokar\cite{AV}, yielded similar results.

These recent studies\cite{AAG,AV} were motivated by and contributed to
a controversy that arose when Golubev and Zaikin (GZ)
claimed\cite{GZ1,GZ2,GZ3,GZ4,GZ5,GZ6,GZ7} to have demonstrated
theoretically that electron-electron interactions intrinsically cause
the decoherence rate $\gammaphi$ to saturate at a nonzero value at low
temperatures, and that this explains the saturation that has been
observed in some experiments\cite{Mohanty}.  In these papers, GZ
proposed a new, exact Feynman-Vernon influence functional for
electrons under the influence of an environment, which takes proper
account of the Pauli principle (as confirmed in
Ref.~\onlinecite{vonDelft04}).  However, the evaluation of this
influence functional is not straightforward, and the approximations
which GZ adopted to this end have been heavily
criticized,\cite{AAG,EH97,AV,KirkpatrickBelitz11,AAV,Imry02,FMSR,AleinerVavilov02,vonDelft03,vonDelft04}, in particular those
pertaining to the terms associated with Pauli blocking.  Very
recently, von Delft has shown\cite{vonDelft04} that if the Pauli
blocking terms are treated somewhat more carefully to include recoil
effects, GZ's approach actually does reproduce the celebrated results
of AAK for the decoherence rate $\gammaphi(T)$, which does not
saturate at low temperatures. The analysis of
Ref.~\onlinecite{vonDelft04} constitutes a formal counterpart to the
present paper I, in which we use partly heuristic arguments to reach
the same conclusions as Ref.~\onlinecite{vonDelft04} in a more
intuitive manner.

\section{Overview of Results}

\label{sec:overviewofresults}

Before embarking on a detailed calculation of the decoherence rate, we
present in this section an overview of the main results and arguments
contained in the present paper, and a short analysis of their various
strengths and weaknesses. It is hoped that the reader will thereby
gain a birdseye view of the problems that typically arise in
calculations of $\gammaphi$, a feeling for what is required to conquer
them, and a glimpse of the type of results obtained.

The weak localization contribution to the magnetoconductivity of a
quasi $d$-dimensional disordered conductor can be written in the
form\cite{AltshulerLarkin82} 
\begin{eqnarray} \delta\sigma_{d}^{\WL} & = &
  -\frac{\sigma_{d}}{\pi\nu\hbar}\int_{\tauel}^{\infty}dt\,
\tC(0,t)\;.
\label{eq:magnetoconductance}
\end{eqnarray}
 Here $\nu= m \kF/(2 \pi^2 \hbar^2)$ 
  is the 3-dimensional density of states per spin at the Fermi surface,
$\tauel$ the elastic scattering time,
  and $\sigma_{d}= 2e^{2}\nu_d D$ is the sample's classical Drude
  conductivity for $d=3$, the inverse square resistance for a $d=2$ film
  of thickness $a$, or the inverse resistance per length for a $d=1$
  wire of cross sectional area $a^{2}$,  with $\nu_d = a^{3-d}\, \nu$
being the effective density of states per spin of the corresponding
dimensionality, and $D= \vF^2 \tauel/3$ 
  the diffusion constant.  $\tC(r,t)$ denotes the
  Cooperon propagator, in the position-time representation, in the
  presence of interactions and a magnetic field.  For $r=0$ it gives the
  probability for an electron propagating along two time-reversed paths
  to return within time $t$ to the starting point without losing phase
  coherence, thus enhancing the backscattering probability and
  reducing the conductivity. In the absence of decoherence and a
  magnetic field it is given by the classical diffusion probability
  density (the {}``diffuson'').

For ease of reference, our notational conventions will mostly follow
those used in Ref. \onlinecite{vonDelft04}.  In particular, various
incarnations of the Cooperon propagator will occur below, related by
Fourier transformation, such as $\tilde C(r,t)$, $\bC_q (t)$, $\tcC
(r,\omega)$, and $\bcC_q (\omega)$, and related versions containing
more than one time or frequency arguments.  Our convention for
distinguishing them notationally, apart from displaying their
arguments, is to use a tilde or bar to distinguish between the
position and momentum representations, and a roman italic or
calligraphic symbol to distinguish between the time and frequency
representations. 

\subsection{Decay Function $F_d (t)$}
\label{sec:decayfunctionoverview}

When the effect of interactions on the full Cooperon $\tilde{C}(r=0,t)$
is calculated within the influence functional approach, one naturally
obtains results of the form 
\begin{eqnarray}
\tC(0,t)\simeq\tCO(0,t)\, e^{-F_d(t)}\;,\quad F_d( t)= 
\smalloneoverhbar \langle
S_{\eff}\rangle_{\textrm{rw}} \;.
\label{eq:firstCapprox}
\end{eqnarray}
Here $\tilde{C}^{0}$ is the bare Cooperon in the absence of
interactions, and $S_{\eff}$ is the so-called effective action. It is
essentially the variance of the fluctuating difference of phases
$\oneoverhbar S_{F}$ and $\oneoverhbar S_{B}$ acquired while
propagating along the two paths, $S_{{\rm eff}} = {1
  \over 2 \hhbar } \left\langle (S_{F}-S_{B})^{2}\right\rangle $.
In the case considered here (linear coupling to Gaussian fluctuations), it is linear in the noise correlator (interaction propagator) and
characterizes the effect of the environment on a pair of time-reversed
trajectories whose interference gives rise to weak localization. Its
average over all random walk trajectories yields the {}``decay
function'' $F_d(t)$, describing the suppression of the Cooperon by
decoherence.  Hence the decoherence time $\tauphi=1/\gammaphi$ can be
defined by the condition\cite{thresholdconstant} $F_d (\tauphi) = 1$.

The decay function $F_d(t)$ turns out [\Sec{eq:generalF}] to be of the
following general form, for trajectories propagating during the
time-interval $[-t/2 , t/2]$ [see Eq.~(\ref{exponentCL})]:
\begin{eqnarray}
F_d(t) & = & \oneoverhbarsq \int_{-t/2}^{t/2}dt_{3}\int_{-t/2}^{t/2}dt_{4}\,
\nonumber \\
 &  & \phantom{.} \hspace{-4mm}
\times \! \int(d\bar{q})
\left[\bP (\qb,|t_{3}-t_{4}|)-\bP (\qb,|t_{3}+t_{4}|)\right] \nonumber \\ 
 &  & \phantom{.} \hspace{-4mm}
\times \! \int(d\wb)\, e^{-i\wb(t_{3}-t_{4})}
\bigl \langle VV \bigr \rangle^\eff_{\qb\wb}
\label{eq:Fgeneral}
\end{eqnarray}
It contains one time-integral for each of the two interfering
trajectories.  Besides, it is a product of a part describing the
diffusive dynamics of the system under consideration (the second line)
and the noise correlator of the effective environment (third
line), which is integrated over all momentum and frequency transfers
$\qb,\wb$. In our notation, 
$\bP (\qb,t')$ is the Fourier
transform of the probability density $\tP (r',t')$ for a random walk
to cover the distance $r'$ in the time $t'$.

 The fact that the second line of Eq. (\ref{eq:Fgeneral}) contains
  a difference between two rather similar expressions reflects the
  fact that the phases picked up along the two trajectories are
  related for fluctuations with sufficiently long wavelengths and/or
  low frequencies, and ensures that such fluctuations do not
  contribute to decoherence.  The $|t_3 - t_4|$ and $|t_3 + t_4|$
  terms correspond to the ``self-energy terms'' and ``vertex
  corrections'' of diagrammatic calcultions of the Cooperon
  self-energy, in which the vertex terms are needed to cancel infrared
  divergences of frequency or momentum integrals.  The simple and
  natural way in which this cancellation arises in the influence
  functional approach is one of the latter's main advantages (the
  other being its physical transparency).

 To calculate $\tilde P(r',t')$, previous works \cite{CS,GZ2} 
have usually averaged over {\bf u}nrestricted {\bf r}andom {\bf w}alks
(urw)   that are not constrained to return to the origin, ignoring the fact
  that all paths contributing to weak localization are closed. We
   show in Sections \ref{sec:rrw-vs-urw} how 
$\tilde P(t',t')$ and its Fourier transform $\bar P (\qb,t')$ 
may be calculated for {\bf c}\emph{losed} {\bf r}andom {\bf
  w}alks (crw) instead of unrestricted ones [Eqs. (\ref{eq:prrw}) and
(\ref{eq:rrw-result-explicit})], and in Section \ref{eq:generalF} how
the resulting more complicated decay functions may be evaluated. For
$d=1$ (but not for $d=2,3$), this improvement leads to a more accurate
result for the numerical prefactor occuring in the decoherence time.
%
[cf. \Eq{eq:finalFrrw}]. The extent of the improvement obtained
with the more accurate result, which is important for quantitative
comparisons with experiment, is checked in Section~\ref{sec:1Dcompare}
by using it to calculate the magnetoconductivity for quasi
1-dimensional conductors with classical white Nyquist noise, and
comparing the result to the celebrated exact {}``Airy function
expression'' of AAK \cite{AAK}. [Using closed random walks also 
turns out to be an essential prerequisite for recovering the results
of AAG from our theory, Sec.~\ref{sec:AAGcompare1}].

The difference between averaging over unrestricted versus closed
random walks can quite generally be summarized by the following
formulas, found in Section~\ref{sec:rrw-vs-urw} (and confirmed in
paper II):
\begin{subequations}
    \label{eq:Frrwurw}
\begin{eqnarray}
  F_d^\crw (t) & = & - {\tilde C^{1} (r=0, t) \over 
    \tilde C^{0} (r=0, t)}  \; , 
    \label{eq:Frrdefine}
\\
\label{eq:Furdefine}
  F_d^\urw (t) & = & - \bar C_{q = \gammaH = 0}^{1} (t) \; .
\end{eqnarray}
\end{subequations} 
Here $\tilde{C}^{1}(r,t)$ is the first order term in an expansion of
the full Cooperon $\tilde{C}(r,t)$ in powers of the interaction, and
$\bar{C}_{q,\gammaH=0}^{1}(t)$ is its momentum Fourier transform in
the absence of a magnetic field. We note that in both cases,
\Eq{eq:firstCapprox} represents the full Cooperon simply as a
reexponentiated version of the first order term (either in momentum or
real space), but since the decay of the real-space Cooperon is
required, the expansion is consistent to leading order in the
interaction only if $F_d^{\crw}(t)$ is used, which is why the
latter gives more accurate results.


  In our paper, we successively present different versions of
  \Eq{eq:Fgeneral}, which are distinct in the effective noise
  correlator $\left\langle VV\right\rangle^\eff_{\qb\wb}$ of the
  environment (classical noise, quantum noise for single particle, or
  quantum noise for many-body situation with Pauli principle). 
They will all, however, be associated with some
  type of Nyquist noise, and factorize as $\smalloneoverhbar \left
    \langle V V \right\rangle^\eff_{\qb\wb} = {\cW_\eff (\wb) \over
    \nu D \bq^2 }$, where $\cW_\eff (\wb)$ will be called
the ``noise spectrum''. This will allow us [with some standard
  approximations, including an average over unrestricted random walks
  (urw)], to reduce \Eq{eq:Fgeneral} to the form
\begin{eqnarray}
  \label{eq:introsimpleFdurw}
  F_{d, \urw} (t) & \simeq & 
  p_d \,  t \int_0^\infty \!\! {d\bar{\omega} } \, 
  {\cW_\eff (\wb) \over  \wb^{2-d/2} } 
  \left[1  - {\sin (\wb t) 
      \over \wb t} \right]    \qqph 
\end{eqnarray}
[the  $p_d$ are given after \Eq{eq:simpleFdurw}].
Note the presence of the infrared cutoff at $\wb \simeq 1/t$
that was mentioned above.

\subsection{Classical white Nyquist noise}
\label{sec:overviewclassical}

If we consider a classical fluctuating potential $V(x,t)$ acting on
the electron (Sec. \ref{sec:Cooperon-decay-for}), then $\left\langle
  VV\right\rangle^\eff_{\qb\wb}$ is given by its symmetric noise
correlator $\langle VV\rangle _{\qb\wb}^\class$. This was used in the
seminal paper of Altshuler, Aronov and Khmelnitskii\cite{AAK}, where
they applied the classical fluctuation-dissipation theorem to obtain
the thermal part of the Nyquist noise, for which $\cW_\eff (\wb) $ is
simply given by the classical noise spectrum, $\cW_\class (\wb) = T$,
leading to a decoherence rate that vanishes at $T=0$ (see also the
semiclassical path integral analysis of Chakravarty and
Schmid\cite{CS} and Stern, Aharonov and Imry\cite{ImryStern}). 

For example, for a
quasi 1-dimensional disordered wire, AAK
found\cite{factorof2,temp=freq}
\begin{eqnarray}
{\frac{1}{\tau_{\varphi,1}^\sAAK}}=\gamma_{\varphi,1}^\sAAK =
\left( T \sqrt \gammaone   \right)^{2/3}
= {T \over g_1 ( L_{\varphi,1}^\sAAK )} \; . 
\qqph
\label{eq:tauhphiAAK}
\end{eqnarray}
Here $g_d (L_\varphi)$ is the dimensionless conductance for a quasi
$d$-dimensional sample at the decoherence length $L_\varphi = \sqrt{ D
  \tauphi}$, given by
\begin{eqnarray}
    g_d (L_\varphi) = {\hbar \sigma_d \over e^2  L_\varphi^{2-d}}
 = \left\{ \begin{array}{cl} \hspace{-1mm}
(\gamma_1 \tauphi)^{-1/2}  ,  &
\gamma_1 = D (e^2 /\hbar \sigma_1)^2 
 , \rule[-3mm]{0mm}{0mm} \\
        g_2 ,   & g_2 = \hbar \sigma_2/e^2 , 
 \rule[-3mm]{0mm}{0mm} \\
(\gamma_3 \tauphi)^{1/2}  ,  & 
\gamma_3 = D (\hbar \sigma_3 / e^2)^2 ,
      \end{array}  \hspace{-0.5cm} \right.
\nonumber \\
    \label{eq:dimensionlessconductance}
  \end{eqnarray}
  for $d=1,2,3$, respectively. $g_d (L_\varphi)$ conveniently lumps
  together all relevant material parameters into a single
  dimensionless quantity, and will be used extensively below.  Since
  good conductors are characterized by having a large dimensionless
  conductance, $g_d(L_\varphi) \gg 1$ [see \Eq{eq:sigma(H)rescaled}
  below], the last equality in \Eq{eq:tauhphiAAK} implies $ T \tauphi
  \gg 1$. This means that for paths of duration $t \simeq \tauphi$, we
  have the inequality $t T \gg 1$, which will be important below.

  As expected, we recover \Eq{eq:tauhphiAAK} when applying our
  influence functional approach to a single particle under the
  influence of classical white Nyquist noise in quasi-1 dimension:
  upon replacing $\cW_\eff (\wb)$ by $\cW_\class (\wb) = T$ in
  \Eq{eq:introsimpleFdurw}, we find 
  \begin{eqnarray}
    F^\class_{1, \urw} (t) & = & c_1^\urw
    (t/\tau_{\varphi,1}^\sAAK)^{3/2}  = 
    (t/\tau_{\varphi,1})^{3/2}  \; , \qquad 
    \label{eq:F-IF-classicalnoiseresult}
  \end{eqnarray}
  with $c_1^\urw = {4 \over 3 \sqrt \pi}$, thus the decay function is
  governed by the same decoherence time $\tau_{\varphi,1}^\sAAK$ as
  obtained by AAK.  For the second equality, we used our convention of
  defining the decoherence time via $F_d (\tau_{\varphi , d}) = 1$, to
  absorb the numerical prefactor into the decoherence time itself,
  yielding $\tau_{\varphi,1} = \tau_{\varphi,1}^\AAK
  (c_1^\urw)^{-2/3}$.  If the calculation is done for closed random
  walks, the result is the same, except that the prefactor changes to
  $c_1^\crw = \sqrt {\pi} / 4$.

\subsection{Quantum noise}
\label{sec:quantumnoiseoverview}

The case of a fully quantum-mechanical environment is more involved.
If one considers the motion of a \emph{{\bf s}ingle} electron in the
presence of {\bf q}uantum {\bf n}oise (sqn) but in absence of a Fermi
sea, one can apply the standard Feynman-Vernon influence functional
approach\cite{FeynmanHibbs64} (Sec.
\ref{sec:DecayQuantumNoiseWithoutPauli}).  In this way, one obtains
Eq. (\ref{eq:Fgeneral}) with $\langle VV \rangle^\eff_{\qb\wb}$
replaced by the \emph{symmetrized} quantum noise correlator $\langle
VV \rangle^\sqn_{\qb\wb} = \frac{1}{2}\langle \{\hat{V},\hat{V}\}
\rangle_{\qb\wb}$.  By detailed balance, this always includes a factor
$\coth (\wb/2T) = 2 n (\wb) + 1$, ($n$ being the Bose function) and
the resulting effective spectrum turns out to be $\cW_\sqn (\wb) =
\toh \wb \bigl[ 2 n(\wb) + 1 \bigr]$, which describes both thermal
$(2n)$ and quantum (+1) fluctuations of the environment.  (In this
paper, temperature is measured in units of frequency, \ie\ $T$ stands
for $k_{\rm B} T / \hbar$ throughout.)  Physically, the quantum
fluctuations incorporate the decoherence due to spontaneous emission
into the environment, which is possible even for an environment at
$T=0$, if the single electron has a finite energy (in a metal, its
energy is typically near $\ve_{\rm F}$).  For such a single electron,
quantum fluctuations thus lead to a finite $T=0$ decoherence rate (but
for a physical model which is distinct from the original one
describing disordered metals, which involve many electrons). The
result for $F_{d, \urw} (t)$ obtained for this type of noise turns out
to coincides with the one obtained by Golubev and Zaikin\cite{GZ3},
if, following them, we introduce an upper frequency cutoff by hand (to
prevent an ultraviolet divergence), and take it to be the elastic
scattering rate.

However, diagrammatic calculations\cite{Abrahams,AAG} {[}summarized
in paper II, Section~\ref{sec:standardDyson}, see
\Eq{eq:gamma0selffirstexp}{]} indicate that the presence of other
electrons cannot be neglected, since the Pauli principle plays an
important role in preserving the coherence of low-lying excitations in
degenerate Fermi systems. Setting up a Dyson equation for the Cooperon
in the momentum-frequency representation and extracting from the
Cooperon self-energy the decoherence rate
$\gamma_{\varphi}^{\varepsilon}$ for an electron with
energy\cite{temp=freq} $\ve$ relative $\ve_F$, one obtains a rate
 where the factor $2n(\wb)+1$ 
is effectively replaced by
\begin{equation}
\label{eq:2n1f-f}
  2n(\wb)+1+f(\ve+\wb)-f(\ve-\wb)\,.\end{equation}
In the literature, this factor often occurs in the form of the
combination {}``$\coth(\wb/2T)+\tanh[(\ve-\wb)/2T]$''.   
The Fermi functions ensure
that processes which would violate the Pauli principle ($\wb\gg{\rm
  max}[T,\ve]$) do not occur, which turns out to eliminate the
ultraviolet divergence mentioned above. However, in contrast to the
influence functional approach it is rather difficult to properly
include vertex corrections in the diagrammatic approach. In fact,
Fukuyama and Abrahams\cite{Abrahams} introduced a low-frequency cutoff
$1/\tau_{\varphi}$ by hand, which then has to be determined
self-consistently. The neglect of vertex corrections also means that
the decay function is always linear in $t$, whereas e.g. in quasi-1D
the classical result is known to grow like $t^{3/2}$ (as emphasized by
GZ in Ref.~\onlinecite{GZ3}). What is needed, evidently, is an
expression for the decay function that keeps both the vertex
corrections and the Pauli principle (and thus is free from infrared
and ultraviolet divergencies, respectively). This is the main goal of
both papers I and II.

In the present paper, we address the question of how to incorporate
the Pauli principle in an influence functional approach. First we
provide a heuristic discussion of decoherence in the presence of a
Fermi sea (Sec. \ref{sec:goldenrule}). If an initial perturbation
creates a coherent superposition between two single-particle states
$\lambda$ and $\lambda'$, the decoherence rate (within a Golden
Rule calculation, without vertex corrections) is given by a sum of
a particle and a hole-scattering rate [see Eq.~(\ref{eq:gammaphi})]:
\begin{equation}
 \Gamma_{\varphi}(\lambda,\lambda')=
\frac{1}{2}[\Gamma_{e}(\lambda)+\Gamma_{h}(\lambda')+
\Gamma_{e}(\lambda')+\Gamma_{h}(\lambda)] \; .
\label{eq:gammaphiFromPandH}\end{equation}
Inserting the usual scattering rates containing Fermi functions
for Pauli blocking, one realizes that incorporating the Pauli principle
effectively means replacing the symmetrized quantum noise correlator
by the following combination (see Eq. (\ref{eq:replacePPfirst})):
\begin{eqnarray}
\label{eq:replacePPfirstPreview}
\left\langle VV\right\rangle^\pp_{\qb\wb} & = &
\toh\bigl\langle\{\hat{V},\hat{V}\}\bigr\rangle_{\wb}+
\\ & & \qquad \nonumber
\bigl[f(\ve+\wb)-f(\ve-\wb)\bigr]\toh\bigl\langle
[\hat{V},\hat{V}]\bigr\rangle_{\wb}\,,
\end{eqnarray}
where we took the energies of the two relevant states to be
nearly identical\cite{wneq0},
 as they will be in a calculation of the zero-frequency
conductivity. This formula corresponds to substituting for
  $\cW_\eff (\wb)$ a ``Pauli-principle-modified'' \weightingFunction\
  $\cW_\pp (\wb)$, given by $\toh \wb$ times \Eq{eq:2n1f-f}, and is
consistent with the results obtained diagrammatically, e.g. by
Fukuyama and Abrahams\cite{Abrahams}.  We then discuss the
consequences of this comparatively simple prescription for adding the
Pauli principle to an influence functional,  and use
it to calculate the energy dependence of $\gamma_{\varphi}$. 

Sec.~\ref{sec:GZcomparison} of the present paper I and all of paper II
are devoted to a justification of this prescription from more rigorous
approaches.  In Sec.~\ref{sec:GZcomparison} we demonstrate that
our heuristic prescription yields a result that is equivalent to that
recently obtained by one of us (JvD) by an analysis\cite{vonDelft04}
based on Golubev and Zaikin's exact influence functional for Fermi
systems\cite{GZ1,GZ2,GZ3,GZ4}.  Their expression for the effective
action $ S_{\eff}$ contains Fermi functions that correctly represent
Pauli blocking. Indeed, it was pointed out \cite{vonDelft04} that
AAK's expressions for the decoherence rate can be recovered from this
approach, by considering the action in momentum-frequency
representation and properly keeping \emph{recoil} effects, i.e.\ the
energy change $\ve\rightarrow\ve\mp\bomega$ that occurs each time the
electron emits or absorbes a noise quantum.

In paper II, we shall show how a diagrammatic
analysis based on an approximate solution of the full Bethe-Salpeter
equation for the Cooperon (including vertex corrections) 
leads to the same results as here.

\subsection{Results for decay function $F^\pp_d(t)$}
\label{sec:mainresultsF(t)}
 
The main novel results of our paper are contained in
\Sec{sec:ResultsDetailsF(t)}, where the decay function is evaluated
explicitly for the case of quantum Nyquist noise for an electron
moving in a Fermi sea of other electrons at thermal equilibrium. Using
\Eq{eq:Fgeneral} with the modified quantum noise correlator
$\left\langle VV\right\rangle^\pp_{\qb\wb}$ of
\Eq{eq:replacePPfirstPreview}, we find [\Sec{sec:e-averaged}] after
averaging over the electron's energy that the decay functions $F_{d,
  \crw}^\app (t) = \langle F^\pp_{d,\crw} (t) \rangle_\ve $ have the
following forms (whose leading terms also follow from
\Eq{eq:introsimpleFdurw}, with $\cW_\pp (\wb)$ as
\weightingFunction):\cite{vonDelft04first}
\begin{subequations}
  \label{eq:finaldecayfunctions}
   \begin{eqnarray}
     \label{eq:previewmostimportantresult}
     F_{1,\crw}^\app (t)
     & = & (t/\tau_{\varphi,1})^{3/2}
     \left[ 1  -  {2^{3/2} |\zeta(\toh)| \over \pi }
{1 \over \sqrt{T t} }  \right]
     \; , 
     \\
     F_{2,\crw}^\app (t) & = & (t/\tau_{\varphi,2})
     \left[{\ln (Tt) - (1-\gamma_{\rm Euler}) \over \ln (T
       \tau_{\varphi,2}) }\right]  \; ,  \qquad 
     \\
     F_{3,\crw}^\app (t) & =&  (t/\tau_{\varphi,3})
     \left[ 1  -  {\pi 2^{3/2}  \over 3 \, \zeta(3/2)}
        {1 \over \sqrt{T t} }  \right] \; .  \qqph 
\end{eqnarray}
\end{subequations}
The leading terms depend on the ``classical decoherence
rates''\cite{temp=freq}
\begin{subequations}
  \label{overviewsubeq:tauphiselfconsistent}
  \begin{eqnarray}
    \label{overvieweq:tauphi1}
    {1 \over \tau_{\varphi,1}} & = &
    ({\textstyle {1 \over 4}}\sqrt{\pi \gamma_1 } \, T )^{2/3}  , \qqph  
    \\
    {1 \over \tau_{\varphi,2}} & = & 
    { T  \over 2 \pi g_2 } \ln (2 \pi g_2)  
\; , \qqph 
\label{eq:overviewtauphi2}
\\
{1 \over \tau_{\varphi,3}}  & = &
{3 \zeta(3/2)  \over  (\pi^3 2^5)^{1/2}} \, 
{T^{3/2} \over \sqrt {\gamma_3} }\; ,
\label{eq:overviewtauphi3}
  \end{eqnarray}
\end{subequations}
which reproduce the results of AAK for classical white Nyquist noise
(except that AAK's numerical prefactors are different, since our way
of defining $\tau_{\phi,d}$ is slightly different from theirs).  The
next-to-leading terms in \Eqs{eq:finaldecayfunctions} generate the
leading quantum corrections to these classical decoherence times.
Extracting the modified decoherence times $\ttau_{\varphi,d}$ from the
condition\cite{thresholdconstant} $F_{d,\crw}^\app (\ttau_{\varphi,d})
= 1$, we find
\begin{subequations}
  \label{subeq:overviewexplicitshifts}
\begin{eqnarray}
  \label{eq:overviewexplicitshifts}
  \ttau_{\varphi,1} & = & \tau_{\varphi,1} \left[ 1 + 
{\tilde b_1 \over \sqrt{ T \tau_{\varphi,1}}} \right] \; , 
  \\
  \ttau_{\varphi,2} & = & \tau_{\varphi,2} \left[ 1 + 
{ \tilde b_2 
\over \ln (T \tau_{\varphi, 2}) }  \right]  \; , 
  \\
  \ttau_{\varphi,3} & = & \tau_{\varphi,3} \left[ 1 + 
{\tilde b_3 
      \over \sqrt{ T \tau_{\varphi,3}}}  \right] \; ,
\end{eqnarray}
\end{subequations}
where $\tilde b_1 = 2^{5/2} |\zeta(\toh)| / (3 \pi) = 0.8767 $,
$\tilde b_2 = 1- \gamma_{\rm Euler} = 0.4228$, and $\tilde b_3 = \pi
2^{3/2} / [3 \zeta (3/2)] = 1.134$. Thus, the next-to-leading terms
are parametrically smaller than the leading ones (confirming the
conclusions of Ambegaokar and Vavilov\cite{AV}) by $g^{-1/2}_1
(L_{\varphi, 1})$ for $d = 1$, or $ 1/\ln g_2 $ for $d = 2$, or
$g^{-1/3}_3 (L_{\varphi, 3})$ for $d = 3$.  Our calculations therefore
conclusively show that in the weak localization regime where $g_d
(L_{\varphi,d}) \gg 1$, AAK's results for $\tau_{\varphi,d}$, obtained
by considering classical white Nyquist noise, remain correct for
quantum Nyquist noise acting on an electron moving inside a Fermi sea
at thermal equilibrium. Nevertheless, since it is not uncommon for
weak localization experiments to reach the regime where the product
$T \tau_{\varphi}$ is only on the order of $10$ (e.g. 
Ref.~\onlinecite{Pierre03}), the corrections discussed here can amount to
an appreciable effect [illustrated in Fig.~\ref{cap:Deviations}
below].

As a check of \Eqs{eq:finaldecayfunctions}, we use them to calculate
[\Sec{sec:AAGcompare1}] the first-order-in-interaction contribution to
the weak localization magnetoconductivity, $\sigma_d^{\WL(1)}$, in the
regime $\gammaphi \ll \gammaH \ll T$, where $\gammaH$ is the magnetic
dephasing rate.  Reassuringly, this reproduces the leading \emph{and}
next-to-leading terms of the corresponding results of AAG\cite{AAG},
obtained via an elaborate perturbative diagrammatic calculation, which
keeps vertex corrections but is restricted to short times. We also
show how to resolve an inconsistency between AAG's way of extracting
the decoherence rate from $\sigma_d^{\WL(1)}$ and the results of AAK.

Finally, we also discuss the energy-dependence of the decoherence rate
(\Sec{sec:Energy-dep-decay-function}).  We calculate explicitly how
the decoherence rate $\gamma_{\varphi,d}$ crosses over to essentially
the energy relaxation rate $\propto \ve^{d/2}$ as $\ve$ is increased
with respect to $T$, and find that the energy scale at which the
crossover happens, namely $T g^{2/3}_1 (L_T)$, $T \ln g_2$ or $T$ for
$d = 1,2$ or $3$, respectively, is parametrically larger than
temperature for $d=1,2$. -- This concludes our overview.

\section{Cooperon decay for classical noise}

\label{sec:Cooperon-decay-for}
\label{sub:Exponential-form-for}

In this section we review how the decay of the Cooperon can be
calculated using influence functionals for the case of classical
noise.  Although this is a standard calculation, we shall cast it in a
form that generalizes straightforwardly to the cases treated in
subsequent sections, namely quantum noise
[\Sec{sec:DecayQuantumNoiseWithoutPauli}] and quantum noise plus Pauli
principle [\Sec{sec:goldenrule}].

\subsection{Definition of Cooperon}
\label{sec:magnetoconductance}

The full Cooperon $\tilde C (r,t)$ appearing in
\Eq{eq:magnetoconductance} for $\delta\sigma^\WL_d$ can be written as
a path integral
\begin{eqnarray} \tC(r,t)& =& 
  \! \int_{r^{F}(-t/2)=0}^{r^F(t/2)=r}{\mathcal{D}}
  r^{F}(t_{3}) \! 
\int_{r^B(-t/2)=r}^{r^{B}(t/2)=0} {\mathcal{D}}
  r^{B}(t_{3}) \rule[-5.5mm]{0mm}{0mm} 
  \nonumber \\
& & \qquad \times  A[r^{F}(t_{3}),r^{B}(t_{3})] \label{eq:PICooperon} \qqph
\end{eqnarray} over pairs of electron paths with opposite start- and
endpoints, to be called forward and backward paths, with amplitude
$A[r^{F}(t_{3}),r^{B}(t_{3})]$.  
The fact that they are time-reversed has been exploited
to denote the start and end times of a path
of duration $t$ by $\pm t/2$ (this yields
time integrals over intervals symmetric
around $t=0$ below, which turns out to be very convenient). 
Semiclassically, the path integral will be
dominated by time-reversed pairs of diffusive paths, i.e.\ 
\begin{eqnarray}
  \label{eq:timereversedpaths}
  r^{F}(t_{3})=r(t_{3})=r^{B}(-t_{3}),   
\end{eqnarray}
and for $\tC(0,t)$, these will
have the same start- and endpoints. 

In the absence of interactions and a magnetic field, the amplitude
$A[r^{F}(\cdot),r^{B}(\cdot )]$ simply equals
$e^{i(S_{0}[r^{F}(\cdot)]-S_{0}[r^{B}(\cdot)]) \shbar}$, where
$S_{0}[r(\cdot)]$ is the free action describing the propagation of a
free electron through a disordered potential landscape.  The
corresponding free Cooperon propagator $\tCO(r,t)$ is thus determined
by the probability density 
for an {\bf u}nrestricted {\bf r}andom {\bf w}alk (in $d$-dimensions)
to reach a volume element $d^3r$ separated from
the initial point by a distance $r$, in time $t$:
\begin{subequations}
\label{subeq:defineCooperon}
\begin{eqnarray} 
  \tP^\urw (r,t) & = & {1 \over a^{3-d}} \, 
    e^{-r^{2}/4D|t|}(4 \pi D|t| )^{-d/2} \; . 
\label{eq:unrestrictedrandomwalk}
\end{eqnarray}
In the presence of a magnetic field $H$  (which, for
$d= 1$ or 2, we shall assume to be perpendicular to the wire or 
plane of the film), the free Cooperon
 is multiplied by a dephasing factor $e^{-t/\tau_{H}}$, where the
magnetic dephasing rate $\gammaH = 1/\tau_{H}$ increases with increasing $H$
($\gammaH = 4DeH/(\hbar c)$ for $d=3$, or $\gammaH = D(eHa)^2/(3c^2
\hbar^2)$ for $d = 1,2$, see \Ref{AltshulerLee80}).
Thus, we have 
\begin{eqnarray}
\tCO(r,t) & = & \theta(t)\,\tP^\urw (r,t) \, e^{-t/\tau_{H}} \; . 
\label{eq:bareCooperon}
\end{eqnarray} 
\end{subequations}
[In contrast, the bare diffuson is magnetic-field independent:
$\tDO(r,t) = \theta(t)\,\tP^\urw (r,t) $.] 

Inserting \Eqs{eq:firstCapprox} and (\ref{eq:bareCooperon})
for the full Cooperon $\tC(0,t)$ into \Eq{eq:magnetoconductance}  
for the magnetoconductivity, the latter can
be written as 
\begin{eqnarray}
  \label{eq:sigma(H)rescaled}
{\delta \sigma_d^\WL  \over \sigma_d} = - 
{2^{1-d} \over \pi^{1+d/2}}  \int_{\tauel}^\infty {dt \over t} \, 
{ e^{-{t/\tauH}} e^{-F_d (t)} \over g_d (L_t)} \; ,   \qph 
\end{eqnarray} 
where $L_t = \sqrt{Dt}$.  For $H=0$, the integral is of order $1/g_d
(L_\varphi)$ (or larger for $d= 2,3$, since $\tauel/\tauphi \ll 1$).
(To see this, change variables to $z = t/\tauphi$ and note that $F(z
\tauphi) \gtrsim 1$ for $z \gtrsim 1$.)  Good conductors, which are
characterized by the fact that the relative change $\delta
\sigma^\WL_d / \sigma_d$ in conductance due to weak localization is
small even at zero magnetic field, therefore have $1/g_d (L_\varphi)
\ll 1$.  This is a well-known and very important small parameter in
the theory of weak localization, which will be used repeatedly below.
(For $d=1$, where it turns out that $1/g_d(L_\varphi) =
(\gamma_1/T)^{1/3}$, this ceases to be a small parameter at sufficiently
small temperatures. This signals the onset of the regime of strong
localization, which is beyond the scope of the present analysis.)

\subsection{Averaging over classical noise}

\label{sec:averagingclassnoise}

Let us now explore how the Cooperon is affected by interactions, or
more generally, by noise fields. Generally speaking, these will cause
the propagation amplitudes for the forward and backward paths to pick
up random phase factors, hence destroying their constructive
interference and causing the Cooperon to decay as function of time.

In the case of purely classical noise, a single-particle description
is exact, and the decay of the Cooperon can readily be evaluated using
path integrals \cite{AAK,CS}. It is instructive to review how this is
done. Let us describe the noise, imagined to arise from some classical
environmental bath, using a classical, real, scalar potential
$V_{j}=V(r_{j},t_{j})$, with correlator \begin{equation}
  - i \hhbar {\mathcal{L}}_{ij}^{\class}\equiv
\left\langle     V_{i}V_{j}\right\rangle^{\class}
=\int(d\kb)\, e^{i\kb
    x_{ij}}\left\langle VV\right\rangle
  _{\bar{q}\bar{\omega}}^{\class}\,
\label{VVFourier}
\end{equation}
(the superscript denotes \textbf{cl}assical, the prefactor $-i \hhbar$
is conventional).  Here we used the shorthand notation
$(d\kb)=(d\qb)(d\wb)$, with $(d\bar{q})=d^{d}\bar{q}/[(2\pi)^{d}
a^{3-d}]$, $(d\bar{\omega})=d\bar{\omega}/(2\pi)$, where
$\kb=(\bar{q},\bar{\omega})$ is our standard notation to be used for
momentum- and frequency-transfers between the electron and the bath,
and $\kb x_{ij} =\qb r_{ij}-\wb t_{ij}$, where we abbreviate
$r_{ij}=r_{i}-r_{j}$, $t_{ij}=t_{i}-t_{j}$ (and, for future use,
$\tildet_{ij}=t_{i} +t_{j}$).  The noise properties can be specified
in terms of the Fourier components of the noise correlator,
$\left\langle VV\right\rangle _{\bar{q}\bar{\omega}}^{\class}$.  It is
symmetric in $\bq$ for homogeneous, isotropic samples.  Moreover, for
classical (but not quantum) noise, it is necessarily also
\emph{symmetric} in frequency,
\begin{eqnarray}
  \label{eq:Vclass-symmetric}
\left\langle VV\right\rangle^\class_{\bar{q} \bar{\omega}}
=\left\langle VV\right\rangle^\class_{\bar{q},-\bar{\omega}} \; ,
\end{eqnarray}
 because $\left\langle     V_{i}V_{j}\right\rangle^{\class}$
is invariant under $t_{ij} \rightarrow t_{ji}$.

 In the presence of a given configuration of the 
potential field $V_{j}$, the propagation amplitude for a pair of
random forward and backward paths, $r^{a}(t_{3})$, with $a=F/B$, is
multiplied by an extra phase factor $e^{i(S_{F}-S_{B})\shbar }$, with
\begin{equation}
i( S_{F}-S_{B}) = - i \int_{-t/2}^{t/2}dt_{3}\,
\sum_{a=F/B}s_{a}V_{3 a}\,,
\label{eq:sfsb}
\end{equation}
where $V_{j a}\equiv V(r^{a}(t_{j}),t_{j})$, and $s_{a}$ stands for
$s_{F/B}=\pm1$. The average of this phase factor over all
configurations of the field $V_{j}$ can be performed without any
approximation if the field is assumed to have a Gaussian
distribution\cite{RPA}, 
\begin{eqnarray}
\left\langle e^{i(S_{F}-S_{B})\shbar }\right\rangle _{V} & = &
e^{-S_{\eff} \shbar}\; ,
\label{exponentAvg}
\label{subeq:Seff}
\end{eqnarray}
where the ``effective action'' $S_{\eff}[r^{F}(\cdot),r^{B}(\cdot)]$ is a
functional of the forward and backward paths and describes the effect
of the environment on the propagating electron,
\begin{subequations}
\label{subeq:Sefftwo}
\begin{eqnarray} 
  & & 
\phantom{.} \hspace{-6mm}
S_{\eff}  =  {1 \over 2 \hhbar}  \int_{-t/2}^{t/2}dt_{3}\,
  dt_{4}\,\sum_{aa'=F/B}s_{a}s_{a'} \, 
  \left\langle     V_{3 {a}}V_{4 {a'}}\right\rangle,
\label{eq:Seff}
\qqph 
\\
& & 
\phantom{.} \hspace{-6mm}
\left\langle     V_{3 {a}}V_{4 {a'}}\right\rangle =   
\int (d \bar k) \, 
e^{i ( \bar q  [r^a (t_3) - r^{a'} (t_4)] - \bomega t_{34})} 
\left\langle     V_{a}V_{ a'}\right\rangle_{\bq \bomega} .
\nonumber \\
\label{eq:Seff2}
\end{eqnarray} 
\end{subequations} 
In the present section~\ref{sec:Cooperon-decay-for}, 
$\langle V_{a}V_{ a'}\rangle_{\bq  \bomega}$ 
is (for all $a$, $a'$) simply equal to the classical
noise correlator $\langle VV \rangle^{\class}_{\bq \bomega}$ of
\Eq{VVFourier}.  (The more general
notation 
will become useful for reusing \Eqs{subeq:Sefftwo} 
[and \Eq{exponentCL} below] in later sections, which
involve more complicated correlators.) 
Note that $S_\eff$ is purely real, because the 
classical correlators 
$\left\langle     V_{3 {a}}V_{4 {a'}}\right\rangle$
are real.

To obtain the effect of the environment on the Cooperon, $e^{-
  S_{\eff} \shbar } $ should be evaluated along and averaged over
time-reversed pairs of paths [\Eq{eq:timereversedpaths}].  The $a=a'$
terms in \Eq{eq:Seff} then correspond to the {}``self-energy''
contributions to the Cooperon decay rate, to adopt terminology that is
commonly used in diagrammatic calculations of the Cooperon decay rate.
These terms describe the decay of the individual propagation
amplitudes for the forward or backward paths, corresponding to decay
of the {}``retarded'' or {}``advanced'' propagators. The $a\neq a'$
terms in Eq.~(\ref{eq:Seff}) correspond to the {}``vertex
corrections'' to the Cooperon decay rate. The self-energy and vertex
terms have opposite overall signs ($s_{F}s_{F}=s_{B}s_{B}=1$ vs.
$s_{F}s_{B}=-1$, respectively). Consequently, the contributions of
fluctuations which are slower than the observation time
(with $\bomega \lesssim 1/t$) and hence
indistinguishable from a static random potential, mutually cancel, and
do not contribute to decoherence. This is already apparent from
\Eq{eq:sfsb}, even before averaging over $V_{j}$: for sufficiently
slow fluctuations, the two terms in $i(S_{F}-S_{B})$ cancel if
$r^{F}(t_{3})=r^{B}(-t_{3})$.

\subsection{Closed versus unrestricted random walks}
\label{sec:rrw-vs-urw}

In order to explicitly evaluate the modification of the full Cooperon
due to the fluctuating potential, we still have to average the factor
$e^{-S_{\eff} \shbar }$ over diffusive paths $r(t)$ (\ie\ random walks) .  An
exact way of performing this average has been devised by AAK
\cite{AAK}, by deriving and then solving a differential equation for
the full Cooperon (which can be done exactly for the quasi
1-dimensional case in the presence of thermal white Nyquist
noise). However,
it is possible to obtain qualitatively equivalent results by a
somewhat simpler approach (also used by GZ): Following Chakravarty and
Schmid (CS)\cite{CS}, we approximate the average over random walks by lifting
the average into the exponent, 
\begin{eqnarray}
  \label{eq:Jensen}
\left\langle
  e^{-S_{\eff} \shbar }\right\rangle _{\textrm{rw}}\simeq e^{-F_d (t)}, 
\qquad F_d (t)= \smalloneoverhbar 
\left\langle S_{\eff}\right\rangle _{\textrm{rw}}
\; .  \qquad   
\end{eqnarray}
The ``decay function'' $F(t)$
will turn out to grow with time (starting from $F(0)=0$) and describes
the decay of the Cooperon [cf.\ \Eq{eq:CooperondecaywithF}].  By
lifting the average into the exponent, we somewhat \emph{overestimate}
the decay of the Cooperon with time, since for any real variable $x$,
the inequality $\langle e^{- x}\rangle_x \ge e^{-\langle x \rangle_x}$
holds independent of the distribution of $x$ (Jensen's inequality).

\begin{figure}
\begin{center}\includegraphics[%
  width=0.90\columnwidth]{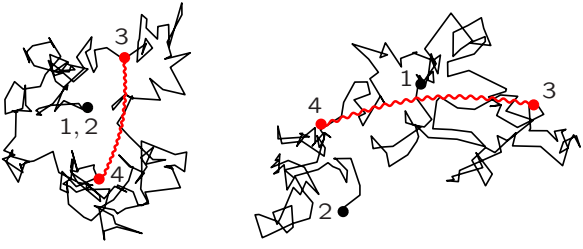}\end{center}
\caption{\label{cap:The-distance-between} Random walks from
  $1\equiv(r_{1},-t/2)$ to $2\equiv(r_{2},+t/2)$, via $3 \equiv
  (r_3,t_3)$ and $4 \equiv (r_4,t_4)$: For given times $t, t_3$ and
  $t_4$, the distance between $r_{3}$ and $r_{4}$ is overestimated
  when the closed random walk (left) is substituted by an unrestricted
  random walk (right). This leads to increased decoherence and an
  underestimate of the weak-localization correction to the
  conductivity (see Fig.~\ref{cap:Magnetoconductance-for-a}, dotted
  line). }
\end{figure}

At the corresponding point in their own work, CS make two further
approximations when evaluating $F_d (t)$: firstly, they do not evaluate
the vertex corrections explicitly, but instead mimic their effect by
dropping (by hand) the contributions of frequency transfers $\wb<1/t$
to the self-energy terms, i.e.\ they introduce a sharp infrared cutoff
in the latter's frequency integrals. Secondly, while averaging the
correlators $\left\langle V_{3 a}V_{4 a'}\right\rangle ^{\class}$ of
\Eq{eq:Seff2} over random walks [\ie\ averaging the Fourier exponents
in Eq.~(\ref{eq:Seff2})], both CS and GZ employ the probability density
$\tP^\urw (r_{34},t_{34})$ for an \textbf{u}nrestricted \textbf{r}andom
\textbf{w}alk to diffusively reach a volume element $d^3r$
removed by a distance $r_{34}=r_{3}-r_{4}$,
in time $|t_{34}|$:
\begin{eqnarray}
\left\langle e^{i\qb(r(t_{3})-r(t_{4}))}\right\rangle _{\textrm{urw}} 
& \equiv &  \int d^3 r_{34}\, \tP^\urw (r_{34},|t_{34}|)\, e^{i\qb r_{34}}
 \nonumber \\
& \equiv & \bP^\urw (\qb , |t_{34}|) 
 =  e^{-D\qb^{2} |t_{34}|}\,. \qqph 
\label{unrestrictedAvg}
\end{eqnarray}
Here and below, position integrals like $\int dr_{34}$ stand for
$a^{3-d} \int d^dr$; the prefactor comes from the integral over the
transverse directions, and it cancels the prefactor of $\tP^\urw$ in
\Eq{eq:unrestrictedrandomwalk}.

The two approximations discussed above are known to be adequate to correctly
capture the functional dependence of the function $F(t)$ on time,
temperature, dimensionless conductance, etc. In the following,
however, we shall be more ambitious, and strive to evaluate the
\emph{numerical prefactor} of $F(t)$ with reasonable accuracy, too.  
To this end, we have to go beyond the two approximations of CS
(dropping vertex terms and doing an {\bf un}restricted average), since
both modify the numerical prefactor by a factor of order one: Firstly,
we shall fully retain the vertex corrections; in effect, we thereby
explicitly evaluate the actual shape of the effective infrared cutoff
function, instead of inserting a sharp cutoff by hand. Secondly, we
shall perform the random walk average more carefully than in
Eq.~(\ref{unrestrictedAvg}), in that we consider only the actually
relevant ensemble of
\emph{closed}\cite{Montambaux04}  random walks of duration $t$,
that are restricted to start and end at the same point in space: 
$r(-t/2)=r(t/2)=0$. 
Thus, we use
\begin{eqnarray}
\nonumber 
\left\langle e^{i\qb(r(t_{3})-r(t_{4}))}
\right\rangle_\crw 
&  \equiv &   \int \! dr_{3} \, dr_{4} \, 
  \tP^\crw_{(0, t)}  (3,4) 
 \,  e^{i\qb r_{34}} 
\\ & \equiv &  
  \bP^\crw_{(0, t)}(\qb , |t_{34}|) \; . 
\qqph 
\label{Uavg}
\end{eqnarray} 
Here $\tP^\crw_{(r_{12},t_{12})} (3,4) $
is the probability density for a \textbf{c}losed \textbf{r}andom
\textbf{w}alk that starts at the space-time point $(r_2, t_2)$ and
ends at $(r_1, t_1)$, to pass through two volume elements around the
intermediate points $(r_4, t_4)$ and $(r_3, t_3)$. For $t_2 < t_{4} <
t_{3}< t_1$ we have
\begin{eqnarray}
\label{eq:prrw}
\lefteqn{    \tP^\crw_{(r_{12}, t_{12})} 
(3,4)} 
\\
& & =\frac{ \tP^\urw (r_{13}, t_{13}) \, 
 \tP^\urw (r_{34},t_{34}) \,
 \tP^\urw (r_{42},t_{42})}{\tP^\urw (r_{12},t_{12})} \;. 
\nonumber
\end{eqnarray} 
The denominator ensures that the integral of
$\tP^\crw$ over $r_{4},r_{3}$ yields 1 {[}as can be seen using $\int
dr_{3}\, \tP^\urw (r_{13},t_{13})\, \tP^\urw (r_{32},t_{32})= \tP^\urw
(r_{12},t_{12})${]}. We shall confirm below that $\tP^\crw$ does
  not depend on $(r_1, t_1)$ and $(r_2, t_2)$ separately, but
  only on the differences $r_{12}$ and $t_{12}$, as anticipated 
on the left-hand side of \Eq{eq:prrw}.

  The probability density $\tP^\crw$ obviously does not depend on the
  magnetic  field.
  Note, though, that \Eq{eq:prrw} does not change if we multiply each
  factor $\tP^\urw (r_{ij}, t_{ij})$ by a dephasing factor $\exp(- t_{ij} / \tau_H
  )$ to obtain a bare Cooperon $\tCO (r_{ij}, t_{ij})$, since these factors
  completely cancel out in \Eq{eq:prrw}.  In the following, we shall
  thus use \Eq{eq:prrw} with $\tP^\urw$ replaced by $\tCO$, since
  this will be convenient when comparing to perturbative expressions
  below, which are formulated in terms of $\tCO$'s. Performing the
  integrals $\int dr_3 dr_4$ of \Eq{Uavg} by Fourier transformation
  and using 
  \begin{eqnarray}
    \label{eq:barecooperonB}
\bC^0_{q}(t)=\theta(t)\,  e^{ - E_q t}  \, , \qquad E_q \equiv  D q^2 +
1 / \tauH   \, ,   
  \end{eqnarray}
for the momentum Fourier transform of the bare Cooperon $\tCO(r,t)$ of
\Eq{eq:bareCooperon}, 
we readily find:
\begin{eqnarray}
 \nonumber 
 \bP^\crw_{(r_{12}, t_{12})}(\qb , t_{34})
& = & 
{\int (dq) \, e^{i q r_{12}} \, 
\bC^0_{q}(t_{13})\,\bC^0_{q-\qb}(t_{34})\,\bC^0_{q}(t_{42})
\over  \tCO (r_{12},t_{12})} 
\\ & = & \label{eq:rrw-result-explicit}
e^{ - D \qb^2 t_{34} 
\left(1 - {t_{34} \over t_{12}} \right) + i \qb r_{12}  {t_{34}
    \over t_{12}} } \, . 
\end{eqnarray}
In the limit $t_{12} \to \infty$ this reduces to $ \bP^\urw (\qb ,
t_{34})$, as expected, provided $t_{34}$ is kept fixed. Note, though,
that the latter condition is \emph{not} appropriate for the evaluation
of the long-time limit of the Cooperon, for which \emph{both} time
differences, $t_{12}$ \emph{and} $t_{34}$, become large.

\subsection{General Form of the Decay Function $F(t)$}
\label{eq:generalF}

Let us now evaluate the decay function $F(t) = \langle S_\eff
\rangle_\rw $, starting from the effective action of
\Eq{subeq:Sefftwo}.  Averaging the latter over time-reversed, 
random walks according to \Eqs{eq:timereversedpaths} and
(\ref{unrestrictedAvg}) or (\ref{Uavg}), the result can be written as
\begin{eqnarray}
F_d (t) & = &  
\oneoverhbarsq
\int_{-{t \over 2} }^{{t \over 2} }  
dt_{3} \int_{-{t \over 2} }^{{t \over 2}} dt_{4}\,
\int (d\bar{q}) \int (d\bar{\omega}) \, e^{-i\bar{\omega}t_{34} }  
\nonumber
\\ 
\label{exponentCL} 
& & 
\times 
\left\langle V V \right\rangle _{\qb\wb}^\eff
\, \delta \bP(\qb; \tau_{34}, \ttau_{34})   \; . 
\end{eqnarray}
For classical noise, where $\langle V_{a}V_{ a'}\rangle_{\bq \bomega}=
\langle VV \rangle^{\class}_{\bq \bomega}$, the ``effective''
environmental noise correlator appearing here likewise stands for the
classical noise correlator, $\langle V V \rangle _{\qb\wb}^\eff =
\langle V V \rangle _{\qb\wb}^\class$.  For the more general case that
the $\langle V_{a}V_{ a'}\rangle_{\bq \bomega}$ depend on $a,a'$, as
will be needed in our treatment of quantum noise below, the effective
noise correlator is found to have the form
\begin{subequations}
\label{subeq:generalpropFFBB}
\begin{eqnarray}
  \label{eq:FFBB=BFFB}
\left\langle V V \right\rangle_{\qb\wb}^\eff  & = & 
\toh 
\left[ \left\langle V_F V_F \right\rangle_{\qb\wb}
+ \left\langle V_B V_B \right\rangle_{\qb\wb} \right] , \qqph
\\
  \label{eq:FFBB=BFFB2}
& = & 
\toh \left[ \left\langle V_B V_F \right\rangle_{\qb\wb}
 + \left\langle V_F V_B \right\rangle_{\qb\wb} \right] ,
\end{eqnarray}
\end{subequations}
where, looking ahead, we used the fact that the first and second lines
are equal for all the types of noise to be considered in this paper.

\Eq{exponentCL} also contains the object
\begin{subequations}
\label{eq:barPPP}
\begin{eqnarray}
  \label{eq:barPPPa}
\delta \bP (\bq; \tau_{34}, \ttau_{34}) & \equiv & 
  \bP(\qb , |t_{34}|)   -
  \bP(\qb , |\tildet_{34}|)   \qqph 
\\
  \label{eq:barPPPb}
& = & e^{-\bq^2 D t \tau_{34}} - e^{-\bq^2 Dt \ttau_{34}} \; , 
\end{eqnarray}
\end{subequations}
which describes the diffusive dynamics of the time-reversed
trajectories. The first term in \Eq{eq:barPPP} arises from self-energy
terms [with $a=a'$ in \Eq{subeq:Sefftwo}], the second from vertex
corrections [with $a \neq a'$]. Here $\tau_{34}$ and $\ttau_{34}$
stand for
\begin{subequations}
\label{subeq:definetau34}
\begin{eqnarray}
  \label{eq:definetau34}
  \tau_{34} & = & {|t_{34}| / t}  \quad \textrm{or} \quad
  \tau_{34} =  \left[ 1 - {|t_{34}| / t} \right] {|t_{34}| / t}
\; , \qqph
\\
  \ttau_{34} & = & {|\tildet_{34}| / t}  \quad \textrm{or} \quad
  \ttau_{34} =  \left[ 1 -  {|\tildet_{34}| / t} \right]
  {|\tildet_{34}| / t}
 \; , \qqph
 \end{eqnarray}
\end{subequations}
($\tildet_{34} = t_3 + t_4$) depending on whether the average over
paths is performed over unrestricted or closed random walks [using
$\bP^\urw (\qb , |t_{34}|)$ from \Eq{unrestrictedAvg} or
$\bP^\crw_{(0, t)}(\qb , t_{34})$ from \Eq{eq:rrw-result-explicit}],
respectively.

Since the time integrals in \Eq{exponentCL} are symmetric, only that
part of $\left\langle V V \right\rangle_{\qb \wb}^\eff$ that is
symmetric under $\bomega \to - \bomega$ contributes to $F_d(t)$;
moreover, if this symmetric part is real, so is $F_d (t)$.

The fact that the decay function is linear in interaction propagators
has an important implication: when expanding both sides of
\Eq{completeAvgRRW} in powers of the interaction, the term linear in
$F_d(t)$ on the right-hand side of \Eq{completeAvgRRW} must equal
$\tC^{1}(0,t)$, the first-order contribution to the full Cooperon
$\tC(0,t)$, implying $-F_d^\crw (t)=\tC^{1}(0,t)/\tCO(0,t)$ [cf.\
\Eq{eq:Frrdefine}].  We added the superscript ``\crw'', because
  this relation turns out to hold only if the average over paths for
  $F_d(t)$ is over closed random walks.
The expression (\ref{eq:CooperondecaywithF}) for $\tC(0,t)$ thus
amounts to a simple reexponentiation of the first order interaction
correction,
\begin{subequations}
\label{subeq:Capprox}  
\begin{equation}
 \tC(0,t)\simeq \tCO(0,t)
\exp\left[\frac{\tC^{1}(0,t)}{\tCO(0,t)}\right]\, , 
\label{Capprox}
\end{equation}
 evaluated in the \emph{position}-time representation (at $r=0$). 
 In contrast, if (following hitherto standard practice\cite{CS,GZ2})
 the average over paths is performed over \emph{un}restricted random
 walks instead, \ie\ if $\bP^\urw$ is used instead of $\bP^\crw$ in
 (\ref{exponentCL}), one finds \(- F_d^\urw (t) = \int dr \,
 \tC^1_{\gamma_H = 0} (r,t) = \bC_{q=\gamma_H=0}^{1}(t)\) [cf.\
 \Eq{eq:Furdefine}].  In this case Eq.~(\ref{eq:CooperondecaywithF})
 yields
\begin{equation}
  \tC(0,t) \simeq \tCO(0,t)\exp
\left[\bC_{\gamma_H = q=0}^{1}(t)\right]\, ,
\label{CapproxUnrestricted}
\end{equation}
\end{subequations} implying that here the first order correction in
the \emph{momentum}-time representation is reexponentiated;
consequently, the left- and right-hand sides of
\Eq{CapproxUnrestricted} are not consistent when expanded to first
order in the interaction. Hence, \Eq{Capprox} can be expected to be
more accurate than (\ref{CapproxUnrestricted}), as will be confirmed
below.

To make further progress with the evaluation of $F(t)$, we now exploit
the fact that for all types of noise to be considered in this paper,
the effective noise correlator  \emph{factorizes} into a
 frequency-dependent \weightingFunction\ $\cW_\eff (\bomega )$, symmetric in
frequency, and a $\bq$-dependent denominator:
\begin{eqnarray}
\smalloneoverhbar \left \langle V V \right\rangle^\eff_{\qb\wb}
\label{eq:spectrumfactorizes}
& = & {\cW_\eff (\wb) \over \nu D \bq^2 }  \; . 
\end{eqnarray}
This fact allows us to proceed quite far with the evaluation of
  \Eq{exponentCL} for $F_d (t)$ without specifying the actual form of
  $\cW_\eff(\wb)$ (which will be done in later sections): after changing
  integration variables to the sum and difference times $\tildet_{34}$
  and $t_{34}$, \Eq{exponentCL} can be written as
\begin{eqnarray}
\label{exponentCL-white}
F_d (t) & = &
2 \! \int_0^t  \!\! dt_{34} \, W_\eff  (t_{34}) 
\! \int_0^{t - t_{34}} \!\!  d\tildet_{34} \,  
\delta \tP_d ( \tau_{34},\ttau_{34}) \; ,  \qqph
\end{eqnarray}
where the kernel
\begin{eqnarray}
\label{eq:defineKernel}
W_\eff (t_{34}) = 
\int (d\bar{\omega}) \, e^{-i\bar{\omega}t_{34} } \cW_\eff (\wb ) \; 
\end{eqnarray}
contains \emph{all} information about the frequency-dependence of the
noise correlator describes the range in time of the
effective interaction, whereas the dimensionless quantity
\begin{eqnarray}
  \delta \tP_d (\tau_{34}, \ttau_{34})  & = & 
  \int (d\bar{q}) \, 
{e^{-\bq^2 D t \tau_{34}} - e^{-\bq^2 Dt \ttau_{34}}
    \over \hhbar 
    \nu D \bq^2 } \qqph 
\label{eq:definedeltatildeP}
\end{eqnarray}
is the same for all types of noise studied below, but depends on the
dimensionality of the sample.  Note that the integrand in
\Eq{eq:definedeltatildeP} is well-behaved for small $\qb^2$ despite of
the $1/\qb^2 $ factor, because the self-energy and vertex
contributions to $\delta \bP (\bq; \tau_{34}, \ttau_{34}) $
[\Eq{eq:barPPP}] cancel each other for momentum transfers smaller than
$D \qb^2 \lesssim 1/t$, thus regularizing the divergence.  This is
precisely as expected for density fluctuations with dispersion $\wb
\simeq D \qb^2$: on time scales of order $t$, fluctuations with
frequencies below $\wb \lesssim 1/t$ appear to be essentially static,
and hence do not contribute to decoherence.

The $\int (d \qb)$ and $\int d \tildet_{34}$ integrals in
  \Eqs{eq:definedeltatildeP} and (\ref{exponentCL-white}) can be
  performed explicitly (see \App{app:cPd}), with a result for $F_d(t)$
  of the form
\begin{eqnarray}
  \label{eq:Fdecayalmostfinal}
F_d (t) =  {t \over 
g_d (L_t)}
\int_0^t d t_{34} \, W_\eff (t_{34}) \, \cP_d (t_{34}/t)  \; ,
\end{eqnarray}
where $L_t = \sqrt{D t}$.  Explicit expressions for the functions
$\cP_d (z)$, which differ for closed or unrestricted random walks, are
given in \App{app:cPd}. However, it will turn out below that we really
only need the leading terms in an expansion of $P_d(z)$ for small
values of its arguments, because the decoherence rate is extracted
from the \emph{long}-time behavior of $F_d(t)$.   The leading and
subleading terms of $\cP_d^\crw$ for closed or $\cP_d^\urw$ for
unrestricted random walks (upper or lower entries, respectively) are
given by:
%
%
\begin{subequations}
  \label{subeq:expandPd}
  \begin{eqnarray}
    \label{eq:expandPd1}
    \cP_1^{\{ \stackrel{\crw}{\scriptscriptstyle \urw} \}}
 ( z ) & = &  
\left\{ \begin{array}{l} \sqrt \pi/2  \\  8/ (3 \sqrt \pi) 
  \end{array} \right\} 
- {4 \sqrt {z/ \pi }} + 
{\cal O} ( z ) \; , 
\qqph \\
\cP_2^{\{ \stackrel{\crw}{\scriptscriptstyle \urw} \}} (z ) 
& = &   - {1 \over \pi} 
\left[ \ln z +
\left\{ \begin{array}{l} 2 + {\cal  O} (z) \\
    1 + {\cal  O} (z \ln z) 
\end{array} \right\}  \right]
\; , 
\\
\cP_3^{\{ \stackrel{\crw}{\scriptscriptstyle \urw} \}} (z ) 
& = &  {1 \over \pi^{3/2}} \left[ {1 \over \sqrt  z}  - 
\left\{ \begin{array}{l} \pi  \\  2 \end{array} \right\} \right]
+  {\cal O} (z^{1/2}) \; .
  \end{eqnarray}
\end{subequations}
Thus, the difference between averaging over closed or unrestricted
random walks turns out to matter for the leading terms of only $\cP_1$
(but not of $\cP_{2,3}$), implying, as we shall see below, that it
matters for the leading long-time behavior of only $F_1(t)$ [but not of
$F_{2,3}(t)$].

\Eq{eq:Fdecayalmostfinal} is a central result on which the following
sections rely.  To find the decay function $F_d(t)$ (and a
corresponding decoherence time), all that remains to be done is to
determine the \weightingFunction\ $\cW_\eff(\wb)$ (which depends on the type
of noise studied, and on whether the Pauli principle is taken care
of), calculate its Fourier transform to obtain the kernel $W_\eff (t_{34})$
of \Eq{eq:defineKernel}, and perform the integral $\int dt_{34}$ in
\Eq{eq:Fdecayalmostfinal}.

We close this section with a comment on the relation of the above
approach to diagrammatic methods, \eg\ the calculation of the Cooperon
$\tC^1 (0,t)$ to first order in the interaction (from which $F_d(t)$
can be extracted). The key to our derivation of
\Eq{eq:Fdecayalmostfinal} was to essentially work in the time domain,
postponing any time integrals until after all momentum and frequency
integrals had been performed, as exemplified by our definitions of
$\tP^\crw $ [\Eq{eq:rrw-result-explicit}, involving $\int (d q)$],
$\delta \tP_d$ [\Eq{eq:definedeltatildeP}, involving $\int (d \bq)$]
and $W(t_{34})$ [\Eq{eq:defineKernel}, involving $\int (d \wb)$].  
In contrast, in diagrammatic calculations, the $\int dt_3 dt_4$ integrals
are performed first (namely when deriving the Feynman rules for the
frequency-momentum), \emph{before} any momentum or frequency
intetrals. However, for the present problem the resulting momentum
integrals then take intractably complicated forms (see
\App{app:integrals}).

Nevertheless, for the case of unrestricted random walks, the leading
asymptotic behavior of $F_{d,\urw}(t)$ can be obtained rather simply
by judiciously neglecting some terms that are subleading in the limit
of large times. As shown in \App{app:integrals}, this leads to the
following expression,
\begin{eqnarray}
  \label{eq:simpleFdurw}
  F_{d, \urw} (t) & \simeq & 
  p_d \,  t \int_0^\infty \!\! {d\bar{\omega} } \, 
  {\cW_\eff (\wb) \over  \wb^{2-d/2} } 
  \left[1  - {\sin (\wb t) 
      \over \wb t} \right] ,   \qqph 
\end{eqnarray}
with $p_1 = \sqrt{ 2 \gamma_1} / \pi$, $p_2 = 1/(g_2 2 \pi)$, $p_3 =
1/(\sqrt {2 \gamma_3} \, \pi^2)$ [for $\gamma_1, g_2, \gamma_3$, see
\Eq{eq:dimensionlessconductance}]. This formula for $F_{d, \urw}(t)$
is less accurate than \Eq{eq:Fdecayalmostfinal}, but perhaps
physically more transparent, since formulated in the frequency domain:
The factor $[1 - \sin (\wb t)/ \wb t]$ acts as infrared cutoff,
suppressing frequencies $\wb \lesssim 1/t$. Due to the factor
$\wb^{d/2-2}$, the integral gets its dominant contribution from small
frequencies of order $1/t$ for $d=1$, gets logarithmic contributions
from both small and large frequencies for $d=2$, and is dominated by
large frequencies for $d=3$.  In particular, for $d=2,3$ an
ultraviolet cutoff is needed at large frequencies to render the
integral well-defined.  As we shall see below, such a cutoff will be
provided by $\cW_\eff (\wb)$.  

\subsection{Comparison with exact classical 1-D result } 
\label{sec:1Dcompare}

To gauge quantitatively the difference
between averaging over closed or unrestricted random walks,
we shall now use the above results to calculate the
magnetoconductivity for a quasi 1-dimensional conductor and classical
white Nyquist noise.  For this particular case, the average $\langle
e^{-S_{\eff}\shbar }\rangle_\textrm{rrw}$ was calculated exactly by
AAK,\cite{AAK} so that the resulting expression for the
magnetoconductivity,\cite{factorof2}
\begin{eqnarray}
  \delta\sigma_{{\textrm{exact}}}^\WL  = 
\frac{e^{2}\sqrt{D\tau^\sAAK_{\varphi}}}{\pi \hhbar}\frac{1}{[\ln{\rm
    Ai}(\frac{\tau_{\varphi}^{\rm AAK}}{\tau_{H}})]'}\,,
\label{eq:Airy}\end{eqnarray}
can be used as benchmark for other approximations.  In
  \Eq{eq:Airy}, Ai is the Airy function,  and 
$\tau_\varphi^\sAAK$ is the decoherence time of \Eq{eq:tauhphiAAK}.
\begin{figure}
\begin{center}\includegraphics[%
  width=0.95\columnwidth]{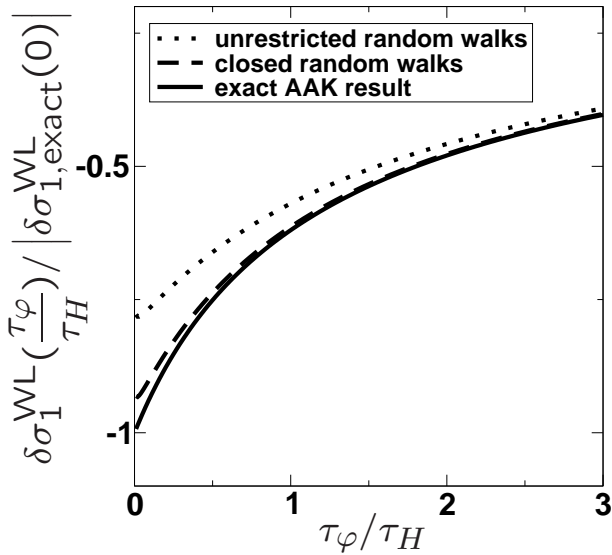}\end{center}
\caption{\label{cap:Magnetoconductance-for-a} The magnetoconductivity
  for a quasi 1-dimensional wire experiencing decoherence by classical
  white Nyquist noise, as function of $\tau_{\varphi,1}^\sAAK /
  \tauH$, comparing AAK's exact Airy function result (\ref{eq:Airy}),
  solid line, against two approximations: the commonly employed
  average over unrestricted random walks in the path-integral
  exponent, {[}\Eq{CapproxUnrestricted}{]}, and the improved version
  obtained by averaging over closed random walks {[}\Eq{Capprox}{]},
  resulting in \Eq{eq:finalFrrw}, with $c_1^\urw$ or $c_1^\crw$,
  respectively.  For $\tau_{\varphi}/\tau_{H}\rightarrow0$, the result
  for unrestricted (closed) random walks equals $0.808$ ($0.964$) of
  the exact value.  For $\tau_{\varphi,1}^\sAAK /\tau_{H}\gg1$, all
  three curves become indistinguishable. }
\label{fig:Airy}
\end{figure}


For the particular case of classical white Nyquist noise, the noise
correlator is given by $ T/ \nu D \qb^2$, so that the effective noise
correlator of \Eq{exponentCL} takes the form:
\begin{equation}
\label{eq:Nyquist}
\smalloneoverhbar 
\left\langle V V\right\rangle _{\qb\wb}^{\eff} \mapsto
\smalloneoverhbar \left\langle VV\right\rangle _{\qb\wb}^{\class}=
\frac{T 
}{\nu   D\qb^{2}} \; . 
\end{equation}
Thus, the weighting function in \Eq{eq:spectrumfactorizes} is
frequency-independent for classical white Nyquist noise, $\cW_\class
(\bomega) = T $, so that the corresponding Kernel is an infinitely
sharp delta function, $W_\class (t_{34}) = T 
\delta (t_{34})$.  The $\int d t_{34}$ integral in
\Eq{eq:Fdecayalmostfinal} for $F_1^\class (t)$ is thus easily
performed, yielding
\begin{eqnarray}
 \label{eq:finalFrrw}
F^\class_1 (t) & = & 
 c_1  { t T  \over  g_1(L_t) }  = 
(t / \tau_{\varphi,1})^{3/2} \; , \quad 
\begin{array}{c} c_1^\crw = {\sqrt \pi  \over  4} \; , 
\rule[-3mm]{0mm}{0mm} \\
c_1^\urw = {4 \over 3 \sqrt \pi} ,
\end{array} \qqph  
\end{eqnarray}
where $\tau_{\varphi,1} = (c_1 \sqrt{\gamma_1} T)^{-2/3}$. Depending
on whether we average over closed or unrestricted random walks, two
different values for the prefactor $c_1$ are obtained. The decoherence
time was obtained by solving $F^\class_1 (\tau_{\varphi,1}) = 1$ for
$\tau_{\varphi,1}$, which reproduces \Eq{eq:tauhphiAAK}, with an
extra\cite{thresholdconstant} $c_1$ in front of $T$.

The fact that $\tau^\crw_{\varphi,1}$ is somewhat
larger than $\tau^\urw_{\varphi,1}$ (the relative factor is
$(16/3 \pi)^{2/3} = 1.423 $)
  can intuitively be understood from
  Fig.~\ref{cap:The-distance-between}: For a given time difference
  $t_{34}$, the distance $r_{34}$ between the points $r_{3}$ and
  $r_{4}$ is overestimated on average when closed random walks are
  replaced by unrestricted random walks. Thus the latter give somewhat
  too much weight to the effect of long wavelength or small momentum
  transfers, which dominate the integral in
  \Eqs{eq:definedeltatildeP}. They hence somewhat overestimate the
  decoherence rate, and consequently underestimate the decoherence
  time and 
magnitude of
  the weak localization correction to the {conductivity} [cf.\
  Fig.~\ref{cap:Magnetoconductance-for-a}].

  The weak localization contribution to the magneto{conductivity}
  $\delta \sigma_1^{\rm WL}$ of a quasi 1-dimensional wire 
    can now be obtained by inserting $F_1^\class (t)$ of
    \Eqs{eq:finalFrrw} into \Eq{eq:sigma(H)rescaled}.
  Fig.~\ref{fig:Airy} compares the results so obtained using
    $c_1^\urw$ (dotted line) and $c_1^\crw$ (dashed line) to AAK's
  Airy function result [\Eq{eq:Airy}, solid line{]}. Firstly and most
  importantly, all three approaches agree fully in their predition for
  $\tauphi$, which acts as the scale on which the magnetoconductivity
  is suppressed as a function of increasing magnetic field (\ie\
  increasing $1/\tauH$). However, the methods differ somewhat in their
  predictions for the magnetoconductivity $\delta \sigma^\WL_1$ at $B=
  0$, which gives the overall magnitude of the weak localization
  effect: Averaging over unrestricted random walks (dotted line)
  yields only qualitative agreement with the exact Airy-function
  result, deviating from it by about 20\% at $B=0$.  In contrast,
  averaging over closed random walks (dashed line) gives excellent
  quantitative agreement with the exact result, yielding practically
  identical results for large and intermediate field strengths, with a
  maximal deviation of less than 4\% in the limit of vanishing
  magnetic field.  It is quite remarkable that such good agreement
  with an exact result can be obtained by means as elementary as the
  above.

\section{Cooperon decay for quantum noise without Pauli principle}

\label{sec:DecayQuantumNoiseWithoutPauli}

Inspired by the simplicity and elegance of the above treatment
  of classical noise fields, we shall explore in this section to what
  extent a quantum bath can be similarly dealt with in a
  single-particle path integral picture: is it possible to identify
  suitably chosen ``effective classical noise'' correlators, such that
  the decay function $F(t)$ is again of the form (\ref{exponentCL-white}),
  with $\langle VV\rangle _{\qb\wb}^{\class}$ replaced
 by   some suitably chosen effective noise correlator
$\langle VV\rangle _{\qb\wb}^{\eff}$? The advantage of
such a formulation would evidently be (i) that the results are sure to
be free of infrared problems, and (ii) that the trajectory averages
could be performed with the same ease as above.

Of course, we know from the outset that a strategy based on mimicking
quantum by classical noise fields can never be exact or complete,
because the correlator $\angle \hat V \hat V\rangle_{\bar{q}
  \bar{\omega}}$ of a quantum noise field differs from that of a
classical noise field in an elementary but fundamental way: In
contrast to $\langle V V\rangle_{\bar{q} \bar{\omega}}^\class$ , which
is symmetric in frequency, $\langle \hat V \hat V\rangle_{\bar{q}
  \bar{\omega}}$ is \emph{asymmetric} in frequency, reflecting the
asymmetry between energy absorption from ($\wb<0$) and emission into
($\wb>0$) the bath, $\wb$ being the change in bath energy.
In particular, the asymmetry manifests itself in the
possibility of spontaneous emission events which
are possible even at $T=0$, and hence strongly affect
the low-temperature behavior. 
Nevertheless, we shall see that when the effective action is evaluated
along time-reversed paths as needed to describe the decay of the
Cooperon, it is again governed by a noise correlator symmetric in
frequency, so that this decay \emph{can} be described by a
suitably-chosen ''effective'' classical noise field.  

\subsection{Definition of quantum noise correlators}
\label{subsec:quantumcorrelators}

We begin by  recalling that for a free bosonic quantum field
$\hat{V}_{j}\equiv\hat{V}(r_{j},t_{j})$, the two correlation functions
\begin{eqnarray}
  \label{eq:V>}
-i  \hhbar  \tilde {\mathcal{L}}_{ij}^>  =   
\bigl \langle \hat V_i \hat V_j \bigr
  \rangle \; ,   \qquad 
-i \hhbar   \tilde {\mathcal{L}}_{ij}^<  =   \bigl \langle \hat V_j \hat V_i \bigr
  \rangle \; ,  
\end{eqnarray}
are not equal (as would be the case for a classical field), but are
related, after Fourier transforming, by the detailed-balance relation
$\bar {\mathcal{L}}_{\qb}^{<}(\wb)=e^{-\beta\wb} \bar
{\mathcal{L}}_{\qb}^{>}(\wb)$.  This implies that the symmetrized and
anti-symmetric correlators, or, equivalently, the Keldysh, retarded
and advanced correlators
\begin{subequations}
\label{subeq:LKRAdefine}
  \begin{eqnarray}
  \label{eq:LKdefine}
  - i  \hhbar \tilde {\mathcal{L}}_{ij}^K & = &   
\bigl \langle \{ \hat V_i , \hat V_j \} \bigr \rangle \; ,  
   \label{eq:LRAdefine} \\
\label{VVasymmetricFDTandLRdef}    
- i \hhbar \tilde {\mathcal{L}}_{ij}^{R/A}  & = & \pm \theta (\pm t_{ij})   
\bigl \langle [ \hat V_i ,  \hat V_j ] \bigr \rangle \; , 
\end{eqnarray}
\end{subequations} 
are related by the fluctuation-dissipation theorem, 
\begin{subequations}
\begin{eqnarray}
-\toh  i \bar {\cal L}^K_{\qb}(\wb) & = &
\textrm{Im} \bar {\cal L}^R_{\qb}(\wb)\,\coth
\Bigl[\frac{\wb}{2T}\Bigr]\;, 
\label{eq:fluct-diss}
\\
\label{eq:symmetrizedalt}
& =   &
 \textrm{Im} \bar {\cal L}^R_{\qb} ( | \wb| )
  \, \bigl[ 2  n(|\wb |)+ 1 \bigr]  \; , \qqph
\end{eqnarray}
\end{subequations}
where $\textrm{Im} \bar {\cal L}^R_{\qb}(\wb)= {1 \over 2 \hhbar}
\bigl\langle[\hat{V},\hat{V}]\bigr\rangle_{\qb\wb}$ is an odd function
of $\wb$. For example, the noise generated by the standard screened
Coulomb interaction, namely quantum Nyquist noise (which we shall
focus on for the remainder of this paper), can be described by taking
\begin{eqnarray}
  \label{eq:LCoulomb}
\textrm{Im} \bar {\mathcal{L}}_{\qb}^{R}(\wb)
& = & {\bomega  \over  2\nu D \bbmq^2 }
\end{eqnarray}
[see, \eg, Refs.~\onlinecite{AAG} or \onlinecite{vonDelft04}].
Using the above relations, the 
quantum noise correlator can be written as 
\begin{eqnarray}
\smalloneoverhbar \bigl\langle\hat{V}\hat{V}\bigr\rangle_{\qb\wb} & = &
 \textrm{Im} \bar {\cal L}^R_{\qb}(|\wb|)
\,  \bigl[2 n(|\wb|)+  2 \theta(\wb)\bigr]\; . 
\label{quantumspectrum}
\end{eqnarray}
In \Eq{quantumspectrum}, the contribution of $n(|\wb|)$ dominates at
frequency transfers smaller than the temperature, for which the number
of activated quanta is large ($n \gg1$). The additional step function
$\theta(\wb)$, responsible for the asymmetry of
$\bigl\langle\hat{V}\hat{V}\bigr\rangle_{\qb\wb}$, contributes even if
the bath is at zero temperature ($n =0$), and hence is sometimes said
to reflect zero-point fluctuations of the bath. It describes the
possibility of spontaneous emission of energy by the electron into the
bath, enabling excited electrons to relax to states of lower energy.
Of course, in a many-body situation, the rates for such relaxation
processes will also contain Fermi functions that Pauli-block them if
no empty final states are available. 
However, we shall defer a detailed discussion of Pauli
blocking to Section~\ref{sec:goldenrule} and completely ignore it in
the present section, which thus applies only to situations for which
Pauli restrictions are irrelevant. The latter would include a purely
single-particle problem, or, for a many-body degenerate Fermi gas, 
an electron that is very highly excited above the Fermi surface, with
plenty of empty states below to decay into.

\subsection{Averaging over quantum noise}
\label{sec:avequantnoise}

It is known \cite{FeynmanHibbs64} that the effects of a quantum noise
field $\hat V (r,t)$ on a quantum particle can be described in terms
of classical (c-number) fields by proceeding as follows: one considers
a path-integral with a Keldysh forward-backward contour, and includes
the noise via phase factors $e^{-i\int dx_{j}dt_{j}V_{jF}\shbar }$ (as
part of $e^{iS_{F}\shbar}$) and $e^{i\int dx_{j}dt_{j}V_{jB} \shbar}$
(as part of $e^{-i S_{B}\shbar }$) that contain two \emph{different}
fields, $V_{jF}=V_{F}(r_{j},t_{j})$ and $V_{jB}=V_{B}(r_{j},t_{j})$,
on the forward- and backward-contour, respectively.  [In the case of
classical noise the Keldysh fields are equal, $V_{F}=V_{B}=V$.]  The
classical field correlators are related to the quantum noise field
correlators $ \bigl \langle \hat V_i \hat V_j \bigr \rangle $ of
\Eqs{eq:V>} by time-ordering along the Keldysh contour:
\begin{subequations}
\label{classicalvsquantumfields}
\begin{eqnarray}
\label{eq:defFF}
 \bigl\langle V_{i F} V_{j F} \bigr\rangle & \equiv & \bigl\langle \hat{T}\hat{V}_i  \hat{V}_j \bigr\rangle 
   =  - {\textstyle {\hbaroverstar{2}}} i  (\tilde {\cal L}^K + \tilde {\cal L}^R + 
\tilde {\cal L}^A)_{ij} \; , \qqph
\\
\label{eq:defBB}
\bigl\langle V_{i B} V_{j B} \bigr\rangle 
& \equiv & \bigl\langle
  \tilde{\hat{T}}\hat{V}_i \hat{V}_j \bigr\rangle 
  =  - {\textstyle {\hbaroverstar{2}}} i (\tilde {\cal L}^K - 
\tilde {\cal L}^R - \tilde {\cal L}^A)_{ij} \; , \qqph
\\
\label{eq:defBF}
 \bigl\langle V_{i  B} V_{j F} \bigr\rangle 
& \equiv & \bigl\langle \hat{V}_i \hat{V}_j \bigr\rangle  \phantom{ \hat{T}}
  =  - {\textstyle {\hbaroverstar{2}}} i (\tilde {\cal L}^K + \tilde {\cal L}^R - 
\tilde {\cal L}^A)_{ij} \; , \qqph
\\
\label{eq:defFB}
\bigl\langle V_{i F} V_{j B} \bigr\rangle 
& \equiv & \bigl\langle \hat{V}_j\hat{V}_i \bigr\rangle  \phantom{ \hat{T}}
  =  - {\textstyle {\hbaroverstar{2}}} 
i (\tilde {\cal L}^K - \tilde {\cal L}^R +
 \tilde {\cal L}^A)_{ij}  \; . \qqph 
\label{eq:keldyshvv}
\end{eqnarray}
\end{subequations} 
The first set of relations in \Eqs{classicalvsquantumfields} follows
from comparing the expansions generated when expanding the factors
$e^{-i\int dx_{j}dt_{j}V_{jF}\shbar}$ occuring in a time-ordered path
integral for the forward path, and $e^{i\int dx_{j}dt_{j}V_{jB}\shbar}$
occuring in a backward path, with the corresponding expansions of the
quantum time evolution operators $\hat{U}_{F}=\hat{T}e^{-i\int
  dx_{j}dt_{j}\hat{V}_{j}\shbar}$ and
$\hat{U}_{B}^{\dagger}=\tilde{\hat{T}}e^{i\int
  dx_{j}dt_{j}\hat{V}_{j}\shbar}$ occuring in Keldysh perturbation theory,
respectively. The second set of relations in
\Eqs{classicalvsquantumfields} are simply standard identities, following from
the definitions (\ref{subeq:LKRAdefine}) of $\tilde
{\mathcal{L}}_{ij}^{K,R,A}$. 

The effective action for a {\bf s}ingle-particle subject to such {\bf
  q}uantum {\bf n}oise (sqn), to be denoted by $S_\eff^\sqn$, is
obtained as for \Eq{exponentAvg}: we now have to perform a Gaussian
average over $V_{F},V_{B}$, again with the action $i(S_{F}-S_{B})$ of
\Eq{eq:sfsb} in the exponent, but now with $V_{j a} \equiv V_a (r^a
(t_j), t_j)$, \ie\ both the paths $r^a (t)$ \emph{and} the fields
$V_a$ differ on the forward and backward contours. The result for
$S^\sqn_\eff$ again has the form of Eqs.~(\ref{subeq:Sefftwo}), but
now $\bigl\langle V_{a}V_{ a'}\bigr\rangle_{\qb \bomega}$ stands for
the Fourier transform of the quantum correlators of
\Eqs{classicalvsquantumfields}. In fact, $e^{-S^\sqn_\eff \shbar}$ is
nothing but the Feynman-Vernon influence functional for a single
particle interacting with a quantum bath.  Note that in contrast to
the effective action for classical noise, $S^\sqn_\eff$ is in general
complex, since the correlators $\bigl\langle V_{i a} V_{j a'}
\bigr\rangle $ are all complex (because $-i \tilde {\cal L}^K_{ij}$,
$\tilde {\cal L}^R_{ij}$ and $\tilde {\cal L}^A_{ij}$ are by
construction purely real).

\subsection{Quantum noise spectrum without Pauli principle}
\label{subsec:symCooperondecay}

To describe the effect of quantum noise on the Cooperon, the influence
functional $e^{- S^\sqn_\eff \shbar}$ has to be averaged over all
time-reversed pairs of closed random walks.  This can be done in
the same way as as in Section~\ref{sec:rrw-vs-urw}.
The result for  $F_d^\sqn (t) \equiv \oneoverhbar \langle
S^\sqn_\eff \rangle_\rw$ has the same 
form as \Eq{exponentCL},
\begin{eqnarray}
F_d^\sqn (t) & = & \oneoverhbarsq
\int_{-{t \over 2} }^{{t \over 2} }  
dt_{3} \int_{-{t \over 2} }^{{t \over 2}} dt_{4}\,
\int (d\bar{q}) \int (d\bar{\omega}) \, e^{-i\bar{\omega}t_{34} } 
\nonumber \\ \label{exponentCLsym} 
& & 
\times 
\left\langle V V \right\rangle_{\qb\wb}^\sqn
\, \delta \bP(\qb; \tau_{34}, \ttau_{34})   \; ,
\end{eqnarray}
where the effective noise correlator in \Eq{exponentCL} now takes the
form [obtained from \Eqs{subeq:generalpropFFBB} and
\Eqs{classicalvsquantumfields}]:
\begin{subequations}
 \label{replaceSymmetrized}
\label{eq:symmetrized} 
\label{symmCorrFDT}
\begin{eqnarray}
  \bigl\langle VV\bigr\rangle _{\qb\wb}^\eff  \mapsto 
  \bigl\langle VV\bigr\rangle _{\qb\wb}^\sqn  =
  - {\textstyle \hbaroverstar{2}}  i \bar {\cal L}^K_{\qb \wb} \, = \,
  \toh \bigl\langle \{ \hat V , \hat V \} \bigr\rangle_{\qb\wb}  .
  \qph 
\end{eqnarray}
For the case of quantum Nyquist noise [\Eq{eq:LCoulomb}], 
$ \oneoverhbar \langle VV \rangle _{\qb\wb}^\sqn$ can be
written in the factorized form 
${\cW_\sqn / \nu D \bq^2 } $
of \Eqs{eq:spectrumfactorizes}, the
corresponding \weightingFunction\ being
\begin{eqnarray}
  \label{eq:sqnWeighting}
  \cW_\sqn (\bomega) = \toh |\bomega| \bigl[ 2 n (|\bomega|) + 1
  \bigr] \; . 
\end{eqnarray}
\end{subequations}  
 Just as for classical noise, this effective noise spectrum is
\emph{symmetric in $\bomega$} and $\bq$ and real [as follows from
\Eq{eq:symmetrizedalt}], which implies that $F^\sqn_d (t)$ is
purely real. In other words, the imaginary part of the effective
action $S^\sqn_\eff$ vanishes upon averaging over time-reversed paths;
the reason is that
(the Fourier transforms of) 
the purely imaginary $\mp \toh i (\bar {\cal L}^R+ \bar {\cal L}^A)$
contributions from \Eqs{eq:defFF} and (\ref{eq:defBB}) cancel each
other when inserted into \Eq{subeq:generalpropFFBB}, as do the
contributions $\mp \toh i (\bar {\cal L}^R- \bar {\cal L}^A)$ from
\Eqs{eq:defBF} and (\ref{eq:defFB}).

The fact that $\langle S_\eff^\sqn \rangle_\crw$ is purely real along
time-reversed paths has the following useful implication: for the
particular purpose of calculating the Cooperon decay, it is possible to mimick the
effect of a quantum-mechanical environment by a purely classical noise
field, if we so wish,
provided its noise correlator is postulated to be given precisely by $
\langle VV \rangle _{\qb\wb}^\sqn $ of \Eq{eq:symmetrized},
\ie\ the \emph{symmetrized} version of the asymmetric quantum noise
correlator  (\ref{quantumspectrum}).  This can be verified
by rewriting the effective action in terms of even and odd
combinations of $V_{F/B}$, namely
\begin{eqnarray} 
  V_{+j} = \toh (V_{Fj} + V_{Bj} ) \; , \qquad\
  V_{-j} = V_{Fj} - V_{Bj} \; .
\label{eq:newV's}\end{eqnarray}
It is then  readily found that the decay of
the Cooperon is governed only by 
the even field $V_{j+}$; indeed,
since their correlators are given by 
\begin{eqnarray}
{\displaystyle \left(\begin{array}{cc} 
\bigl\langle V_{+i}V_{+j}\bigr\rangle & 
\bigl\langle V_{+i}V_{-j}\bigr\rangle  
\rule[-3mm]{0mm}{0mm} \\
\bigl\langle V_{-i}V_{+j}\bigr\rangle 
& \bigl\langle V_{-i}V_{-j}\bigr\rangle 
\end{array}\right)
=-i \hhbar \left(\begin{array}{cc}
    \toh \tilde \LL_{ij}^{K} & \tilde \LL_{ij}^{R} \rule[-3mm]{0mm}{0mm} \\
    \phantom{\toh} \tilde \LL_{ij}^{A} & 0\end{array}\right)} , \qph 
\label{eq:tridiagonalKeldysh-LLb}
\end{eqnarray}
we see that the symmetrized noise correlator  $\langle V V
\rangle^\sqn_{\bq \bomega}$ governing the Cooperon decay function
$F^\sqn_d (t)$ is equal to the correlator $\langle V_+ V_+ \rangle_{\bq
  \bomega}$ of the even field $V_{+j}$, whereas the correlators $\langle
V_\pm V_\mp \rangle_{\bq \bomega}$ involving the odd field play no
role in determining $F^\sqn_d (t)$.

\subsection{The effect of spontaneous emission}
\label{sec:spontemission}

It is instructive to analyse the differences between the classical
\weightingFunction\ $\cW_\class (\bomega) = T$ and the quantum case
$\cW_\sqn (\bomega)$ of \Eq{eq:sqnWeighting}.  Since both are
symmetric in $\bomega$, the main qualitative difference between them
is the presence in the quantum case of \emph{spontaneous emission}, 
leading to an extra contribution that does not vanish at zero temperature.
Although spontaneous emission only enhances the scattering rate for
transitions downward in energy, the preceding analysis shows that
 the
asymmetric quantum noise spectrum may just as well be replaced by
its symmetrized version. Physically, this is possible because both
the upward and downward transitions are equally effective in contributing
to dephasing (if they are allowed), and thus it is only their sum that
matters for the dephasing rate. Schematically, we have

\begin{eqnarray}
\Gamma_{\varphi}&=&{1 \over 2}(\Gamma_{\uparrow}+\Gamma_{\downarrow}) \nonumber \\
&\propto&{1\over 2}(n+(n+1)) \nonumber \\
&=&{1\over 2}([n+{1\over 2}]+[n+{1\over 2}]) , \label{eq:updownrates}
\end{eqnarray}
where $n=n(|\wb|)$ is the Bose occupation number for the frequency transfer $\wb$ under
consideration.

This procedure is possible 
for a single, excited electron without Fermi
sea, or, in a many-body situation, for an electron so highly excited
above the Fermi sea that Pauli restrictions on the available final
states are negligible. In such a case, spontaneous emission processes evidently
  persist down to zero temperature and will
  thus cause the decoherence rate to remain finite even at $T=0$ [see
  also Ref.~\onlinecite{FMSR}].  Indeed, this can be seen explicitly
  from \Eq{eq:simpleFdurw} for $F_{d,\urw}^\sqn (t)$: replacing
  $\cW_\eff (\wb)$ therein by the single-particle quantum noise
  spectrum at zero temperature, $\cW^{(T=0)}_\sqn (\bomega) = \toh
  |\wb|$, and introducing an upper cutoff $\int_0^{\wbu} d \wb$ to
  regularize the ultraviolet divergence that then arises, one readily
  finds that $ F_{d,\urw}^{\sqn} (t) = {1 \over d} p_d \, \wbu^{d/2}
  t$, implying a finite decoherence rate at zero temperature:
\begin{eqnarray}
  \label{eq:FdurwT=0}
  \gamma_{\varphi,d}^{\sqn  (T=0)} = 
  {\textstyle {1 \over d}} p_d \, \wbu^{d/2}  \; . 
\end{eqnarray}

\section{Decoherence and the Pauli principle}
\label{sec:goldenrule}

The scenario discussed at the end of the previous section will of
course change as soon as Pauli blocking becomes relevant: consider a
many-body situation, and a noise mode whose frequency $\bomega$ is
much larger than both the temperature \emph{and} the excitation energy
of the propagating electron. In such a case we expect that spontaneous
emission really \emph{would} be severely inhibited by the lack of
available final states.  The present section is devoted to offering a
heuristic understanding of these effects.  (Previous, more formal,
approaches\cite{GZ2,vonDelft04} for dealing with Pauli blocking
effects in a functional integral context are briefly reviewed in
Section~\ref{app:GZrelation}.)  Remarkably, we shall find that the
decay of the Cooperon can again be described by a classical field with
a symmetrical noise spectrum, but now containing an extra term to
describe Pauli blocking, that turns out to block
spontaneous emission. Putting it differently, the Pauli blocking term counteracts the effects of the vacuum fluctuations of the environment. 

\subsection{Early attempts  to include the Pauli principle 
in path-integral calculations of decoherence}

\label{sub:Previous-approaches}

The importance of Pauli blocking was recognized early on in the theory
of decoherence in weak localization\cite{AAK}. The simplest heuristic
strategy to cope with this problem seems to be to derive the classical
spectrum by applying the classical fluctuation-dissipation theorem
(FDT) to the linear response correlator. The  result
is equivalent to replacing the
$\bigl[ n(|\wb|) + 1\bigr]$ in Eq.~(\ref{eq:symmetrized}) by 
its low-frequency limit $T/|\wb|$.  This approach works well
for the case of Johnson-Nyquist noise\cite{AAK}, which has a relatively large
weight at low frequencies, so that these dominate anyway. 
Note, though, that 
more care has to be excercised for super-Ohmic baths, such as phonons.

The question of how to include the Pauli principle has received
surprisingly little attention in the early decoherence literature. The most
concrete suggestion was due to Chakravarty and
Schmid\cite{CS,CSfootnote}. For unexplained {}``general reasons'' ,
i.e. probably in view of the perturbation-theoretic
treatments\cite{Abrahams}), they proposed the following replacement
as a way of incorporating the Pauli principle for the decoherence
of thermally distributed electrons:
 \begin{eqnarray}
 \label{subeq:CSreplacement}
2 n(|\wb|)+1 & \mapsto & 2 n(|\wb|)+2 f(|\wb|)
\label{eq:ChakSchmidtguess}
 \\
\coth \left[\frac{|\wb|}{2T} \right] & \mapsto &  
\coth\left[\frac{|\wb|}{2T}\right] -
\tanh\left[\frac{|\wb|}{2T}\right] =
{2 \over \sinh \bigl[\frac{|\wb|}{T}\bigr]} 
\nonumber 
\end{eqnarray}
At low frequencies, $|\wb|\ll T$, this yields the same factor
$T/|\wb|$ as the application of the classical FDT; moreover, it also
provides an exponential cutoff at energy transfers $\wb$ larger than
the temperature, thereby accounting for the absence of thermally
excited bath modes at these high frequencies. The fact that
\Eq{eq:ChakSchmidtguess} does \emph{not} include spontaneous emission
at all (and actually vanishes at $T=0$), is of no concern if we
consider the decoherence of an electron picked from a thermal
distribution, i.e.\ within $T$ of the Fermi energy, for which
spontaneous emission would have been Pauli-blocked anyway. However,
the absence of spontaneous emission \emph{would} be a concern when
describing highly-excited, non-thermal electrons, which, for a
zero-temperature bath, would incorrectly be predicted not to relax at
all.  In other words, what is missing in \Eqs{subeq:CSreplacement} is
any reference to the energy $\ve$ of the propagating electron.

In the following sections we shall reanalyze these issues, but take
care to include Pauli blocking throughout. Our conclusions turn out to
qualitatively confirm the heuristic rule (\ref{subeq:CSreplacement}) for
thermal electrons, but are quantitatively
more precise, and will also show how it should be generalized to
deal with highly excited, non-thermal ones.

\subsection{Electron and hole decay rates}

In order to gain intuition about the effects of the Pauli principle,
we shall first discuss the perturbative calculation of {}``golden
rule'' decoherence rates. Although this is only a first-order
calculation in the interaction, the result is expected to be revealing
nevertheless, since we know from \Eq{completeAvgRRW} for $F(t)$ that
the decay function is needed only to linear order in the interaction,
too.

\begin{figure}
\begin{center}\includegraphics[%
  width=0.95\columnwidth]{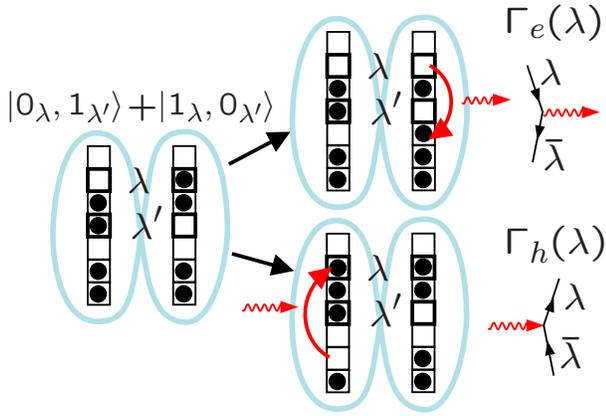}\end{center}
\caption{\label{cap:Sketch-of-twofig} Sketch of two many-particle states
forming a coherent superposition, and two possible scattering processes
contributing to $\Gamma_{e}(\lambda)$ and $\Gamma_{h}(\lambda)$,
by emission and absorption, respectively. }
\end{figure}

Consider a degenerate system of fermions under the influence of a
fluctuating environment that leads to transitions between the
single-particle levels. The environment (e.g. a bath of harmonic
oscillators) is described by a fluctuating potential $\hat{V}$ that
couples to the fermions via some single-particle operator, and the
correlator $\bigl\langle\hat{V}\hat{V}\bigr\rangle_{\wb}$ will
determine the decoherence rate. For brevity of notation, we do not
consider a spatial dependence, $\hat{V}(r,t) \to \hat V(t)$, and we
shall assume the single-particle operator to connect any two levels
with equal matrix element (which is set to unity in the following). 
The generalization to arbitrary coupling is straightforward. The
golden rule decay rate for an electron in level $\lambda$ to be
scattered into any other level is given by
{[}Fig.~\ref{cap:Sketch-of-twofig}{]}
\begin{eqnarray}
\nonumber
\lefteqn{\Gamma_{e}(\lambda)  = \staroverhbar{2 \pi}
 \int_{- \infty}^\infty \! (d\wb)
\bigl\langle   \hat{V}\hat{V}\bigr \rangle 
_{\wb} \bigl[ 1-f(\ve_{\lambda}-\wb) \bigr] D(\ve_{\lambda}-\wb) }
\\
\label{eq:Gammael-a}
\\ \label{eq:Gammael-b}
& = &  
\int_0^\infty \! (d\wb)\,2 \textrm{Im}\bar {\cal L}^R(\wb)
\Bigl\{ \bigl[ 1 + n (\wb) \bigr] \,  \bigl[ 1 - f(\ve_\lambda - \wb
)\bigr] \Bigr. \nonumber 
\\
& & \Bigl. \times D(\ve_{\lambda}-\wb) 
+ n (\wb)  \, \bigl[ 1 - f(\ve_\lambda + \wb ) \bigr]  \, 
  D(\ve_{\lambda}+ \wb) 
\Bigr\}
\,.   \qph 
 \nonumber 
\end{eqnarray}
where $D(\ve)$ is the density of single-particle levels.  
In the second equality,
the first and second terms describe emission and absorption
processes, whereby the electron in the level $\lambda$ is scattered to
a lower- or higher-lying empty level, respectively.  At $T=0$ where
$n(\wb)=0$, the only surviving process is spontaneous emission, and if
$\ve_{\lambda}$ approaches the Fermi energy,
$\ve_{\lambda}\rightarrow0$, this process is suppressed, too, by the
Fermi function $\bigl[1-f(-\wb)\bigr]$. 

Below we shall also need the rate for an initially empty state
$\lambda$ to be filled, \ie\ the decay rate of a hole, given by:
\begin{eqnarray}
\label{subeq:Gammahole}
\label{eq:Gammahole-a}
\label{eq:Gammahole-b}
\Gamma_{h}(\lambda) & =  & \staroverhbar{2 \pi} \int_{- \infty}^\infty 
\! (d\wb) 
\bigl \langle
  \hat{V}\hat{V}\bigr\rangle_{\wb} \, 
f(\ve_{\lambda}+\wb)D(\ve_{\lambda}+\wb) \qqph
\\ \nonumber 
& = & 
\int_0^\infty \! (d\wb)\,2 \textrm{Im}\bar {\cal L}^R(\wb)
\Bigl\{ \bigl[ 1 + n (\wb) \bigr] \, f(\ve_\lambda + \wb ) 
\Bigr. 
\\ \nonumber 
& & \Bigl. \times 
  D(\ve_{\lambda}+\wb) 
+ n (\wb)  \, f(\ve_\lambda - \wb )  \, 
  D(\ve_{\lambda}- \wb) 
\Bigr\}
\,. 
\end{eqnarray}
In the second equality, the first and second terms describe emission
and absorption processes, whereby an electron from a higher- or
lower-lying level is scattered into the empty hole level $\lambda$,
respectively. Again we see that at $T=0$, only spontaneous emission is
possible, and if $\ve_{\lambda}\rightarrow0$, this process is
suppressed, too. 

\subsection{Golden rule decay rates for coherent superpositions}
\label{subsec:goldenrule}

In order to calculate the decoherence rate, we have to consider a
somewhat more complicated situation.  Suppose we are interested in the
linear response of the system to some perturbation (as is the case for
the {conductivity} calculation in the weak localization problem). If
the perturbation scatters an electron from level $\lambda$ to level
$\lambda'$, then the resulting state is of the form
$(1+\epsilon\psi_{\lambda'}^{\dagger}\hat{\psi}_{\lambda})\bigl
|\Phi\bigr\rangle $, where $\epsilon\ll1$ is small. Contributions of
this type can occur if $\bigl |\Phi\bigr\rangle $ has one electron in
$\lambda$ but none in $\lambda'$; apart from this restriction, $\bigl
|\Phi\bigr\rangle $ is some Slater determinant with arbitrary
distribution of fermions over the other single-particle levels, and we
will perform a thermal average over such states in the end.
Effectively, we have thus created a coherent superposition of two
many-particle states, which, for brevity, we shall call $\bigl
|1_{\lambda},0_{\lambda'}\bigr\rangle + \epsilon \bigl
|0_{\lambda},1_{\lambda'}\bigr\rangle $.  These are formed from an
initial state with unoccupied $\lambda$ and $\lambda'$ by inserting a
single extra particle into a coherent superposition $c^\dagger_\lambda
+ \epsilon c^\dagger_{\lambda'}$ of these two levels (Fig.
\ref{cap:Sketch-of-twofig}).  Our task is to calculate the decay rate
of this coherent superposition, which under appropriate assumptions
(discussed below) corresponds to the {}``decoherence rate''.
Although we shall eventually need only the case $\ve_\lambda =
\ve_{\lambda'}$ (see Ref.~\onlinecite{wneq0}), we shall for clarity 
distinguish the indices $\lambda$ and $\lambda'$ througout
this subsection. 

The coherent superposition will be destroyed by any process that leads
to a decay of one or the other many-particle state. This includes not
only an electron leaving $\lambda$ or $\lambda'$, but also an electron
entering the respective unoccupied state ($\lambda'$ or $\lambda$).
The total decay rate for the coherent superposition therefore is the
sum of four contributions (after thermal averaging over the electron
distribution): 
\begin{equation} 
\Gamma_{\varphi}(\lambda,\lambda')=
  \toh \Big[ \Gamma_{e}(\lambda)+\Gamma_{h}(\lambda')+\Gamma_{e}(\lambda')
  +\Gamma_{h}(\lambda) \Bigr] \, .\label{eq:gammaphi}
\end{equation}
 The first two terms give the decay rate for the state $\bigl
|1_{\lambda},0_{\lambda'}\bigr\rangle $, while the latter two refer to
$\bigl |0_{\lambda},1_{\lambda'}\bigr\rangle $. The factor $\toh$ comes
about because decoherence is due to the decay of wave functions rather
than populations (the same is seen in usual master equation
formulations of decoherence of systems with a discrete Hilbert space).

In writing down Eq.~(\ref{eq:gammaphi}), we have assumed that all
of the decay processes lead to decoherence. However, one may think
of situations where \Eq{eq:gammaphi}
would \emph{overestimate} the decoherence rate:
For example, an electron traveling two different paths may scatter
a phonon on both of these paths (similar to decay processes making
the electron leave $\lambda$ and $\lambda'$). But the interference
is destroyed only if the wavelength of the phonon is sufficiently
short to be able to distinguish the two paths from each other, since
otherwise the information about the path of the electron is not revealed
in the scattering process. In fact, disregarding this possibility
amounts to neglecting vertex corrections in the diagrammatic calculation. 
Whether such an approximation is justified depends (among other things)
on the operator whose expectation value is to be calculated in the
end. 
This operator should connect the two states 
$\bigl |1_{\lambda},0_{\lambda'}\bigr\rangle $
and $\bigl |0_{\lambda},1_{\lambda'}\bigr\rangle $, in order to be
sensitive to the coherence of the state. It is necessary to specify
this operator for each particular microscopic model, as well as the
operator of the initial perturbation and the details of the system-bath
coupling. 
However, such details are not important in the present section, since
our aim here is merely to display the simple generic features of the
golden rule decoherence rate in a fermion system.  

It is illuminating
to relate the calculation presented above to the decay of the
single-particle retarded Green's function, $G_{\lambda}^{R}(t)$, which
appears in diagrammatic calculations. According to the definition
of $G^{R}_\lambda$, we have to consider the decay of the following overlaps:
\begin{eqnarray}
  i \hhbar G_{\lambda}^{R}(t) & \equiv & \theta(t)\bigl\langle 
\bigl\{ 
\hat{\psi}_{\lambda}(t),\hat{\psi}_{\lambda}^{\dagger}(0)
\bigr\} \bigr\rangle \nonumber \\
  & = & \theta(t)\left[\bigl\langle
    \hat{\psi}_{\lambda}^{\dagger}\hat{U}(t)\Phi_{0}
    \bigl |\hat{U}(t)\hat{\psi}_{\lambda}^{\dagger}\Phi_{0}
    \bigr. \rangle \right.\nonumber \\
  &  & \left.+\bigl\langle \hat{U}(t)\hat{\psi}_{\lambda}\Phi_{0}
    \bigl |
    \hat{\psi}_{\lambda}\hat{U}(t)\Phi_{0}\bigr. \bigr\rangle \right],
\label{eq:GRdecay}
\end{eqnarray}
where $\bigl |\Phi_{0}\bigr\rangle $ denotes the ground state at $T=0$
(or some state over which a thermal average is to be performed).  Here
$\hat{U}(t)$ is the full time evolution operator, including the
coupling to the environment. The first overlap can decay by two
processes: Either the ket-state changes during the time $t$ by the
particle leaving the initial level $\lambda$, or the bra-state changes
by a particle entering $\lambda$ (before the time $t$). Thus, the
decay rate for the first term is
\begin{eqnarray}
  &  & \Gamma_{e}(\lambda)+\Gamma_{h}(\lambda)
  =\staroverhbar{2\pi}\int_{-\infty}^{\infty}(d\wb)\,
  \textrm{Im}\bar {\cal L}^R(\wb)\times\nonumber \\
  &  & \left[\coth\Bigl(\frac{\wb}{2T}\Bigr)+
    \tanh\Bigl(\frac{\ve_{\lambda}-\wb}{2T}\Bigr)\right]
  D(\ve_{\lambda}-\wb)\,. 
\label{eq:cosh+tanh}
\end{eqnarray}
By similar reasoning, the decay rate for the second term of 
\Eq{eq:GRdecay} is given by
the same expression.  Note in particular that a combination
``$\coth+\tanh$'' arises \Eq{eq:cosh+tanh}; this combination is well
known from diagrammatic calculations of the decoherence rate in weak
localization (see \eg\ Ref.~\onlinecite{Abrahams}).  At large positive
energy transfers $\wb\gg\ve_{\lambda},\, T$ (emission of energy into
the environment), this factor vanishes, due to the Pauli blocking of
final states.  Likewise, large negative energy transfers (absorption
of energy) are also forbidden, because the environment does not
contain thermal quanta needed for that process.

Likewise, the decay rate of $G_{\lambda'}^{A}(t)$ is found to be
$\Gamma_{e}(\lambda')+\Gamma_{h}(\lambda')$.  Thus, (in the absence of
vertex corrections) the decay rate for
$G_{\lambda}^{R}G_{\lambda'}^{A}$ coincides with the total decoherence
rate $\Gamma_{\varphi}(\lambda,\lambda')$ of \Eq{eq:gammaphi}, as
expected. For the purpose of calculating the decoherence rate, for
which we need the long-time behavior of the Cooperon $\tC(0,t)$ for
$Tt \gg 1$, it suffices to take the electron and hole energies
equal\cite{wneq0}, $\ve_\lambda = \ve_{\lambda'}$, which will be done
henceforth.

In a related context, 
an explicit (non-diagrammatic) calculation of the decoherence
rate has been performed for a fermionic ballistic Mach-Zehnder
interferometer. This calculation 
is devoid of the complications introduced by impurity averaging
but displays all the features regarding the Pauli principle which have
been discussed here\cite{MachZehnder}.

\subsection{Pauli-blocked noise correlator 
$\bigl\langle \hat{V} \hat{V}  
  \bigr\rangle^{\rm \pp}_{\qb \wb} $}
\label{sec:symmPauli-mod}

The discussion of decoherence and weak localization in
Section~\ref{sec:Cooperon-decay-for} had been purposefully restricted
to situations where the Pauli principle played no role; in a many-body
situation, this corresponds \eg\ to a highly excited electron state,
with $\varepsilon_\lambda$ so large that $f(\ve_{\lambda}\pm\wb)=0$.
Taking the density of states $D (\ve)$ to be constant henceforth, we
then note from \Eqs{eq:Gammael-a} and (\ref{eq:Gammahole-a}) that for
$f = 0$ the decoherence rate calculated in the previous subsection
indeed depends on the symmetrized correlator only,
\begin{equation}
\Gamma_{e}(\lambda)+\Gamma_{h}(\lambda)
\propto
\int_{-\infty}^\infty (d\wb) \, \toh \bigl\langle 
\{ \hat{V},\hat{V} \} \bigr\rangle _{\wb} \; , 
\label{eq:simplysym}
\end{equation}
in agreement with the conclusion of \Eqs{exponentCLsym} and
(\ref{eq:symmetrized}).

Moreover, we observe from Eqs. (\ref{eq:Gammael-a}) and
(\ref{eq:Gammahole-a})  that in the present context,
``switching on'' the Pauli principle
(i.e. permitting $f\neq0$) amounts to replacing $\toh \bigl\langle
\{ \hat{V},\hat{V} \} \bigr\rangle _{\wb}$ in \Eq{eq:simplysym} by
\begin{eqnarray}
  \label{eq:replacePPfirst}
  \toh  \bigl\langle \{ \hat{V},\hat{V} \} 
  \bigr\rangle _{\wb}  
+ \bigl[ f(\ve_{\lambda}+\wb)
  -f(\ve_{\lambda}-\wb) \bigr]
 \toh \bigl\langle 
  [\hat{V},\hat{V} ]\bigr\rangle _{\wb}.  \; 
\end{eqnarray}
This yields a ``Pauli-blocked'' noise spectrum [compared to
\Eq{eq:symmetrized}], which is, however, still symmetric and
non-negative.

Now, since the decay functions $F(t)$ discussed in earlier sections
and the golden rule analysis presented above are both linear in the
interaction propagator, the lessons learnt from the golden rule about
Pauli blocking should be directly relevant for $F(t)$, too. 
Thus, we formulate the following hypothesis:
if the first-order correction to the Cooperon $\tilde C^1$ were
calculated by a proper many-body technique, the result
for $F(t) = - \tC^1 (0,t)/\tC^0 (0,t)$ will have the
same form as \Eq{exponentCLsym},
\begin{eqnarray}
F^{\pp}_d (t) & = & 
\oneoverhbarsq \int_{-{t \over 2} }^{{t \over 2} }  
dt_{3} \int_{-{t \over 2} }^{{t \over 2}} dt_{4}\,
\int (d\bar{q}) \int (d\bar{\omega}) \, e^{-i\bar{\omega}t_{34} } 
\nonumber \\ 
& & 
\times 
\bigl \langle  V  V \bigr \rangle_{\qb\wb}^\pp
\: \, \delta \bP(\qb; \tau_{34}, \ttau_{34})   \; , 
\label{exponentCLpaulimodified}
\end{eqnarray}
except that the effective noise spectral function now has the
following form, modified by the Pauli principle (pp):
  \begin{subequations}
    \label{subeq:theGreatReplacementa}
    \begin{eqnarray}
      \label{theGreatReplacementa}
      \lefteqn{\bigl\langle V V  
        \bigr\rangle^\eff_{\qb \wb}   \mapsto 
        \bigl\langle V V  
        \bigr\rangle^\pp_{\qb \wb} }
      \\
      & &  \phantom{.} \hspace{0cm} \equiv 
      \toh  \bigl\langle \{ V, V \}   \bigr\rangle _{\qb \wb}  
      + \bigl[ f(\ve +\wb)
      -f(\ve -\wb) \bigr]
      \toh \bigl\langle 
      [\hat{V},\hat{V} ]\bigr\rangle _{\qb \wb}\, 
      \nonumber 
      \\
      & & \phantom{.} \hspace{0cm}   = 
      \hhbar \, \textrm{Im} \bar  {\mathcal{L}}_{\qb}^{R}(\wb)
      \Bigl\{
      \coth \Bigl[{\bomega \over 2T}\Bigr] \, 
      + \toh (\textrm{th}_- - \textrm{th}_+) \Bigr\} .  \qqph
      \label{theGreatReplacement}
    \end{eqnarray}
  \end{subequations} 
    Here  $\ve$ is the initial energy with which an
    electron starts off on its diffusive trajectory, and $\textrm{th}_\pm
    \equiv {\tanh[(\ve \pm \wb)/ 2T]} $.  The combination ``$\coth+\tanh$''
    arising in \Eq{theGreatReplacement} is well known from diagrammatic
    calculations of the decoherence rate in weak localization (see \eg\
    Ref.~\onlinecite{Abrahams}).

\begin{figure}
\begin{center}\includegraphics[
  width=0.90\columnwidth]{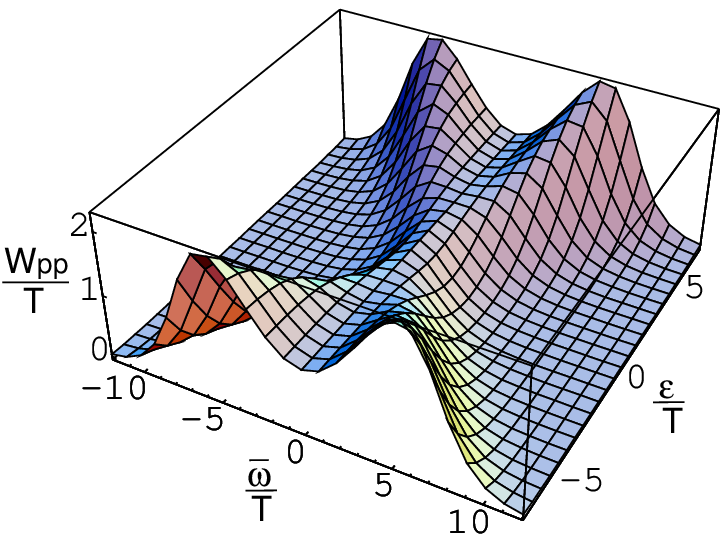}\end{center}
\caption{\label{fig:Wpp} The \weightingFunction\ $\cW_\pp (\wb)$ of
  \Eq{eq:weightingpp} as a function of frequency and electron energy.
  It has the properties $\cW_pp (0) = 1$, $\cW_\pp \to 0$ for $\wb \gg
  \max\{T, \ve \}$, and in the limit $T \to 0$, behaves as $\toh |\wb|
  \theta (|\ve| - |\wb|)$. }
\end{figure}

    For the case of quantum Nyquist noise [\Eq{eq:LCoulomb}], 
 $\oneoverhbar \langle V V \rangle^\pp_{\qb \wb} $ can 
    be written in the factorized form 
$\cW_\pp (\wb) / 
\nu D \bq^2 $ of \Eq{eq:spectrumfactorizes}, 
    with a corresponding Pauli-principle-modified \weightingFunction,
    \begin{eqnarray}
      \cW_\pp (\bomega) & = & 
      {|\bomega| \over 2 } \Bigl\{ 2 n(|\wb |)  + 1 
      + f(\ve + |\wb|) - f(\ve - |\wb|) \Bigr\} ,
      \nonumber \\
      \label{eq:weightingpp}
    \end{eqnarray}
shown in Fig.~\ref{fig:Wpp}.
It has the very important property that it cuts off the contribution
of all frequencies $
|\bomega| \gtrsim \max\{T,\ve\}$. In particular, for an electron at
the Fermi surface $(\ve = 0)$, the factor in curly brackets reduces to
the combination $2 n(|\wb|) + 2 f(|\wb|)$ anticipated by Chakravarty
and Schmid [\Eq{eq:ChakSchmidtguess}], cutting off all frequencies
$|\bomega| \gtrsim T$.  At $T=0$, the \weightingFunction\
  $\cW_\pp (\bomega)$ reduces to $\toh |\wb| \theta( |\ve| - |\wb|)$.
  Moreover, at $\ve = 0$ and $T=0$, it yields $\{0 + 1 + 0 - 1 \} =
0$, \ie\ \emph{the new Pauli terms precisely cancel the spontaneous
  emission term} discussed in Section~\ref{sec:spontemission}, as
announced at the beginning of Section~\ref{sec:goldenrule}.  Thus,
this \weightingFunction\ can be expected to lead to a decoherence rate
that \emph{vanishes} for  sufficiently small
temperatures. We shall see below that this is indeed
the case.

The hypothesis that the use of $\langle \hat{V}\hat{V}
\rangle^\pp_{\qb \wb} $ in our formula for $F(t)$ is the
appropriate way to incorporate the Pauli principle into an influence
functional approach will be shown to be correct in subsequent parts of
this work: In Section~\ref{sec:Simplified-influence-action}, this
replacement will be justified within the context of the functional
integral analysis of decoherence of Ref.~\onlinecite{vonDelft04}.
Moreover, the Bethe-Salpeter analysis of paper II likewise turns out
to lead to a decay function [\EqII{eq:exponentBetheSalpeter}]
involving precisely the Pauli-principle-modified weighting
function of \Eq{subeq:theGreatReplacementa}; in particular, the
diagrammatic calculation of $\tC^1(0,t)$ performed there confirms
explicitly that $F(t)^\pp_d = - \tC^{1} (0,t)/ \tC^{0}(0,t)$.

\section{Results for Decay Function $F_d^\pp (t)$}
\label{sec:ResultsDetailsF(t)}

\subsection{Energy-averaged Decay Function $\langle F_d^\pp(t)
  \rangle_\ve$}
\label{sec:e-averaged}

Since the correlator $\langle V V \rangle^\pp_{\qb \wb} $ 
occuring in $F^\pp (t)$ 
depends on the initial energy $\varepsilon$ with which 
an electron starts off on its diffusive trajectory, an average 
of the function $e^{-F^{\pp}_d (t)}$ over this
energy still has to be performed, using 
the usual derivative of the Fermi function:
\begin{eqnarray} 
\label{eq:energyaveragedefine}
\langle ... 
\rangle_\ve \equiv \int d  \ve [-f' (\ve)] \:  ... \; . 
\end{eqnarray}
We shall simplify the
calculation by lifting the energy average into the exponent:
\begin{eqnarray}
  \label{eq:Eaverageexponentearly}
  \left 
    \langle e^{-F_d^{\pp} (t)} 
  \right \rangle_\ve 
  \simeq e^{- \langle  F^{\pp}_d (t) \rangle_\ve } \; ,
\end{eqnarray}
thereby again somewhat \emph{over}estimating the actual decoherence
rate [cf.\ discussion after \Eq{eq:Jensen}].  The energy average of
the decay function, $ F^\app_d (t) \equiv \langle F^{\pp}_d
(t)\rangle_\ve$, has the same form as \Eq{exponentCLpaulimodified},
but now with an energy-averaged \weightingFunction:
\begin{eqnarray}
\cW_\app (\bomega)  \equiv 
\bigl \langle \cW_\pp (\bomega) \bigr \rangle_\ve
  & = & 
T \left( { 
 \wb / 2T \over \sinh [
\wb / 2T] } 
\right)^2 \; ,  \qqph
  \label{eq:energyaverageof<VV>early}
\end{eqnarray}
which exponentially suppresses the contribution of all frequencies
$
|\bomega| \gtrsim T$, as anticipated above. The fact
that $\cW_\app (\bomega) \le \cW_\class (\bomega) = T$ 
for all frequencies,  but in particular for $\bomega \gtrsim T$,
implies that the effective energy-averaged Pauli-blocked noise 
is somewhat less noisy than classical noise; thus, 
the decoherence times  for the former can be expected to be
somewhat longer than for the latter, as will indeed be found below.

Evaluating the Fourier transform of $\cW_\app (\bomega)$ (by closing
the $\int d \bomega$-integral in \Eq{eq:defineKernel} along a
semicircular path in the complex plane), we find the kernel
%
\begin{eqnarray}
  \label{eq:W(t)eaverage}
 W_\app (t_{34}) = \pi T^2 w (\pi T t_{34}), \;
\quad  w(z) =  {z \coth z  - 1 \over \sinh^2 z} \; , 
\qph 
\end{eqnarray}
where $w(z)$ is a positive, peak-shaped function with weight
$\int_{-\infty}^\infty dz \, w(z) = 1$. Thus, $W_\app (t_{34})$ equals
$T$ times a broadened delta function of width $1/T$, and in the limit
$T \to \infty$, we recover $W_\app (t_{34}) \to W_\class (t_{34})$.
Inserting \Eq{eq:W(t)eaverage} into \Eq{eq:Fdecayalmostfinal}, we
obtain
\begin{eqnarray}
  \label{eq:FWaveragez}
  F^\app_d (t) = {T t \over 
g_d (L_t)} 
\int_0^{\pi T t} \!\! dz \, w (z) \,
 \cP_d^\crw   \left({z \over \pi T t }\right)    
\; . \qqph
\end{eqnarray}
Since the decoherence time is defined by asking when $F(\tauphi) =
{\cal O}(1)$, and $g_d (L_\varphi) \gg 1$ [cf.\ discussion after
\Eq{eq:sigma(H)rescaled}], \Eq{eq:FWaveragez} implies that $T \tauphi
\gg 1$. Thus, to determine $F(t)$ for times $t\simeq \tauphi$, we may
take the limit $Tt \gg 1$ in \Eq{eq:FWaveragez}, obtaining for the
leading and next-to-leading terms:
\begin{subequations}
  \label{subeq:F(t)g(L_t)}
  \begin{eqnarray}
    \label{eq:F1g(L_t)}
F^\app_1 (t) & = &  {T t \over g_1(L_t)}
  \left[ c_1  -  {b_1  \over \sqrt{T t} }  \right]
\; ,
\\
    \label{eq:F2g(L_t)}
F^\app_2 (t) & = &  {Tt  \over g_2(L_t) }
\bigl [ c_2 \ln (T t ) -  b_2 \bigr] \; , 
\\
    \label{eq:F3g(L_t)}   
    F^\app_3 (t) &= & {( T t )^{3/2} \over g_3(L_t)}
    \left[ c_3   - {b_3 \over \sqrt{ T t} }  \right] \; .
  \end{eqnarray}
\end{subequations} 
Here the prefactors $c_1^\crw $ 
and $ c_1^\urw $ 
are the same as in \Eq{eq:finalFrrw}, reproducing the results obtained
there for classical white Nyquist noise. The prefactors $c_2 = 1/(2
\pi)$ and $c_3 = \int_0^\infty dz \, w(z) / (\pi \sqrt{z}) = 
{3 \zeta(3/2)  / \sqrt{\pi^3 2^5} } = 0.2488$
are independent of whether the average over paths is performed over
closed or unrestricted random walks, because the same is true for the
leading terms of $\cP_2$ and $\cP_3$ in \Eqs{subeq:expandPd}.
Hence, in contrast to $d=1$, averaging over closed instead of
unrestricted random walks yields no increase in accuracy for the
leading terms of $F_d^\app (t)$ for $d = 2$ and 3.  It does make a
difference for the subleading terms, for which we find $b_1^\crw =
b_1^\urw = (2\pi)^{-1/2} \, |\zeta(\toh)| = 0.5826 $, $b_2^\crw = 
(1-\gamma_{\rm Euler})/(2 \pi) = 0.06729$,
$b_2^\urw = -\gamma_{\rm Euler}/(2 \pi) = - 0.09187 $,
$b_3^\crw = \toh \pi^{-1/2}$, 
$b_3^\urw =  \pi^{-3/2}$.

For unrestricted random walks, the leading terms of $F_{d,\urw}^\app
(t)$ can also be obtained with remarkable ease from its frequency
representation (\ref{eq:simpleFdurw}): replacing $\cW_\eff (\wb)$ by
$\cW_\app (\wb)$ and evaluating the leading contributions to the
integral in the limit $Tt \gg 1$, one readily recovers the leading
terms of \Eqs{subeq:F(t)g(L_t)} [including the correct prefactors
$c_d$].

Let us now calculate the full decoherence times $\ttau_{\varphi,d}$
(including next-to-leading order corrections).  For $t \simeq
\ttau_{\varphi,d}$, the next-to-leading terms in
\Eqs{subeq:F(t)g(L_t)} are parametrically parametrically smaller than
the leading ones by $g^{-1/2}_1 (L_{\varphi, 1})$ for $d = 1$, or $
1/\ln g_2 $ for $d = 2$, or $g^{-1/3}_3 (L_{\varphi, 3})$ for $d = 3$.
Therefore, we write $\ttau_{\varphi,d} = \tau_{\varphi,d}(1 +
\delta_d)$, where $\delta_d \ll 1$ is a small correction induced by
the next-to-leading terms, and first determine $\tau_{\varphi,d}$,
by setting $b_d \to 0$ and $\delta_d \to 0$. Then the
condition\cite{thresholdconstant} $F_d^\app (\tau_{\varphi,d}) = 1 $ yields
the the following selfconsistency relations and solutions,
\begin{subequations}
\label{eq:AAKrates}
  \label{subeq:tauphiselfconsistent}
  \begin{eqnarray}
    \label{eq:tauphi1}
    \gamma_{\varphi,1} & = & {c_1 T  \over g_1 (L_{\varphi,1})} \; 
    \qquad  \, \Rightarrow  \; 
    \gamma_{\varphi,1}=
    (c_1 \sqrt{\gamma_1} \, T )^{2/3} \!\! , \qqph  
    \\
    \gamma_{\varphi,2} & = & {c_2 T  
      \ln \left( T  \tau_{\varphi ,2} \right) 
      \over g_2 (L_{\varphi,2})} \; 
    \Rightarrow  \; 
    \gamma_{\varphi,2} =
    { c_2 T  \over  g_2 } \ln (g_2/c_2)  
\; , \qqph 
\label{eq:tauphi2}
\\
\gamma_{\varphi,3} & = & {c_3 T \over g_3 (L_T)} \quad 
 \qquad \Rightarrow  \; 
\gamma_{\varphi,3}=
{c_3 T^{3/2} \over \sqrt {\gamma_3} }\; ,
\label{eq:tauphi3}
  \end{eqnarray}
\end{subequations}
where $\gamma_1, g_2, \gamma_3$ are defined in
\Eq{eq:dimensionlessconductance}, and $L_T = \sqrt{D/T}$.  These
results 
reproduce those first derived by AAK for classical white Nyquist
noise.  They can be used to write the decay functions $F_d^\app (t)$
in the form (\ref{eq:finaldecayfunctions}) cited in the overview
[Sec.~\ref{sec:mainresultsF(t)}].

\begin{figure}
\begin{center}\includegraphics[%
  width=0.99\columnwidth]{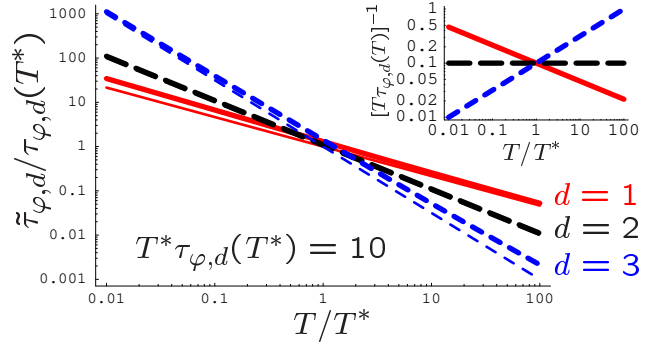}\end{center}
\caption{\label{cap:Deviations} Comparison of the commonly employed
  leading approximation for the decoherence time ${\tau}_{\varphi,d}$
  (thin lines) and the full decoherence time ${\tilde
    \tau}_{\varphi,d}$ (thick lines), as a function of temperature,
  for dimensions $d=1$ (full red lines), $d=2$ (black, long dashed),
  $d=3$ (blue, short dashed). The temperature $T^*$ has been
  (arbitrarily) chosen to give $T^* {\tau}_{\varphi,d}(T^*)=10$ in all
  cases (this amounts to fixing the values of the material parameters
  $\gamma_1$, $g_2$ and $\gamma_3$).  The magnitude of the correction
  is governed by the small parameter $[T \tau_{\varphi,d}]^{-1}$,
  who's $T$-dependence is shown in the inset: In $d=1$ ($d=3$) the
  corrections become smaller (and the weak localization theory is
  applicable) towards higher (lower) temperatures, where $[T
  {\tau}_{\varphi,d}(T)]^{-1} \ll 1$. In $d=2$, the correction amounts
  to a numerical constant factor (shift on the log-scale, hardly
  visible in this figure). Note, in particular, the slow decay of the
  correction in the case of $d=1$, where the correction falls off only
  like $T^{-1/6}$ [compare Eqs.  (\ref{eq:AAKrates}) and
  (\ref{eq:explicitshifts})].  }
\end{figure}

Next we work out the corrections to the decoherence times due to the
next-to-leading terms in \Eqs{subeq:F(t)g(L_t)}.  Reinstating $b_d
\neq 0$ and solving the condition $F_d^\app (\ttau_{\varphi,d}) = 1$
for $\delta_d (\tau_{\varphi,d})$, we find
\begin{subequations}
  \label{subeq:explicitshifts}
\begin{eqnarray}
  \label{eq:explicitshifts}
  \delta_1  & = &  {2 \over 3 } 
  {b_1 / c_1 \over \sqrt{ T \tau_{\varphi,1}}} \; , 
  \quad 
  \delta_2 = { b_2 / c_2  \over \ln (T \tau_{\varphi,2}) }   \; , 
  \quad
  \delta_3 = { b_3 /  c_3  \over \sqrt{ T \tau_{\varphi,3}}}  \; ,
  \nonumber \\
\end{eqnarray}
\end{subequations}
leading to \Eqs{overviewsubeq:tauphiselfconsistent} for
$\ttau_{\varphi,d}$.  As anticipated above, the correction factors
$\delta_d$ are parametrically small, being of order $g^{-1/2}_1
(L_{\varphi, 1})$ for $d = 1$, or $ 1/\ln g_2 $ for $d = 2$, or
$g^{-1/3}_3 (L_{\varphi, 3})$ for $d = 3$. We thus arrive at the most
important conclusion of this paper: the \emph{leading quantum
  corrections} to the classical results for the decoherence rates and
decay functions are \emph{parametrically small in the regime where
  weak localization theory is applicable}.  (The same qualitative
conclusion was arrived at by Vavilov and Ambegaokar\cite{AV} several
years ago by somewhat more indirect means). Nevertheless, note that
the next-to-leading corrections are still parametrically larger than
all futher subleading corrections, that could arise, \eg, from
calculating $F_d^\app(t)$ to second order in the interaction
propagator, or from including crossterms between weak localization and
interaction corrections (as considered diagrammatically in \Ref{AAG}),
since such corrections are all smaller than the leading ones by at
least $1/ g_d (L_\varphi)$. The leading and next-to-leading
approximations to the decoherence times are plotted for all three
dimensions in Fig. \ref{cap:Deviations}.  As it is not uncommon for
weak localization experiments to reach to the regime where the product
$T \tau_{\varphi}$ is only on the order of $10$ (e.g. Ref.
\onlinecite{Pierre03}), we emphasize that the corrections discussed
here can amount to an appreciable effect. We also remark that, in
$d=1$, the relative size of the correction only falls off very slowly
with increasing temperature (like $T^{-1/6}$).

The temperature dependence predicted for the corrected decoherence
times $\ttau_{\varphi,d} (T)$ can be compared to experiment by
proceeding as follows: Express $F_d(t)$ of \Eqs{subeq:F(t)g(L_t)} as a
function of the parameters $(t/ \ttau_{\varphi,d})$ and $T
\ttau_{\varphi, d}$, by inserting $\tau_{\varphi,d} =
\ttau_{\varphi,d} [1 - \delta_d (\ttau_{\varphi,d})]$.  Calculate the
magnetoconductivity $\sigma^\WL_d (H)$ numerically from
\Eq{eq:sigma(H)rescaled}, and for given $T$, adjust the parameter
$\ttau_{\varphi,d}$ such that the numerical curve as a function of
magnetic field best fits the measured curve. Repeat for various $T$,
and compare the function $\ttau_{\varphi,d} (T)$ obtained by this
fitting procedure to the function $\tau_{\varphi,d} (1 + \delta_d)$
predicted above. [If magnetic impurities are suspected to be present,
insert a factor $e^{- t/ \tau_{{\rm m}}}$ into \Eq{eq:sigma(H)rescaled}
  and treat the magnetic scattering time $\tau_{\rm m}$ as a fit
  parameter.  Spin-orbit scattering is not included in our analysis,
  but the corresponding generalization should be straightforward.]

To end this section, some remarks on the role of an ultraviolet cutoff
seem to be in order at this point: for quantum noise in the absence of
Pauli blocking, an ultraviolet cutoff always has to be introduced to
arrive at a finite result for the decoherence rate, to reqularize the
contribution of spontaneous emission processes which occur at all
frequencies (see our discussion in \Sec{sec:spontemission}). In the
full theory, Pauli blocking counteracts spontaneous emission and
introduces via $\cW_\app (\wb)$ an effective ultraviolet cutoff at
frequency transfers of order $T$.  Remarkably, for $d=1$ (but not for
$d=2,3$), the leading result for the decoherence rate can nevertheless
be correctly obtained by simply employing the classical white noise
spectrum $\cW_\class (\wb) = T$ (which contains no Pauli blocking, but
no spontaneous emission either ) over the \emph{full} frequency range
up to arbitrary frequencies.  The reason is
that for $d= 1$, the dominant contribution to decoherence for
time-reversed diffusive paths of duration $t$ comes from frequencies
$\wb \sim D \qb^2 \sim 1/ t $ [cf. \Eq{eq:definedeltatildeP}], which
in the limit $1 \ll T t $ of present interest implies $\wb \ll T$; but
for these frequencies, the \weightingFunction\ $\cW_\pp (\wb)$ reduces
simply to $T$, which equals the classical \weightingFunction\
$\cW_\class (\wb) $.

  In contrast, for $d=2,3$ the large-frequency regime makes a
  logarithmic contribution for $d=2$, and dominates for $d=3$,
  requiring an ultraviolet cutoff to be present in the theory.  AAK had
  to introduce such an ultraviolet cutoff by hand for the cases
  $d=2,3$, because they considered only classical white Nyquist noise,
  whose instantaneous kernel $W_\class (t_{34}) = T \delta (t_{34})$
  involves \emph{no} upper frequency cutoff, which is unphysical.
  Indeed, when one attempts to use it in \Eq{eq:Fdecayalmostfinal},
  the $\int dt_{34}$ integral would be ill-defined, since the $\delta
  (t_{34})$ would produce a $\ln (0)$ or $1/\sqrt{0}$ in $\cP_2$ or
  $\cP_3$ of \Eqs{subeq:expandPd}, respectively. Equivalently, using
  $\cW_\class (\wb) = T$ in \Eq{eq:simpleFdurw}, the frequency
  integral would be ultraviolet-divergent for $d = 2,3$.
  AAK cured this problem by introducing, by hand, an
  ultraviolet cutoff at the scale of the temperature (taking $|\qb|
  \lesssim \sqrt {T/ D}$), and adding, by hand, a term to the
  decoherence rate describing the effect of electron-electron
  collisions with large energy transfers $\wb \gtrsim T$, which had
  been calculated earlier \cite{AltshulerAronov81} within the
  framework of a kinetic equation.  (GZ's work implicitly questions
  this approach, in that they used $1/\tauel \; (\gg T)$ as upper
  cutoff.) 

  Satisfactorily, an ultraviolet cutoff of precisely the type used by
  AAK arises automatically in our treatment of quantum Nyquist noise
  in combination with the Pauli principle, in the form of the
  energy-averaged \weightingFunction\ $\cW_\app (\wb)$: it regularizes
  the large-frequency behavior of \Eq{eq:simpleFdurw}, without the
  need to consider processes with large energy transfers separately.
  Equivalently, in the time domain, it results in $W_\app (t_{34})$
  being a broadened $\delta (t_{34})$-function of width $1/T$. Thus,
  in a very natural (and perhaps somewhat more elegant) manner, we
  have confirmed the validity of AAK's use of temperature as an
  ultraviolet cutoff. Moreover, our explicit treatment of this
  cutoff was essential for accurately calculating the
  next-to-leading terms for the decay function and decoherence times.

  \subsection{Comparison with Magnetoconductivity of AAG}
\label{sec:AAGcompare1}

It is instructive to check the use of the Pauli-blocked correlator
$\bigl\langle \hat{V} \hat{V} \bigr\rangle^\pp_{\qb \wb} $ introduced
above against the results of Ref.~\onlinecite{AAG} (AAG).  There, the
conductivity was calculated diagrammatically for the limit of a
moderately strong magnetic field (for $\gammaphi \ll \gammaH \ll T$).
In this regime the trajectories relevant for weak localization are so
short ($t \lesssim \tauH \ll \tauphi$) that the effects of interaction
on weak localization are still small, so that it suffices to calculate
$\delta\sigma^{\rm WL}_d$ to first order in the interaction. At
the same time, the condition $1/T \ll \tauH$ ensures that the premise
for our calculations of $\tauphi$ in previous sections, namely $1/T
\ll \tauphi$, still holds.  AAG thus calculated the conductivity
diagrammatically to first order in the interaction and including all
contributions of order $1/g^2$ (including not only weak localization
terms, but also interaction corrections and cross terms involving
both). Among these $1/g^2$ terms, AAK identified the one that
decreases most rapidly with magnetic field (largest power of $\tauH $)
as the one relevant for decoherence, and proposed to extract $\tauphi$
from it.  They found that this term has the following form [Eq.~(4.5)
 of Ref.~\onlinecite{AAK}, rewritten in terms of quantities
introduced above],
\begin{eqnarray}
\delta \sigma^{\WL (1)}_{d,\sAAG}
& = & 
-\frac{\sigma_d}{\pi\nu \hhbar}
\int (d\bar{\omega}) (d\bar{q}) (dq) \, 
\staroverhbarsq{2} \, \Bigl \langle 
\bigl \langle \hat V \hat V\bigr \rangle_{\qb\wb}^{\pp} \Bigr \rangle_\ve
\label{eq:AAGresult}
\\
& & 
\times \Big[
\bcC^0_{\bmq - \bbmq} (0 ) \, |\bcC^0_\bmq (\bomega)|^2 
\, - \, 
\left[\bcC^0_\bmq (0 )\right]^2 \,  \bcC^0_{\bmq - \bbmq} ( \bomega) 
\Bigr] \; .   \nonumber 
\end{eqnarray}
AAG evaluated the integrals
using dimensional regularization, finding the
following results\cite{temp=freq} [first two terms
of their Eqs.~(4.13)\cite{AAGtypo2}; we also evaluate their general
result (4.11) for $d=3$, sending $(3-d)^{-1} \to \ln (\tauH/\tauel)$
in the limit $d \to 3$, as appropriate for their dimensional
regularization scheme]:
  \begin{subequations}
\label{eq:AAKresultfinal2d}
\begin{eqnarray}
{  \delta \sigma^{\WL(1)}_{1,\sAAG} \over \sigma_1 } & = & 
  { (T \tauH)^{3/2} 
  \left[ 1 - {|\zeta (\toh) |  \over
      \sqrt{\pi T \tauH / 2}} + \dots \right] 
\over 4 \pi g_1(L_H) g_1 (L_T) }
 \; ,  \qqph 
\label{eq:AAKresultfinal1d}
\\
{\delta \sigma^{\WL(1)}_{2,\sAAG} \over \sigma_2 } & = & 
{  T \tauH  \over 4 \pi^3 g^2_2 } 
\Bigl[ \ln (T \tauH)  - 1  + \dots \Bigr]  \; ,
\label{eq:AAKresultfinal12}\\
{\delta \sigma^{\WL(1)}_{3,\sAAG} \over \sigma_3 } &=&
{ T \tauH 
\left[ {3 \zeta({3 \over 2}) \over (\pi^7 2^9)^{1/2}} - 
{\ln (\tau_H / \tau_{\rm el}) \over 8\pi^3 
\sqrt{T \tauH} } + \ldots \right] 
\over g_3(L_H) g_3 (L_T)} 
\; . \qqph
\label{eq:AAKresultfinal3d}
\end{eqnarray}
\end{subequations}
Here $\dots$ indicates subleading terms with a weaker
$\tauH$-dependence. Moreover, AAG showed that the contribution from
cross terms between interaction corrections and weak localization
(their Eq.~(4.7) for $\delta \sigma^\sAAG_{\rm CWL}$) produce a
$\tauH$-dependence \emph{weaker} than both the leading and
next-to-leading terms of \Eqs{eq:AAKresultfinal2d}.

\Eq{eq:AAGresult} can be reproduced from the analysis presented above,
by calculating the magnetoconductivity using our first order
expression for the Cooperon.  To this end, use $\tC^{1, \ve } = -
\tC^0  F^{\pp}_{d,\crw} $ [cf.\ \Eq{eq:Frrdefine}] in
\Eq{eq:magnetoconductance} for the magnetoconductivity and average
over $\ve$:
\begin{subequations}
\label{subeq:deltaSigmaEpsilon}
\begin{eqnarray}
\label{deltaSigmaEpsilon}
{ \delta\sigma^{\rm WL(1)}_d  \over \sigma_d} 
& = &
\frac{1}{\pi\nu \hbar }
\int_\tauel^{\infty}dt \, \tilde C^0 (0,t) 
\bigl \langle   F^{\pp}_{d, \crw} (t) 
\bigr \rangle_\ve  \, , 
\\
\label{deltaSigmaEpsilont-integral}
& = & 
\frac{2^{1-d}}{ \pi^{1 + d/2}}
\int_{{\tauel\over \tauH}}^{\infty}dx \, 
{e^{-x} \over x^{d/2}} 
{  \bigl \langle   F^{\pp}_{d, \crw} (x \tauH) 
  \bigr \rangle_\ve  \over g_d (L_H)} \,  \qqph 
\end{eqnarray}
\end{subequations}
(where $L_H = \sqrt {D \tauH}$).  Now substitute
\Eq{exponentCLpaulimodified} for $F^{\pp}_{d,\crw} (t) $ into
\Eq{deltaSigmaEpsilon}, represent the $\delta \bP$ occuring therein
via \Eq{eq:barPPPa} for closed random walks, with the $\bP$ in
\Eq{eq:barPPPa} standing for $\bP^\crw_{(0, t)}(\qb , t_{34})$,
represented by the first line of \Eqs{eq:rrw-result-explicit}; then
Fourier transform the three Cooperons occuring in its numerator to the
frequency domain and finally perform all time integrals
[\EqII{subeq:definePfull} is helpful in this regard]; the result is
found to be identical to \Eq{eq:AAGresult}, \ie\ $\delta\sigma^{\rm
  WL(1)}_d = \sigma^{\WL (1) }_{d, \sAAG}$. The same conclusion can be
reached by comparing AAG and our results \emph{after} all necessary
integrals have been performed: inserting \Eqs{subeq:F(t)g(L_t)} for
$F^\app_d (t)$ into \Eq{deltaSigmaEpsilont-integral} for $
\delta\sigma^{\rm WL(1)}_d $ and performing the time integral, we
recover precisely AAG's results (\ref{eq:AAKresultfinal2d}) for
$\delta \sigma^{\WL(1)}_{d,\sAAG}$.  Thus, \emph{our theory is
  consistent with the calculation of AAG}.  Note, in particular, that
the next-to-leading terms of AAG's results for $\delta
\sigma^{\WL(1)}_{d,\sAAG}$ are also correctly reproduced in this
manner; in our approach, they are generated by the subleading
contributions (the $b_d$-terms) to our decay functions $F_d^\app (t)$.
This is a very useful consistency check. It illustrates firstly that
our calculation of the next-to-leading corrections to the decoherence
rate is correct, and secondly that the latter do not contain any
contributions from the cross terms between weak localization and
interaction corrections (which we did not calculate).

AAG proposed to extract the decoherence times $\tau_{\varphi,d}$ from
their final results for $\delta\sigma^{\rm WL(1)}_{d, \sAAG}$
[\Eqs{eq:AAKresultfinal2d}]. To this end, a choice has to be made
about the functional dependence on time of the full Cooperon $\tC
(0,t)$ (which AAG did not calculate explicitly), or, in our scheme,
about the shape of the decay function $F_d (t)$.  Different choices
for $F_d (t)$ imply different ``definitions'' of $\tauphi$, with
different functional dependencies on temperature and magnetic field.
In their Eq.~(3.2), AAG chose to define $1/\tau_{\varphi,d}^\sAAG$ as
a contribution to the ``Cooperon mass'' [in the sense of
\EqII{eq:definegammaphiagain}], which implies that they assumed simple
exponential decay for the Cooperon, $e^{-t(1 / \tauH +
  1/\tau_{\varphi,d}^\sAAG)}$, thus effectively making the choice
$F_{d}^\sAAG (t) = t/\tau_{\varphi,d}^\sAAG$. Inserting this into
\Eq{deltaSigmaEpsilon} one finds
\begin{eqnarray}
  \label{eq:AAGdeftauphi}
{ \delta\sigma^{\rm WL(1)}_d  \over \sigma_d} 
& = &
\frac{2^{1-d} \Gamma (2- {d \over 2})}{ \pi^{1 + d/2} g_d (L_H)}
{\tauH \over \tau_{\varphi,d}^\sAAG}
%
\qqph 
\end{eqnarray}
[reproducing\cite{AAGtypo}  AAG's (4.3)]. When equated to
the leading terms of \Eqs{eq:AAKresultfinal2d}, this yields for $d=1,2$
\begin{subequations}
  \label{subeq:AAGwrongtauphi}
\begin{eqnarray}
  \label{eq:AAGwrongtauphi}
  \tau_{\varphi,1}^\sAAG & = & {2  g_1(L_H) \over T } ,
  \quad \
  \tau_{\varphi,2}^\sAAG  =  {2 \pi g_2 \over T \ln (T \tauH)},
 \end{eqnarray}
reproducing\cite{AAGtypo} AAG's (4.9), while for $d=3$ we obtain
\begin{eqnarray}
  \label{eq:eq:AAGwrongtauphi3}
    \tau_{\varphi,3}^\sAAG  =  {\sqrt{\pi^3 2^5} \over 
    3 \zeta ({3  \over 2} )} \, {g_3 (L_T)  \over T } .
  \qph  
\end{eqnarray}
\end{subequations}
Now, for $d=3$, \Eq{eq:eq:AAGwrongtauphi3} reproduces the classical
result of AAK [\Eq{eq:tauphi3}].  However, for $d=1,2$
\Eqs{eq:AAGwrongtauphi} for $ \tau_{\varphi,d}^\sAAG $ depend on
magnetic field and hence are inconsistent with AAK's results
[\Eqs{eq:AAKrates}] for $\tau_{\varphi,d}^\sAAK$, which are
magnetic-field independent, since AAK chose to define $1/\tauphi$ as
the decoherence rate which the Cooperon would have in the absence of a
magnetic field.
The reason for this inconsistency is that for $d=1,2$, the Cooperon
decay is not purely exponential in time [cf.\ \Eqs{subeq:F(t)g(L_t)}],
so that the usual strategy of adding inverse decay times to determine
the total decay rate of two independent decay mechanisms cannot be
used (as emphasized in the paper by AAK, after Eq.~(32) of \Ref{AAK}).

AAK's magnetic-field-independent results for the decoherence time can
be extracted from AAG's result for $\delta \sigma^{\WL(1)}_{d,\sAAG}$
only if the correct functional form for the decay function is used.
Indeed, inserting the leading terms of \Eqs{subeq:F(t)g(L_t)} for
$F_{1,2}^\app (t)$
into \Eq{deltaSigmaEpsilon}, we find
\begin{subequations}
  \label{eq:sigmajvdproperall}
\begin{eqnarray}
  \label{eq:sigmajvdproper}
  {\delta\sigma^{\rm WL (1)}_1 \over \sigma_1}
 &  = &  
 { 1 \over g_1 (L_H) } {\tau_H^{3/2} 
   \over (\pi \tau_{\varphi, 1})^{3/2}}  \; ,
 \\
 \label{eq:sigma2correctguess}
 {\delta\sigma^{\rm WL(1)}_2  \over \sigma_2}   & = & 
 {1  \over 2 \pi^2 g_2 } \,{ \tauH \over  \tau_{\varphi,2}}
 {\ln (T \tauH) \over  \ln (T \tau_{\varphi,2})} \; .
\end{eqnarray}
\end{subequations}
When equated to \Eqs{eq:AAKresultfinal2d}, this yields 
\begin{eqnarray}
  \label{eq:jvdtauphi=AAK}
\tau_{\varphi,1} = \left(T \sqrt { \gammaone \pi }/4\right)^{-1},
\quad
\tau_{\varphi,2} = g_2 /[T c_2 \ln (T \tau_{\varphi,2})], \qph
\end{eqnarray}
implying $ \gamma_{\varphi,2} = { c_2 T \over g_2 } \ln (g_2/c_2) $
and thus reproducing AAK's results (with proper prefactors included),
as given by the rightmost equations of
(\ref{subeq:tauphiselfconsistent}). 

To summarize this subsection, we conclude that, satisfyingly, our
results for the decay functions $F_d^\app (t)$ provide a bridge
between the work of AAK and AAG: they allow AAK's results for
$\tau^{\sAAK}_{\varphi,d}$, obtained by treating classical white
Nyquist noise nonperturbatively, thereby achieving results free from
infrared problems, to be extracted from AAG's results for $\delta
\sigma^{\WL (1)}_{d,\sAAG}$, obtained by treating fully quantum noise
perturbatively, thereby incorporating Pauli blocking and obtaining
results free from ultraviolet problems. The fact that our approach is
able to make such a connection between two sets of established
results, one nonperturbative but classical, the other quantum but
perturbative, may be regarded as a strong indication that our method
is fundamentally sound.

\subsection{Energy dependence of the decay function}
\label{sec:Energy-dep-decay-function}

Instead of averaging over the energy $\hbar \ve$, it is also
interesting to calculate the dependence of the decoherence rate
$\gamma_\varphi^{\ve, T}$ on both temperature \emph{and} energy, as
would be relevant for an electron injected into a disordered metal
with a definite energy (\eg\ in a geometry such as that used by
Pothier \emph{et al.}  \cite{Pothier97}). To the best of our
knowledge, this energy-dependence has not been studied before. We
shall now obtain it by analysing the energy-dependence of the decay
function of \Eq{exponentCLpaulimodified}, for the
case of closed random walks.

\begin{figure}
  \begin{center}\includegraphics[%
    width=0.90\columnwidth]{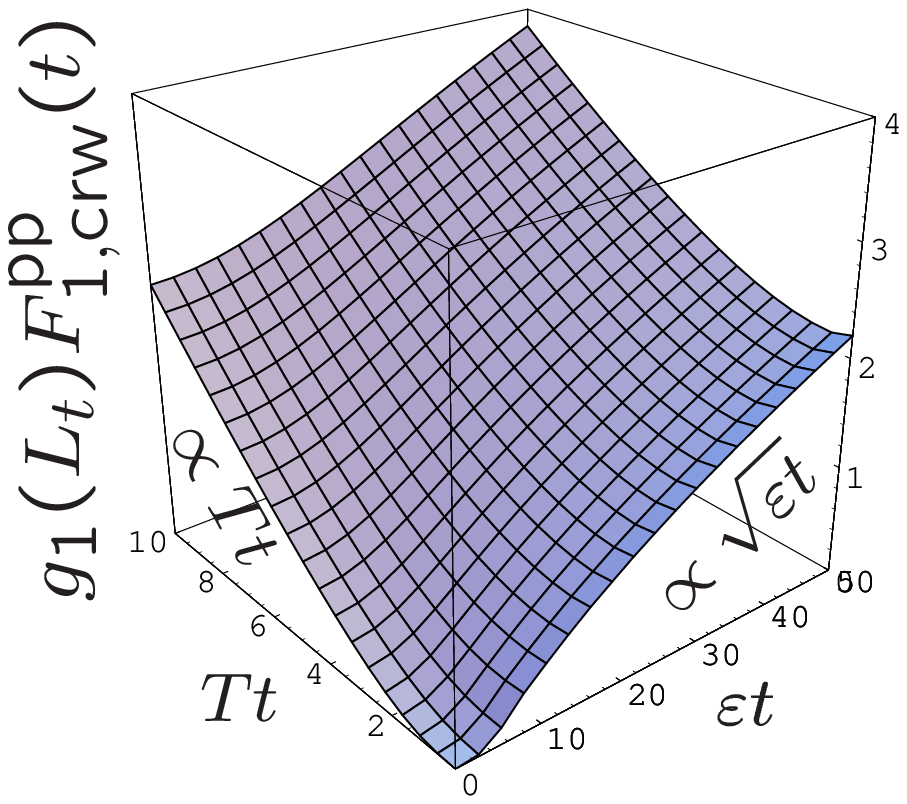}\end{center}
\begin{center}\includegraphics[%
  width=0.90\columnwidth]{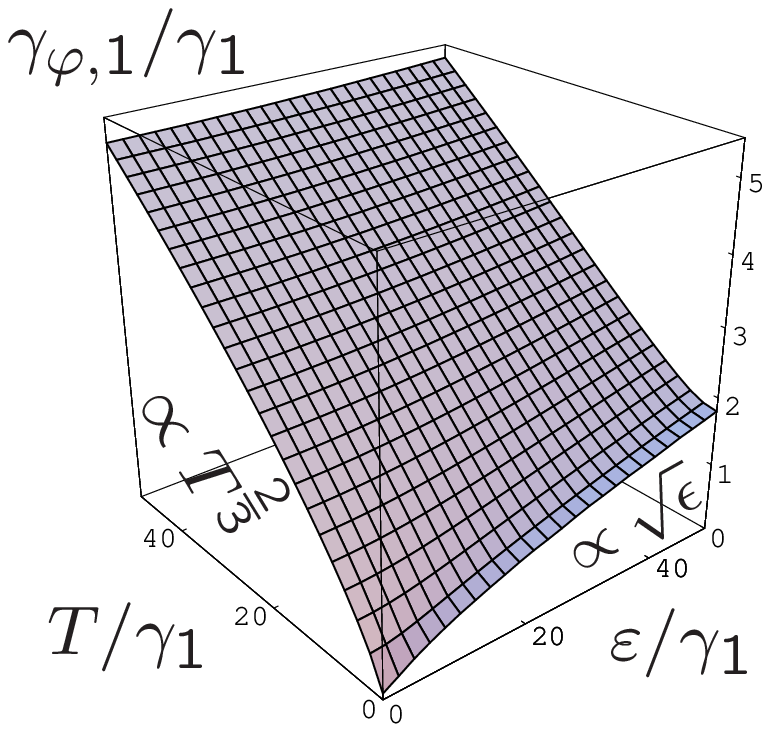}\end{center}
\caption{\label{fig:FET(t)} (a) $g_1 (L_t) F_{1,\crw}^\pp (t)$
  [\Eq{eq:FppETT(t)}] as function of the parameters $\ve t$ and $T t
  $.  (b) The energy- and temperature-dependent
  decoherence rate $\tau_{\varphi,1}$, defined from the condition
  $F^\pp_{1,\crw} (\tau_{\varphi,1}) = 1 $, as function of $T/
  \gammaone$ and $\ve / \gammaone$.}
\end{figure}

To this end, we need the Fourier transform of 
the Pauli-principle-modified \weightingFunction\ $\cW_\pp (\wb)$ 
of \Eq{eq:weightingpp}, which  can  be calculated by closing
the $\int d \bomega$ integral [\Eq{eq:defineKernel}]
along a semicircular contour in the
complex plane. The result can be written 
as $W_\pp (t_{34}) = \pi T^2 w (\ve t_{34}, \pi T t_{34})$, where
\begin{eqnarray}
  \label{eq:WppET(t)}
w(y, z) & = &
{ \cos y\cosh z + (y / z) 
\sin y \sinh z - 1 \over 2 \sinh^2 z} \; . \qqph
\end{eqnarray}
Inserting this into \Eq{eq:Fdecayalmostfinal} for $F(t)$ we 
obtain the expression
\begin{eqnarray}
  \label{eq:FppETT(t)}
F^\pp_{d, \crw} (t) \!  & = & \!\!  { Tt  \over 
g_d(L_t)}
\int_0^{\pi Tt} \!\!\! dz \, w (z x/\pi, z) \, 
 \cP_d^\crw   \! \left({z \over \pi T t }\right)   
, \qqph 
\end{eqnarray}
which shows that $F^\pp_d(t) g_d (L_t)$ is a function of the
parameters $\pi T t$ and $x=\ve/T$, or of $\pi Tt$ and $\ve t$.  
Fig.~\ref{fig:FET(t)}(a) shows the latter functional 
dependence for $d=1$.
Moreover, Fig.~\ref{fig:FET(t)}(b) shows the corresponding energy- and
temperature-dependent decoherence rate $\tau_{\varphi,1}$, defined
from the usual condition $F^\pp_{1,\crw} (\tau_{\varphi,1}) = 1$.

To extract the decoherence rate analytically from $F_d^\pp (t)$, we need
its  asymptotic behavior for large times. 
We find 
that the dominant behavior of $F_d (t)$ for 
either $Tt \gg 1$ or $\ve t \gg 1$ (or both), but
arbitrary ratios of $\ve/T$, is given by the following
expressions:
\begin{subequations}
  \label{subeq:Fcrossover}
  \begin{eqnarray}
    \label{eq:Fcrossover1}
F^\pp_{1, \crw} (t) & = &  { T t \over g_1(L_t)}
  \left[ c_1  + { \cF_1 ({\ve /  T})  \over ( T t)^{1/2}} 
\right] , \qqph 
\\
    \label{eq:Fcrossover2}
F^\pp_{2, \crw} (t) & = &  { T t \over g_2(L_t)}
  \left[ c_2 \ln (Tt)  +  \cF_2 ({\textstyle {\ve / T}}) 
\right],  
\\
    \label{eq:Fcrossover3}
F^\pp_{3, \crw} (t) & = &  { (Tt)^{3/2} \over g_3(L_t)}
  \left[     \cF_3 ({\textstyle {\ve / T}}) 
- {1 \over 2 \sqrt{\pi T t}} 
\right] \; . 
  \end{eqnarray}
\end{subequations}
The crossover functions $\cF_d(\ve/T)$ are defined 
by the relations 
\begin{eqnarray}
  \label{eq:definecalF}
\cF_d (x) = \int_0^\infty d \tilde z \, 
w(\tilde z x / \pi, \tilde z)
\left\{ 
\hspace{-1.5mm} 
\begin{array}{c}
-4 \sqrt{\tilde z} / \pi 
\rule[-2.5mm]{0mm}{0mm} \\
- [\ln (\tilde z /\pi) +2 ]/ \pi
\rule[-2.5mm]{0mm}{0mm} \\
1 / (\pi \sqrt{\tilde z}) 
\end{array} \; 
\hspace{-1.5mm} \right\}
\end{eqnarray}
for $d = 1,2,3$, respectively. They 
have the properties
  \begin{eqnarray}
    \label{eq:calFcrossover}
    \cF_d(x) = \left\{ 
\begin{array}{ll}
\rule[-3mm]{0mm}{0mm}
\cF_d(0) + {\cal O}(x^2 ) &
\quad \textrm{for} \quad x \ll 1 \; ,  \\
\tilde c_d \, x^{d/2} &
\quad \textrm{for} \quad x \gg 1  \; ,
\end{array} 
\right. 
  \end{eqnarray}
  with $\cF_1 (0) = -0.6826$, $\cF_2 (0) = - 0.1161$, $\cF_3 (0) =
  0.2145$ and $\tilde c_1 = \sqrt 2 / \pi$, $\tilde c_2 = 1 / (4
  \pi)$, $\tilde c_3
  = 
  1/(3 \sqrt 2 \pi^2)$.  They govern the crossover of the behavior of
  the decay functions from the regimes of small to large ratios $\ve/T$.
 
  For $\ve/T \ll 1$, $F_d^\pp(t)$ has precisely the same form as
  $F_d^\app (t)$ of \Eqs{subeq:F(t)g(L_t)}, with the same prefactors
  $c_1$ and $c_2$; only the prefactor of $F_3^\pp (t)$ is slightly
  different, namely $c_3 = \cF_3 (0) = 0.2145$.  Therefore, the
  decoherence rates have the same form
  (\ref{subeq:tauphiselfconsistent}) as that derived from $F_d^\app
  (t)$.

In contrast, for $\ve/ T$ large enough that the large-$x$ behavior of
$\cF_d (x)$ dominates the behavior of $F_d^\pp (t)$, 
\Eqs{subeq:Fcrossover} reduce to
\begin{eqnarray}
\label{eq:Fdpplargee/T}
F_d^\pp (t)   & = & 
{\tilde c_d \, (\ve t)^{d/2} \over  g_d (L_t) } = t / \tau_{\varphi , d}
\; , \quad 
\gamma_{\varphi,d} = {\tilde c_d \, \ve \over g_d (L_\ve)} \; . \qqph 
\end{eqnarray}
More explicitly, the corresponding decoherence rates and 
the crossover scales above which they apply are given by
\begin{eqnarray}
\nonumber 
\gamma_{\varphi , 1}   & = & 
\tilde c_1 \, (\gamma_1 \ve)^{1/2}  
\, , \: \quad \, \textrm{for} \;
\ve \gtrsim  (T^4 \!/\gamma_1)^{1/3} = T g^{2/3}_1 (L_T) \; ,
\\
\nonumber 
\gamma_{\varphi , 2}   & = & 
\tilde c_2 \, \ve / g_2 \, , \qquad \quad  \textrm{for} \;
\ve \gtrsim T   \ln g_2   \; ,
\\
\label{eq:gammaphilargee/T}
\gamma_{\varphi , 3}   & = & 
\tilde c_3 \, (\ve^3/ \gamma_3)^{1/2} \, , 
\; \: \textrm{for} \; 
\ve \gtrsim T \; .
\end{eqnarray}
 Thus, for sufficiently high energies
the decoherence rate has the same functional form as the inelastic
energy relaxation rate in quasi $d$ dimensions
\cite{AltshulerAronov81}.  It is interesting
to note that for $d=1,2$ the
crossover scales above which $\ve $ has to lie for
these relations to hold are parametrically much larger than
temperature. 

The crossover behavior of $\tau_{\varphi,d}$ between the regimes of
small and large $\ve/T$ can be determined explicitly, if desired, from
\Eqs{subeq:Fcrossover}, using the usual relation $F_d^\pp
(\tau_{\varphi,d}) = 1$.

Note that for unrestricted random walks, the leading terms of
  $F_{d,\urw}^\pp (t)$ can again easily be obtained from its frequency
  representation (\ref{eq:simpleFdurw}), by replacing $\cW_\eff (\wb)$
  therein by $\cW_\pp (\wb)$ [\Eq{eq:weightingpp}], which suppresses
  frequencies $\wb \gtrsim \max \{T, \ve \}$.  Evaluating the leading
  terms of \Eq{eq:simpleFdurw} in the limits $\ve/T \to 0$ and $Tt \gg
  1$, we recover the leading ($\ve/T = 0)$ terms of
  \Eqs{subeq:Fcrossover} for $F_{d, \urw}^\pp (t) $, with correct
  prefactors $c_d$, and in the limits $T/ \ve \to 0$, $\ve t \gg 1$,
  we recover \Eqs{eq:Fdpplargee/T}, with correct prefactors $\tilde
  c_d$.  Since the derivation of \Eq{eq:simpleFdurw} involved dropping
  some subleading terms, the results which it will produce for the
  crossover behavior between the regimes of small and large ratios of
  $\ve/T$, and the subleading terms of $F_d (t)$, will be
  quantitatively different from those of the more accurate
  \Eq{eq:FppETT(t)}.  However, qualitatively the
  crossover behavior is very similar.

  The fact that the functional dependence of $\tau_{\varphi,1,2}$ on
  $\ve$ for $\ve/ T \ll 1$ is different than its dependence on $T$ for
  $\ve/T \gg 1$, whereas for $\tau_{\varphi,3}$ it is the same, can be
  understood very nicely from the frequency representation
  (\ref{eq:simpleFdurw}) of $F_{d,\urw}^\pp$: for $T/\ve \gg 1$,
  decoherence is dominated by high frequencies $\wb \sim T$ only for
  $d=3$; for $d=2$, the contribution from low frequencies of order
  $\wb \sim 1/t$ is as important as those from $\wb \sim T$, and for
  $d =1 $, the contribution from $\wb \sim 1/t$ dominates. Thus, the
  infrared cutoff matters for $d = 1$ and $2$, but not for $d=3$. This
  is reflected in the fact that the first set of relations for
  $\gamma_{\varphi,d}$ in \Eqs{subeq:tauphiselfconsistent} involve
  selfconsistency relations only for $d = 1$ and $2$, but not for
  $d=3$. In contrast, in the opposite regime of $\ve / T \gg 1$,
  decoherence is dominated by frequency transfers of order $\ve$ not
  only for $d=3$, but also for $d=1,2$, so that the infrared cutoff is
  never important [to see this explicitly, use the $T=0$ version of
  $\cW_\pp (\wb)$, namely $\toh |\wb| \theta ( |\ve| - |\wb|)$, in
  \Eq{eq:simpleFdurw}].  Accordingly, none of the relations
  $\gamma_{\varphi,d} = \tilde c_d \ve /g_d (L_\ve)$
  [\Eq{eq:Fdpplargee/T}] involves a self-consistency condition, for
  every $d$.

\section{Relation to the work of Golubev and Zaikin}
\label{sec:GZcomparison}
\label{app:GZrelation}

To close this paper, we shall now put the use of the Pauli-blocked
noise correlator $\bigl\langle \hat{V} \hat{V} \bigr\rangle^\pp_{\qb \wb}
$ introduced in \Eq{subeq:theGreatReplacementa} on a firmer footing,
by summarizing how it follows from the analysis of
Refs.~\onlinecite{GZ2} and \onlinecite{vonDelft04}.  In
Ref.~\onlinecite{GZ2}, Golubev and Zaikin developed an influence
functional formalism and derived an effective action that explicitly
and correctly included the Pauli principle.  Indeed, a careful (if
anfractuous) reanalysis of GZ's approach by von Delft
\cite{vonDelft04} has shown that one can, in fact, fully recover
Keldysh diagrammatic perturbation theory from it (by starting from the
initial, exact path integral expression for the influence functional,
and properly including fluctuations).  However, when evaluating their
effective action along time-reversed paths, GZ did not adequately
account for the effects of recoil (as will be explained below). In
Ref.~\onlinecite{vonDelft04} [see App.~B.6.3], von Delft showed that
the effects of recoil can be accommodated in the effective action
by ``dressing'' the interaction correlators $\tilde {\cal
  L}^{R}_{ij} (\bomega)$ and $\tilde {\cal L}_{ij}^{A} (\bomega)$ (in
the position-frequency representation) by suitably chosen ``Pauli
factors'' 
\begin{equation}
\label{eq:Paulifactors}
[1-2f(\ve \mp \wb)] = \tanh[(\ve \mp \bomega)/2T] \equiv
\textrm{th}_\mp \; , 
\end{equation}
Instead of recapitulating the (lengthy) derivation of the latter
conclusion, we shall now offer a plausibility argument for it, based
on the requirement of consistency with Keldysh perturbation theory.

\subsection{Describing Pauli blocking by dressed
interaction propagators}

\label{sec:Simplified-influence-action}

One way to see that the original Feynman-Vernon influence functional $e^{-S_\eff \shbar }$,
with $S_\eff$ given by \Eqs{subeq:Sefftwo} and
(\ref{classicalvsquantumfields}), cannot directly be used in a
many-body situation is that its expansion in powers of $S_\eff \shbar
$ does \emph{not} reproduce Keldysh perturbation theory, because the
latter contains Pauli factors, while the former does not. In Keldysh
perturbation theory, the diagrams relevant for the calculation of the
Cooperon have the property that each occurrence of an electron Keldysh
Green's function
\begin{eqnarray}
  \label{eq:GKeldysh}
  \G^{K}_{ij}(\ve \mp\wb)
= \textrm{th}_\mp \: [\G^{R}-\G^{A}]_{ij} (\ve \mp \wb)
\end{eqnarray}
with $\textrm{th}_\mp = 1-2f(\ve \mp \bomega) = \tanh[(\ve -
\wb)/2T]$, is \emph{always} accompanied by either a retarded or an
advanced (but never a Keldysh) interaction propagator attached to one
of its two ends, $\tilde {\cal L}^{R/A}_{j\bar j}(\bomega)$ or $\tilde
{\cal L}^{R/A}_{\bar i i} (\bomega)$.  Since Pauli factors enter in
Keldysh perturbation theory only via $\tilde G^K$, this means that for
the diagrams of present interest, \emph{every} occurence of a Pauli
factor is \emph{always} accompanied by an $\tilde {\cal L}^{R/A}
(\bomega)$ propagator.  To be consistent with this fact, the effective
action in the influence functional approach must evidently contain the
\emph{same} combinations of Pauli factors and propagators $\tilde
{\cal L}^{R/A} (\bomega)$. This is not the case, however, for the
$\tilde {\cal L}^{R/A} (\bomega)$ occuring in
\Eqs{classicalvsquantumfields}. Thus, these propagagors have to be
``dressed'' by Pauli factors if we are to achieve consistency with
Keldysh perturbation theory.

\begin{figure}
\begin{center}\includegraphics[%
  width=0.95\columnwidth]{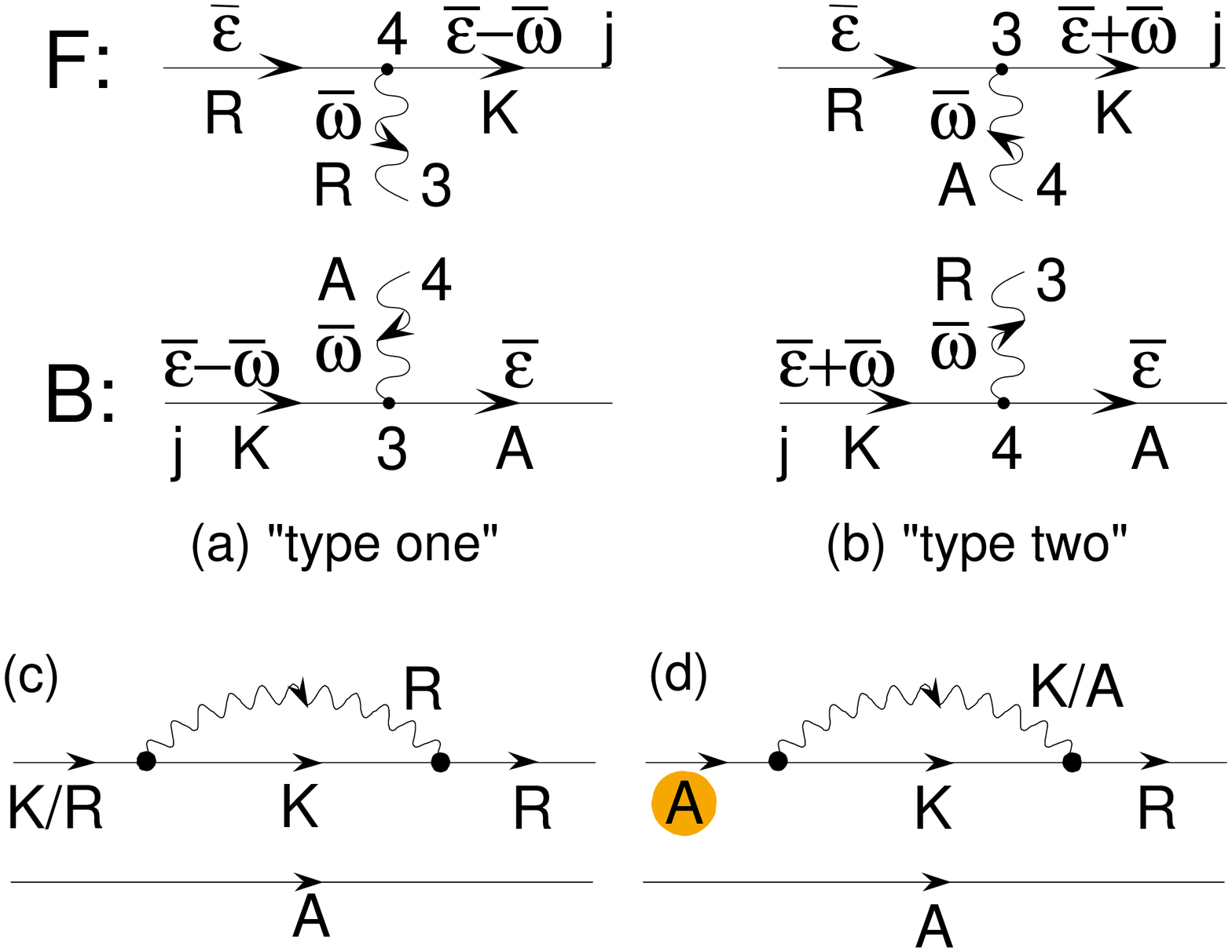}\end{center}
\caption{\label{cap:Standard-Keldysh-rules} (a) Vertices of ``type
  one'' and (b) of ``type two'' arising in Keldysh perturbation
  theory; the accompanying Keldysh Green's functions are $\tilde G^K
  (\ve \mp \wb)$, respectively, producing Pauli factors $\tanh [(\ve
  \mp \wb)/2T] $ that dress the associated interaction propagators
  $\bcL^{R}$ and $\bcL^A$ [\Eq{subeq:generalizedruleofthumbA}].  (c,d)
  Standard Keldysh rules generate diagrams such as both (c) and (d),
  but the latter vanishes upon impurity averaging, since it contains
  $G^{A}$ on the forward propagator. }
\end{figure}

The details of achieving consistency require two types of vertices to
be distinguished (and keeping track of this strictly within the
influence functional approach is the main technical accomplishment of
Refs.~\onlinecite{GZ2,vonDelft04}): For vertices of ``type one''
[Fig.~\ref{cap:Standard-Keldysh-rules}(a)], the
arrows of the $\tilde {\cal L}^{R/A}$ and $\tilde G^K$ correlators
point in the \emph{same} direction (i.e.\ both away from or both
towards the same vertex), in which case Keldysh perturbation theory
produces the combinations:
\begin{subequations}
\label{subeq:generalizedruleofthumbA}
\begin{eqnarray}
  \label{eq:generalizedruleofthumbA}
  \tilde {\cal L}^R_{ 3_a  4_F} (\bomega)
\, \tilde G^K_{j_F  4_F} (\ve - \bomega)
 &  \mapsto &  
 \textrm{th}_- \,  
\tilde {\cal L}^R_{3_a  4_F} (\bomega) \; , 
\\
  \label{eq:generalizedruleofthumbAb}
\quad 
\tilde {\cal L}^A_{3_B  4_{a'}} (\bomega) \, 
\tilde G^K_{3_B j_B} (\ve - \bomega)
 &  \mapsto &  
\textrm{th}_- \, 
 \tilde {\cal L}^A_{3_B 4_{a'}}  (\bomega) \; . \qqph 
\end{eqnarray}
For vertices of ``type two'' 
[Fig.~\ref{cap:Standard-Keldysh-rules}(b)], 
the arrows point in \emph{opposite} directions (one toward, the
other away from the same vertex), which gives the combinations:
\begin{eqnarray}
\label{eq:generalizedruleofthumbBd}
\tilde {\cal L}^A_{3_F 4_{a'}} (\bomega) \, 
\tilde G^K_{ j_F  3_F} (\ve + \bomega)
& \mapsto &
 \textrm{th}_+ \, 
\tilde {\cal L}^A_{3_F 4_{a'}} (\bomega) \; . \qqph 
\\
  \label{eq:generalizedruleofthumbB}
  \tilde {\cal L}^R_{ 3_a  4_B} (\bomega)
\, \tilde G^K_{ 4_B j_B} (\ve + \bomega)
& \mapsto   &  
\textrm{th}_+ \,  
 \tilde {\cal L}^R_{ 3_a  4_B} (\bomega) \; , 
\end{eqnarray}
\end{subequations}
The expressions on the right-hand sides of
\Eqs{subeq:generalizedruleofthumbA} indicate how the $\tilde
L{}^{R/A}_{3_a 4_{a'}}$ propagators in the effective action are to be
dressed by Pauli factors (while $\tilde {\cal L}^{K}_{3_a 4_{a'}}$
remains unmodified).  Thus, the Pauli-modified effective action has the
same form as \Eq{subeq:Sefftwo},
\begin{subequations}
  \label{subeq:SefftwoPP}
\begin{eqnarray} 
  & & 
  \phantom{.} \hspace{-6mm}
  S_\eff^\pp  =  {1 \over 2 \hhbar} \int_{-t/2}^{t/2}dt_{3}\,
  dt_{4}\,\sum_{aa'=F/B}s_{a}s_{a'}
  \left\langle     V_{3 {a}}V_{4 {a'}}\right\rangle^\pp ,
\label{eq:SeffPP}
\qqph 
\\
& & 
\phantom{.} \hspace{-6mm}
\left\langle     V_{3 {a}}V_{4 {a'}}\right\rangle^\pp =   
\int (d \bar k) \, 
e^{i ( \bar q  [r^a (t_3) - r^{a'} (t_4)] - \bomega t_{34})} 
\left\langle     V_{a}V_{ a'}\right\rangle_{\bq \bomega}^\pp ,
\nonumber \\
\label{eq:Seff2PP}
\end{eqnarray} 
\end{subequations}
and the correlators $\left\langle V_{3 {a}}V_{4
    {a'}}\right\rangle^\pp$ have a form similar to
\Eqs{classicalvsquantumfields}, but they are dressed according to the
right-hand sides of \Eqs{subeq:generalizedruleofthumbA}:
\begin{eqnarray}
\nonumber 
\left\langle     V_{F}V_{ F}\right\rangle_{\bq \bomega}^\pp 
& = &
-{\textstyle {\hbaroverstar{2}}} i \Bigl[ 
\bar {\cal L}^{K}_\bbmq (\bomega)  
+ 
\textrm{th}_- \, 
\bar {\cal L}^{R}_\bbmq (\bomega) \, 
+
\textrm{th}_+ \, 
\bar {\cal L}^{A}_\bbmq (\bomega) 
\Bigr] \; ,  
\\
\nonumber 
\left\langle     V_{B}V_{ B}\right\rangle_{\bq \bomega}^\pp 
& = &
- {\textstyle {\hbaroverstar{2}}} i \Bigl[ 
\bar {\cal L}^{K}_\bbmq (\bomega)  
- 
\textrm{th}_+ \, 
\bar {\cal L}^{R}_\bbmq (\bomega) \, 
-
\textrm{th}_- \, 
\bar {\cal L}^{A}_\bbmq (\bomega) \Bigr] \; ,  
\\
\nonumber 
\left\langle     V_{B}V_{ F}\right\rangle_{\bq \bomega}^\pp 
& = &
- {\textstyle {\hbaroverstar{2}}} i \Bigl[ 
\bar {\cal L}^{K}_\bbmq (\bomega)  
+ 
\textrm{th}_- \, 
\bar {\cal L}^{R}_\bbmq (\bomega) \, 
-
\textrm{th}_- \, 
\bar {\cal L}^{A}_\bbmq (\bomega) 
\Bigr] \; ,  
\\
\nonumber 
\left\langle     V_{F}V_{ B}\right\rangle_{\bq \bomega}^\pp 
& = &
- {\textstyle {\hbaroverstar{2}}} i \Bigl[ 
\bar {\cal L}^{K}_\bbmq (\bomega)  
- 
\textrm{th}_+ \, 
\bar {\cal L}^{R}_\bbmq (\bomega) \, 
+
\textrm{th}_+ \, 
\bar {\cal L}^{A}_\bbmq (\bomega) 
\Bigr] \; .  
\\
  \label{subeq:LAA'FT}
\end{eqnarray}
In this way, a single-particle influence functional formalism 
with suitably Pauli-dressed interaction propagators
is able to mimick the essential features of 
the Keldysh many-body formalism.

It should be emphasized that the possibility of using such a dressing
recipe is a feature peculiar to the Keldysh diagrammatics of 
disordered systems, 
as opposed to a generic interacting many-fermion model (see
Fig.~\ref{cap:Standard-Keldysh-rules}): When inserting interaction
lines into the two-particle propagator, standard Keldysh
rules\cite{RammerSmith} in general also lead to contributions like
$G^{K}(\ve-\wb) \bar {\mathcal{L}}^{K}(\wb)$, which would spoil the
above prescription of not dressing $\bar {\mathcal{L}}^K$.  However,
for disordered systems, such diagrams vanish to leading order, since
instead of containing the usual retarded/advanced electron propagators
$G^{R/A}$ on the forward/backward contour [as \eg\ in
Fig.~\ref{cap:Standard-Keldysh-rules}(c)], they contain $G^{A/R}$
there [as \eg\ in Fig.~\ref{cap:Standard-Keldysh-rules}(c)], and after
disorder averaging,
$\left\langle G^{R/A}G^{R/A}\right\rangle _{{\rm imp}}\approx0$.

The decay function $F_d^\pp (t) = \oneoverhbar \langle S_\eff^\pp
\rangle_\rw$, can be calculated in complete analogy to
Sections~\ref{sec:avequantnoise} and \ref{subsec:symCooperondecay}:
Averaging as usual over time-reversed pairs of random walks using
\Eq{exponentCL}, and calculating the effective noise correlator
$\langle V V \rangle_{\qb\wb}^\eff$ occuring  therein using
\Eqs{subeq:generalpropFFBB} and \Eqs{subeq:LAA'FT}, we now find that
the dressed $\tilde {\cal L}^{R/A}$ factors do \emph{not} drop out (in
contrast to what happens in \Eq{replaceSymmetrized} when calculating
$\langle V V\rangle_{\qb\wb}^\sqn$ with undressed
propagators). Instead, the effective noise correlator is now given by
$\langle V V \rangle_{\qb\wb}^\eff \mapsto \langle V V
\rangle_{\qb\wb}^\pp$ of \Eq{subeq:theGreatReplacementa}, and
the decay function (\ref{exponentCL}) takes the form of $F_d^\pp (t)$
of \Eq{exponentCLpaulimodified}.  Thus, \emph{by using dressed propagators
we recover precisely the Pauli-blocked correlator 
$\langle VV \rangle _{\qb\wb}^\pp$ that we had conjectured
on heuristic grounds in Section~\ref{sec:symmPauli-mod}}. 

Further justification for using the latter noise correlator will be
offered by the Bethe-Salpeter analysis of paper II, which leads to a
decay function involving precisely the same Pauli-blocked noise
correlator [cf.\ \EqsII{eq:exponentBetheSalpeter}].  To be somewhat
more precise: the Bethe-Salpeter analysis of paper II improves upon
the analysis of paper I by keeping track of the energy difference
$\omega$ (neglected in paper I, see Ref.~\onlinecite{wneq0}) between
the energies $\ve$ and $\ve - \omega$ of the particle and hole
trajectories, which is why the effective interaction propagator
$\bcL^{\deco}_{{\cal E} \omega, \bbmq} (\bomega)$ of
\EqII{eq:define:Ldeco} depends on $\omega$. The latter propagator can
actually be obtained from \Eq{subeq:generalpropFFBB} for $\langle V V
\rangle_{\qb\wb}^\eff$ by the methods of the present
section, 
namely by setting $\ve \to (\ve - \omega)$ in the Pauli-factors
associated with the backward contour when calculating the dressed the
correlators $\langle V_a V_{a'}\rangle^\pp_{\qb \wb}$, \ie\ in
\Eqs{eq:generalizedruleofthumbAb} and
(\ref{eq:generalizedruleofthumbB}), with corresponding changes in
\Eqs{subeq:LAA'FT}.  However, once the $\omega$-dependence is
retained, the expression for $F_d (t)$ needs another Fourier integral
$ \int d \omega e^{-i \omega t}$, \ie\ \Eqs{exponentCL} or
(\ref{exponentCLpaulimodified}), and our intuitive determination of
the factor $\delta \bP$ therein, become inadequate. A more complete
expression for $F_d (t)$, including this additional integral, is given
by \EqII{eq:C1tauexplicit}.  However, in the long-time limit $\max\{T,
\ve \} t \gg 1$ of present interest, for which $\omega / \max\{ T, \ve
\} \ll 1$, \EqII{eq:C1tauexplicit} reduces to 
\EqII{eq:exponentBetheSalpeter}, which reproduces our
\Eq{exponentCLpaulimodified}.

Finally, having been alerted to the necessity of keeping track of
energy transfers, we note that the vertex diagrams transfer energy
between the forward and backward contours, thereby changing the energy
of both. Strictly speaking, it is thus not fully correct to assign
definite, fixed, energies $\ve$ and $\ve - \omega$ to the upper and
lower contours, since a succession of vertex insertions will produce
an accumulation of frequency transfers. This effect is neglected in
the influence functional approach. This is admissable, essentially
because the dominant contribution of vertex insertions comes from the
infrared regime $\bomega \sim 1/t$, so that the $\bomega$-dependence
in the remaining parts of the diagram may be dropped (they only
generate terms subleading in time).  This point is discussed in more
detail at the end of Sec.~\ref{sec:2ndorderexpansion} of Paper II (and
also in Sec.~B.6.2 of \Ref{vonDelft04}).

\subsection{Importance of recoil}
\label{sec:norecoil}

To conclude this section, let us point out that the effective action
$S_\eff^\GZ$ derived by GZ is essentially the same as our $S_\eff^\pp$
[\Eq{subeq:SefftwoPP}]; the only difference is that in their approach,
the dressed correlators emerge in the position-time representation of
\Eq{classicalvsquantumfields} [as opposed to the $\qb \wb$
representation of our \Eqs{subeq:LAA'FT}], in which each $\tilde {\cal
  L}^{R/A}_{ij}$ is dressed by a Pauli factor $[1 - 2 \tilde \rho
]_{\bar i i}$, or $[1 - 2 \tilde \rho ]_{j \bar j}$, where $\tilde
\rho_{ij} = \langle \psi^\dagger_j (t) \psi_i (t) \rangle$ is the
single-particle density matrix. They wrote their effective action as
$S^\GZ_\eff = i S_R + S_I$, in which $S_R$ contains Pauli factors and
$S_I$ does not.  When averaging over time-reversed pairs of
trajectories, GZ used unrestricted random walks, $F^\GZ_d (t) =
\oneoverhbar \langle S_\eff^\GZ \rangle_\urw$. Moreover, they employed
a position-momentum representation in which they represented the Pauli
factors as $\bigl[1 - 2 f\bigl(\ve(t)\bigr)\bigr]$.

Up to this point, their approach is essentially equivalent to
  ours.  Differences arise at their next step: in order to perform the
  average over paths, GZ assumed the energy of the diffusing electron
  to remain constant along its path \cite{GZexplicitcite}, arguing
  that its collisions with the static impurities are elastic, and
  hence replaced $f\bigl(\ve(t)\bigr)$ by $f (\ve)$. As shown by von
Delft \cite{vonDelft04}, this assumption is equivalent to making the
replacement $\ve \pm \bomega \to \ve$ in the dressed propagators of
our \Eqs{subeq:LAA'FT}, although this is by no means obvious in GZ's
formalism. In other words, they (unwittingly) neglected the $\pm
\bomega$ in the Fermi functions occuring in the Pauli factors, and
thereby neglected the \emph{recoil} experienced by the diffusing
electron upon interacting with its environment and emitting or
absorbing a noise quantum. [The fact that GZ neglect recoil was first
pointed out by Erikson and Hedegard \cite{EH97}.] As a result, GZ's
terms affected by Pauli blocking all cancel each other, mistakenly
causing $\langle i S^R \rangle_\urw$ to vanish, so that the resulting
decay function $F^\GZ_d(t) = \oneoverhbar \langle S_I \rangle_\urw$ is
identical to one for a \emph{single} particle (no Pauli blocking)
under the influence of quantum noise.  Indeed $F^\GZ_d (t)$ can
\cite{GZresultVertex} be brought into the form of our 
$F^\sqn_{d,\urw}   (t)$ [\Eq{exponentCLsym}].
 As discussed in Section~\ref{sec:spontemission},
 the resulting decoherence rate is the same as would be found for the
 physical situation of a single, highly excited electron moving
 through a disordered sample very high above the Fermi surface, or in
 the absence of a Fermi sea, and interacting with a quantum
 environment.  Such an electron can lose its coherence even at zero
 temperature by spontaneous emission, uninhibited by Pauli blocking,
 which is why GZ obtained a nonvanishing decoherence rate at $T=0$.
 Indeed, using $\wbu = 1/ \tauel$ as upper cutoff in \Eq{eq:FdurwT=0}
 for the decoherence rate of a single particle experiencing quantum
 Nyquist noise at $T=0$, we recover precisely GZ's zero-temperature
 decoherence rate for all three values of $d$, including numerical
 prefactors: $\gamma_{\varphi,d}^{\sqn (T=0)} = \gamma^{\GZ
   (T=0)}_{\varphi,d}$ [\ie\ our \Eq{eq:FdurwT=0} reproduces the $T=0$
 values of GZ's Eqs.~(77) and (81) of Ref.~\onlinecite{GZ2}].

\section{Conclusion}
\label{sec:conclusion}

In this paper, we have shown why it is essential to incorporate the
Pauli principle in a description of decoherence by a
quantum-mechanical environment, and how this can be achieved in an
influence functional description of weak localization. We have
explicitly demonstrated how Pauli blocking counteracts the effects of
spontaneous emission to ensure that the decoherence rate vanishes for
sufficiently small temperatures and energies.

At the beginning of this paper, we offered a review of the influence
functional approach for a single electron in a classical environment,
pointing out along the way that quantitative improvements can be
achieved by performing trajectory averages with respect to
\emph{closed} (as opposed to unrestricted) random
walks.\cite{Montambaux04}. We then explained how to extend the
influence functional methodology to the case of a quantum environment,
taking due account of the Pauli principle.

In our approach, a fully quantum-mechanical environment may be treated
in complete analogy to the much simpler case of classical noise. To
this end, the quantum noise spectrum is modified in a well-defined way
that involves Fermi functions, in order to take the place of the
classical noise spectrum in the calculation of the Cooperon decay
function (and the resulting decoherence time). We have shown how this
replacement can be motivated heuristically using a transparent
physical picture for the decay of a superposition of two many-body
states, which shows that the basic idea of the present approach is
general enough to be extended to situations different from weak
localization. In limits where the Pauli principle is ineffective, our
theory reduces to the Feynman-Vernon influence functional, with a
nonvanishing decoherence rate even at zero temperature. In contrast,
for electrons propagating near the Fermi surface Pauli blocking is
very important: it serves to essentially suppress the decohering
effects of quantum fluctuations, confirming the results of Altshuler,
Aaronov and Khmelnitskii, which were derived by keeping only the
classical (thermal) part of the fluctuations.  Moreover, our approach
has also enabled us to calculate quantitatively the leading
corrections to the decoherence rate, and to discuss in detail the
energy-dependence of the decoherence rate (and the Cooperon decay
function), also for energies higher than the temperature.

Paper II will be devoted to substantiating these results with the full
machinery of Keldysh diagrammatic perturbation theory.


\begin{acknowledgments} 
  We thank I. Aleiner, B. Altshuler, M. Vavilov, I. Gornyi, and in
  equal measure D. Golubev and A. Zaikin, for numerous patient and
  constructive discussions.  Moreover, we acknowledge illuminating
  discussions with J. Imry, P. Kopietz, J. Kroha, A.  Mirlin, G.
  Montambaux, H.  Monien, A. Rosch, I.  Smolyarenko, G. Sch\"on,
  P. W\"olfle and A.  Zawadowski.  Finally, we acknowledge the
  hospitality of the centers of theoretical physics in Trieste, Santa
  Barbara, Aspen, Dresden and Cambridge, and of Cornell University,
  where some of this work was performed.  F. M.  acknowledges support
  by a DFG scholarship (MA 2611/1-1), V.A.\ and R.S.\ support by NSF
  grant No. DMR-0242120.

  \emph{Note added:---} After completing the bulk of this work,
  we became aware that ideas similar to those presented in
  Section~\ref{sec:Cooperon-decay-for}  have
  recently been presented in a new book by Akkermans and
  Montambaux\cite{AkkermansMontambaux04}.
\end{acknowledgments}

\appendix

\section{Evaluation of $\cP(z)$}
\label{app:cPd}

In this appendix we derive \Eqs{eq:Fdecayalmostfinal} for $F_d(t)$
and (\ref{subeq:expandPd}) for $\cP_d(t)$, starting from
\Eqs{exponentCL-white}. 

The latter contains the
function $  \delta \tP_d (\tau_{34}, \ttau_{34})$,
defined in \Eq{eq:definedeltatildeP}), which 
may be calculated as follows:
We render the momentum integral in \Eq{eq:definedeltatildeP} Gaussian
using the integral representation
\begin{eqnarray}
  \label{eq:makeintgaussian}
{e^{-\bq^2 D t \tau} \over \bq^2 D} =
{e^{- \bq^2 D t} \over \bq^2 D} -
t \tau \int_{1/\tau}^1 dx \, e^{- \bq^2 D t \tau x} \; ,
\end{eqnarray}
then perform the Gaussian $\int (d \bq)$ integral, and 
finally the auxiliary $\int dx$ integrals, obtaining
(with $L_t = \sqrt{D t}$):
\begin{eqnarray}
\label{eq:P1}
\delta \tP_d (\tau, \ttau)  & = & 
{2^{2-d} \over \pi^{d/2} g_d (L_t) } 
\left\{ \begin{array}{l}
\sqrt{\ttau} - \sqrt{\tau} \rule[-2mm]{0mm}{0mm} \\
\toh \ln (\ttau/ \tau) \rule[-2mm]{0mm}{0mm} \\
{ 1 \over \sqrt{\tau}} - {1\over \sqrt{\ttau}} 
\end{array}
\right\}
\; \textrm{for} \; 
\left\{ 
\begin{array}{l}
d=1 \rule[-2mm]{0mm}{0mm} \\  d = 2 \rule[-2mm]{0mm}{0mm} \\ d = 3
\end{array}
\right\} \! . \nonumber
\\  \label{subeq:Pbar}
\end{eqnarray}
Inserting these expressions into \Eq{exponentCL-white} 
and performing the $d \tilde t_{34}$ integral results in 
\Eq{eq:Fdecayalmostfinal} for
the decay function $F_d(t)$, 
where for closed random walks we obtain
\begin{subequations}
\label{subeq:cPdresults}
\begin{eqnarray}
\cP_1^\crw (z) & = & {{1 \over  \pi^{1/2}}} \left[
 (2z \! - \! 3) \bigl[ (1 \! - \! z)z \bigr]^{1 \over 2} + 
{\textstyle {1 \over 2}} \cos^{-1}{(2z \! - \! 1)} \right] , 
\rule[-2mm]{0mm}{0mm}
\nonumber \\
\cP_2^\crw (z) & = &
{{1 \over  \pi}} \Bigl[ 
 2(z-1) - \ln z \Bigr] , 
    \label{eq:calPcrw}
 \\
\cP_3^\crw (z) & = & {{1 \over  \pi^{3/2}}} \left[
\bigl[ (1 \! - \! z)/z \bigr]^{1 \over 2}
- \cos^{-1}{(2z \! - \! 1)} \right] , \nonumber
\end{eqnarray}
(with $\cos^{-1}(2z-1) \in [\pi, 0]$ for $z \in [0,1]$),
whereas for unrestricted random walks we get
\begin{eqnarray}
\cP_1^\urw (z) & = & {{1 \over  \pi^{1/2}}} \left[
{ \textstyle {8\over 3}}(1 \! - \! z)^{3\over 2} - 4(1 \! - \! z)
z^{1\over 2} \right]  \; , 
\nonumber \\
\cP_2^\urw (z) & = & {{1 \over  \pi}} \left[
(1-z) \Bigl( \ln\bigl[(1-z)/ z\bigr] -1 \Bigr) \right] , \qqph
  \label{subeq:Punrestricted} \\
\cP_3^\urw (z) & = & {{1 \over  \pi^{3/2}}} \left[
(1 \! - \! z)/ z^{1\over 2} - 2 (1-z)^{1\over2} \right] . 
\nonumber 
\end{eqnarray}
\end{subequations}
Expanding $\cP_d(z)$ for small values of its argument
yields \Eqs{subeq:expandPd} of the main text.

\section{Frequency representation for $F_{d,\urw} (t)$}

\label{app:integrals}

In this appendix we discuss the derivation of \Eq{eq:simpleFdurw},
which expresses  $F_{d,\urw} (t)$ in terms of a frequency integral.
 We shall begin, however, more
generally, by considering the case of closed random walks, starting
from \Eq{exponentCL} for $F_{d, \crw} (t) $, with $ \langle VV
\rangle^\eff_{\qb\wb}$ given by \Eq{eq:spectrumfactorizes}. Setting
$t_4 \to - t_4$ in the vertex term of \Eq{eq:barPPPa} for $\delta
\bP^\crw$ yields the combination $[ e^{-i\bar{\omega}t_{34} } -
e^{-i\bar{\omega}\tildet_{34} } ] \bP^\crw_{(0, t)}(\qb , t_{34})$.
Expressing the latter factor through the first line of
\Eqs{eq:rrw-result-explicit}, we obtain
  \begin{eqnarray}
    \label{eq:Fdt3t4first}
    F_{d,\crw}  (t) & = &  
    \int {(d q) (d\bar{q})  (d\bar{\omega}) } \, 
     {\cW_\eff (\wb) \over  \hbar \nu D \bq^2 }{  K_{q \qb \wb}^t \over 
     \tC^0 (0,t)  }  \; , \qqph    
\end{eqnarray}
where we have defined the kernel [with $\Delta = (E_{q-\qb} - E_q )$]
\begin{eqnarray}
\nonumber 
    K_{q \qb  \wb}^t & = & 2 \!
    \int_{-{t \over 2} }^{{t \over 2} }  \!\!
    dt_{3} \int_{-{t \over 2} }^{t_3} \!\! dt_{4} 
    \left[ e^{-i\bar{\omega}t_{34} }  - e^{-i\bar{\omega}\tildet_{34}
      }  \right] e^{- E_q t  - \Delta t_{34}} 
\\
 \label{eq:Ktqqwresults}
& = & 2 e^{- E_q t} 
 \left[ {\Delta t \over \Delta^2 + \wb^2} \left(1  - {\sin (\wb t)
       \over \wb t} \right) 
\right.
\\  \nonumber
& & \qquad \quad \left. + \;  { e^{-t (\Delta + i \wb)} - 1
\over (\Delta + i \wb)^2 } \; - \; 
{  e^{-t \Delta } - e^{i \wb t} 
\over \Delta^2 + \wb^2 } \right] .
\end{eqnarray}
The corresponding expressions for $F_d^\urw(t)$ have the same form,
but [in accord with \Eqs{eq:Frrwurw}] without the $\int (dq)$ integral
and the factor $1/\tC^0$, and with $E_q=0$ in the integrand, so that
$\Delta \mapsto \Delta_0 \equiv D\qb^{2}$, $K_{q \qb \wb}^t / \tC^0
\mapsto K_{0 \qb \wb}^t$.

To proceed further, the \weightingFunction\ $\cW_\eff (\wb)$
has to be specified. For the case of classical white Nyquist noise, where
it is given by $\cW_\class (\wb) = T$, the frequency
integrals can be performed explicitly by contour methods, 
yielding 
\begin{eqnarray}
\nonumber 
  F^\class_{d,\crw} (t) & = &  
  T t \int {(dq) (d\qb) \over \tC^0 (0,t) } { e^{-E_{q}t}  \over 
     \hbar \nu   D \qb^2 }
  \left\{
    1 - \frac{1 - e^{-\Delta t} }
    {\Delta t}\right\} ,  
\\
 \label{subeq:C1s}
\\
F^\class_{d,\urw} (t) & = &  T t 
 \int (d\bar{q})\,\frac{1}{\hbar \nu   D \qb^{2}}\left\{
  1 - \frac{1- e^{-\Delta_0 t}}
{\Delta_0 \, t}\right\} . \label{C1clq=0}
\nonumber 
\end{eqnarray}
For $F^\class_{d,\urw} (t)$, this result stems purely from the first
line of \Eq{eq:Ktqqwresults}, since the two terms in its second line
each (separately) give zero.  [Note that \Eqs{subeq:C1s} are actually
obtained most easily by performing the $\int (d \wb)$ integral in
\Eq{eq:Fdt3t4first} before the time integrals.]  Both expressions for
$F^\class_d (t) $ are free of infrared divergences as $\qb \to 0$ [in
accord with the discussion after \Eq{eq:definedeltatildeP}], but for
$d=2,3$, they are ultraviolet divergent, as discussed at the end of
\Sec{sec:e-averaged}.  For $d=1$, the momentum integrals can be
performed by first rendering them Gaussian, using the integral
representation
$(1-e^{-b})/ b = \int_0^1 dx \, e^{-bx} , $
and performing the auxiliary $dx$ integrals last, whereupon
\Eqs{eq:finalFrrw} are recovered.

For non-trivial choices of $\cW_\eff (\wb)$, such as $\cW_\sqn$ or
$\cW_\pp$ for quantum noise, the $\int (d \wb)$ integral has to be
performed last. To make progress with the momentum integrals, let us
make some simplifications (which the main text manages to avoid):
Firstly, we shall henceforth study only the case of unrestricted
random walks, $F_{d,\urw}(t)$ [whose asymptotic large-$t$ behavior
differs from that of $F_{d,\crw}(t)$ only for $d=1$, as shown in the
main text].  Secondly, we shall retain only the term in the first line
of \Eq{eq:Ktqqwresults} for $K_{0 \qb \wb}^t $, which dominates the
behavior for large $t$, and drop the ``subleading'' terms in the
second line (which yield zero for $\cW_\class$, as mentioned above).
Performing the $\int (d \qb)$ integral in \Eq{eq:Fdt3t4first} using
  \begin{eqnarray}
    \label{eq:definePtilded}
    {2 \over \pi} \int {(d \qb)  \over  \hbar \nu D \bq^2 }
    \left[ { \Delta_0  \over \Delta_0^2 + \wb^2} \right] 
    = p_d \, \wb^{d/2 - 2} \; , 
  \end{eqnarray}
  we obtain \Eq{eq:simpleFdurw} of the main text, with
 $p_1 = \sqrt{ 2 \gamma_1} / \pi$, $p_2 = 1/(g_2 2 \pi)$, $p_3 =
1/(\sqrt {2 \gamma_3} \, \pi^2)$ [for $\gamma_1, g_2, \gamma_3$, see
\Eq{eq:dimensionlessconductance}]. 
%


\begin{thebibliography}{10}
\bibitem{thresholdconstant} In the condition $F_d (\tau_{\varphi,d}) = 1$
    which we use to define $\tau_{\varphi,d}$, our choice of 1 as threshold
    constant is somewhat arbitrary, leading to a corresponding
    arbitrariness in the numerical prefactor of the resulting
    $\tauphi$; this arbitrariness does not, however, affect the
    function $F_d(t)$ itself, or the magnetoconductivity derived
    therefrom.
\bibitem{factorof2} Ref.~\onlinecite{AAK} of AAK is known to contain
  various incorrect factors of two.  They can be eliminated by
  replacing $\eta/2 \to \eta$ in AAK's Eq.~(6), and rederiving all
  subsequent equations. For example, one needs $\toh \omega (t_1 \pm
  t_2) \to \omega (t_1 \pm t_2)$ in Eq.~(9) and $T \to 2T$ in
  Eqs.~(19) and subsequent equations, \eg\ (20), (22), (23), (26) and
  (28), so that their $d=1$ decoherence time ($\tau_N$) and
  magnetoconductivity ($\Delta \sigma$) take the forms given in this
  paper (or by Eqs.~(2.38b) and (2.42c) of Ref.~\onlinecite{AAG},
  respectively).
\bibitem{temp=freq} 
In this paper, temperature is measured in units of
  frequency, \ie\ $T$ stands for $k_{\rm B} T / \hbar$ throughout;
  likewise, although $\ve$ will often be referred to as ``excitation
  energy'', it stands for a frequency.
\bibitem{vonDelft04first} The leading term of 
    \Eq{eq:previewmostimportantresult} has also been derived in
    Ref.~\onlinecite{vonDelft04}.
\bibitem{RPA} Two examples of noise with a Gaussian distribution are
  classical white Nyquist noise, or the Coulomb interaction, treated
  in the random phase approximation.
\bibitem{CSfootnote}See the discussion after Eq. (9.14) of Ref. 
  \cite{CS}; the sign in front of $f$ is wrong.
\bibitem{AAGtypo2} In AAG's Eq.~(4.13b), the factor $\bigl[\ln (T
  \tauH) + 1 \bigr]$ contains a typo: it should be $\bigl[\ln (T
  \tauH) - 1 \bigr]$ (private correspondence with I. Aleiner).
\bibitem{AAGtypo} The result $\tau_{\varphi,1}^\sAAG = 2 / (T
  \sqrt{\gammaone \tauH})$ corresponds to AAG's Eq.~(4.9b) after
  correcting for two typos in \Ref{AAG}: their Eq.~(4.3b) needs an
  extra factor of $1/2$ on the right-hand side, implying that the
  $1/4$ in their Eq.~(4.9b) should be replaced by $1/2$.
\bibitem{GZexplicitcite} 
This assumption is stated in Ref.~\cite{GZ2},
  in the first sentence after Eq.~(68): the Fermi function ``depends
  only on the energy and not on time because the energy is conserved
  along the classical path.''
\bibitem{GZresultVertex} By averaging GZ's decay function $f_d (t)$,
  defined in Eq.~(22) of Ref.~\onlinecite{GZ3}, over time-reversed
  pairs of unrestricted random walks using \Eq{unrestrictedAvg}, it
  can be brought into the form of our $F^\sqn_{d,\urw} (t)$
  [\Eq{exponentCLsym}].  Moreover, setting $F^\sqn_{1,\urw} (\tauphi)
  = 1$ in our \Eq{eq:simpleFdurw}, with $\cW_\eff (\wb)$ replaced by
  $\cW_\sqn(\wb) $ of \Eq{eq:sqnWeighting}, we immediately recover
  GZ's defining relation for $1/\tau_{\varphi,1}$, as given in
  Eq.~(76) of Ref.~\onlinecite{GZ2} (except that they inserted the
  infrared cutoff by hand and determined it selfconsistently, because
  in Ref.~\onlinecite{GZ2} they did not evaluate the vertex
  contributions explicitly; the latter were included fully in
  Ref.~\onlinecite{GZ3}).
\end{thebibliography}
\end{document}